\documentclass[a4paper,11pt]{article}

\usepackage{latexsym}
\usepackage{amsfonts} 
\usepackage[mathscr]{euscript}
\usepackage{amsmath,mathrsfs,bm,amssymb,color}

\title{\textsf{
Quantum Griffiths  inequalities
}}

\date{\empty}
\author{
Tadahiro Miyao\\ 
 {\it Department of Mathematics,}
{\it Hokkaido University,}\\
{\it Sapporo 060-0810, Japan}\\
E-mail:
 miyao@math.sci.hokudai.ac.jp
}

\newcommand{\one}{{\mathchoice {\rm 1\mskip-4mu l} {\rm 1\mskip-4mu l}
{\rm 1\mskip-4.5mu l} {\rm 1\mskip-5mu l}}}
\newcommand{\h}{\mathfrak{H}}
\newcommand{\he}{\mathfrak{H}_{\mathrm{ext}}}

\newcommand{\ex}{\mathrm{e}}

\newcommand{\D}{\mathrm{dom}}

\newcommand{\Fock}{\mathfrak{F}}

\newcommand{\la}{\langle}
\newcommand{\ra}{\rangle}
\newcommand{\Tr}{\mathrm{Tr}}

\newcommand{\slim}{\mbox{$\mathrm{s}$-$\displaystyle\lim_{n\to\infty}$}}

\newcommand{\BbbR}{\mathbb{R}}
\newcommand{\BbbN}{\mathbb{N}}
\newcommand{\BbbZ}{\mathbb{Z}}

\newcommand{\BbbC}{\mathbb{C}}
\newcommand{\vepsilon}{\varepsilon}
\newcommand{\vphi}{\varphi}

\newcommand{\Cone}{\mathfrak{P}}
\newcommand{\Conee}{\mathfrak{P}_{\mathrm{ext}}}

\newcommand{\no}{\nonumber \\}

\newcommand{\bphi}{{\boldsymbol \phi}}
\newcommand{\bomega}{\boldsymbol \omega}

\newcommand{\mb}{\mathbf}

\newcommand{\He}{H_{\mathrm{ext}}}
\newcommand{\The}{\tilde{H}_{\mathrm{ext}}}
\newcommand{\CK}{\mathfrak{K}_{\Lambda, +}}
\newcommand{\Ne}{N_{\mathrm{e}}}
\newcommand{\al}{\mathfrak{A}}

\newcommand{\Br}{\succeq}

\begin{document}

\newtheorem{define}{Definition}[section]
\newtheorem{Thm}[define]{Theorem}
\newtheorem{Prop}[define]{Proposition}
\newtheorem{lemm}[define]{Lemma}
\newtheorem{rem}[define]{Remark}
\newtheorem{assum}{Condition}
\newtheorem{example}{Example}
\newtheorem{coro}[define]{Corollary}

\maketitle

\begin{abstract}
We present a general framework of  Griffiths  inequalities for quantum systems.
Our approach is based on   operator inequalities associated with
 self-dual cones  and provides a  consistent  viewpoint  of the Griffiths
 inequality.
As examples, we discuss the quantum Ising model,    quantum rotor
 model,   Bose--Hubbard model,   and Hubbard model.   
We present a model-independent 
structure that  governs the correlation inequalities. 
\end{abstract}

\section{Introduction}
Ever since its formulation by Lenz \cite{Ising},
the Ising model has been  the most fundamental model to illustrate the phenomenon  of phase 
transitions.
Let $\Lambda$ be a finite subset of $\BbbZ^d$. The system's Hamiltonian is given by the function
\begin{align}
H_{\Lambda} (\sigma)=-\sum_{x,y\in\Lambda}J_{xy} \sigma_x\sigma_y
\end{align} 
for each $ \sigma=\{\sigma_x\}_{x\in \Lambda}\in \{-1, +1\}^{\Lambda}$.
$J_{xy}$ is a non-negative coupling constant. 
The expectation value of  the function $f: \{-1,  +1\}^{\Lambda}\to \BbbR$ is 
\begin{align}
\la f\ra_{\beta}=\sum_{\sigma\in \{-1,
 +1\}^{\Lambda}}f(\sigma)\, \ex^{-\beta H_{\Lambda}(\sigma)}\Big/ Z_{\beta},
\end{align} 
where $Z_{\beta}$ is the normalization constant
$Z_{\beta}=\sum_{\sigma\in \{-1, +1\}^{\Lambda}} \ex^{-\beta H_{\Lambda}(\sigma)}$.
In \cite{Griffiths1}, Griffiths discovered  the following  famous
inequalities\footnote{To be precise, this general formulation was
established  by Kelly
and Sherman \cite{KS}.}:
\begin{itemize}
\item First Griffiths  inequality:
\begin{align}
\la \sigma_A\ra_{\beta} \ge 0
\end{align}  for each
      $A\subseteq \Lambda$, where $\sigma_A=\prod_{x\in A}\sigma_x$.
\item Second Griffiths inequality: 
\begin{align}
\la \sigma_A\sigma_B\ra_{\beta}\ge \la
      \sigma_A\ra_{\beta}\la \sigma_B\ra_{\beta} \label{IsingQGII}
\end{align} 
 for each $A,
      B\subseteq \Lambda$.
\end{itemize} 
Since Griffiths' discovery,  a large number of rigorous studies  on the
Ising ferromagnets  has been successfully  undertaken by applying his inequalities.  
The fact that  Griffiths inequalities are  so useful indicates that 
they  express the  essence of   correlations in the Ising system.
Therefore,  it is natural to ask whether similar inequalities hold true  for
 other models.
Studying this problem means  trying to seek a model-independent or  universal property  of
the notion of correlations.
 Griffiths   inequalities already hold true for some  classical
models, e.g.,  the  plane rotor model. This suggests that our
problem is certainly  meaningful.
Ginibre took the first important step 
toward providing a general framework  for  Griffiths  inequalities \cite{Ginibre}.
However,  we know  of only a few concrete examples of  quantum (i.e., noncommutative) models that  satisfy 
Griffiths  inequalities \cite{CL,DLP, Gallavotti,  KP, Percus}.
Our goals here are as follows:
\begin{itemize} 
\item[(a)] To  present a general method for  constructing  Griffiths
inequalities  for classical and  quantum systems. 
\item[(b)] According to (a),  to highlight a universal property    of correlations. 
\end{itemize} 
To this end, we advance the technique of
 operator inequalities associated with  self-dual cones.

We already know that the quantum Ising
 and  rotor models    satisfy Griffiths
inequalities.
Thus,  these two models  can be regarded as role models for our purpose.
A  standard approach   to proving the Griffiths inequality for these
systems
is to reduce the $d$-dimensional quantum systems to the corresponding $d+1$-dimensional 
classical systems using the Trotter--Kato product formula 
\cite{BG, CI, DLP, KP}.  However,  since known proofs of  the quantum Griffiths
inequalities rely on  the results of  classical systems,
it is difficult to extend these proofs to quantum models that  cannot
 be  reduced to classical ones.
Considering this situation, we  take  the following steps:
\begin{itemize}
\item[(i)] We prove the Griffiths  inequality for the quantum Ising  and  
	   rotor models  using  a method of  operator inequalities and understand 
	   common mathematical structures underlying both models.
\item[(ii)]We  seek similar  structures in other models from  our
	   viewpoint  of operator inequalities and construct the
	   Griffiths  inequality by analogy.  
\end{itemize} 
By carrying out these steps,  we construct quantum Griffiths  inequalities
for  the Bose--Hubbard  and Hubbard models.
We note that the  proposed method can be applied to many  other models, e.g.,
the Su--Schrieffer--Heeger (SSH) model,   Holstein--Hubbard model,  and
Fr\"ohlich model\footnote{The problem of the quantum Heisenberg model
 is still open.}.
Although we   present a few concrete  applications of our results  here, we expect these
 inequalities  to play important roles in   statistical physics just as
 the   original Griffiths  inequalities did for  the Ising system.
We also remark that  some of results  in this paper can be proved by probabilisitc
approaches, e.g., random walk representations. However, we believe that
the 
proposed method can be applicable to a wider class of quantum models and
clarify new aspects of the quantum Griffiths inequality.
Finally,  we emphasize the following:  from the  viewpoint of operator inequalities, we can find a common mathematical 
structure from among the several models mentioned above.
This universal structure enables us to construct the Griffiths
 inequality for   each model.  From this fact, we expect
 to 
 obtain  a model-independent or general  expression of the notion of
  correlation from our viewpoint, see Section
 \ref{CRemark} for details.
\medskip\\

This paper is organized as follows. 
In Section \ref{Sec1}, we introduce  a useful operator inequality induced by
self-dual cones. 
Using this, we develop   a  general theory of the Griffiths
inequality for quantum systems.
In the following sections, 
we will demonstrate  how our  operator inequalities are effective for the study
of  correlation functions for quantum models. 

In Section \ref{GeneralRP}, we reformulate  reflection positivity
from the  viewpoint of our operator inequalities. We then describe how we
construct the Griffiths inequality using  reflection positivity.
 This construction and the one in Section \ref{Sec1} are complementary to
 each other. 

In Sections \ref{Sec2} and \ref{SecQM}, we discuss the quantum Ising  and
 rotor models, respectively. These sections provide  not only  something  of a
warm-up  but also   important clues for finding a common structure
underlying the Griffiths inequality.
Readers can learn how to use the operator inequalities through these
sections as well.

Sections \ref{Sec3} and  \ref{Sec4} are devoted to advanced
applications of the 
abstract  theory established in Sections \ref{Sec1} and \ref{GeneralRP}.
We construct the Griffiths inequality for the Bose--Hubbard
model (Section \ref{Sec3}),  and Hubbard model (Section\ref{Sec4}).
We emphasize that 
our constructions are natural modifications and extensions  of the
methods discussed  in Sections \ref{Sec2} and \ref{SecQM}.

In Section \ref{CRemark}, we present  concluding remarks.
In Appendix  \ref{Appendix}, we  collect useful propositions
concerning our operator inequalities. 
These propositions will be used repeatedly  in this  study. 

\begin{flushleft}
{\bf Acknowledgements:}
This work was supported by KAKENHI(20554421).
I would be grateful to  the anonymous referee for useful comments.
\end{flushleft}

\section{General theory }\label{Sec1}
\setcounter{equation}{0}
\subsection{First inequality}
Let $(\h, \la \cdot|\cdot\ra)$ be a  complex Hilbert space and  $\Cone$ be a convex cone in 
$\h$. The dual cone $\Cone^{\dagger}$ of  $\Cone$ is defined as 
\begin{align}
\Cone^{\dagger}=\big\{x\in \h\, \big|\, \la x| y\ra\ge 0 \ \forall y\in \Cone\big\}.
\end{align} 
We say that $\Cone$ is {\it self-dual} if
\begin{align}
\Cone=\Cone^{\dagger}.
\end{align}  
Henceforth, we always assume that $\Cone$ is self-dual.
 Each element $x$ in $\Cone$ is called positive w.r.t. $\Cone$ and
written as  $x\ge 0$ w.r.t. $\Cone$. 

\begin{define}{\rm 
Let $\mathscr{B}(\h)$ be the set of all bounded linear operators on $\h$.
Let $A\in \mathscr{B}(\h)$. If $Ax \ge 0$ w.r.t. $\Cone$ for all
 $x\in \Cone$,  then we say that  $A$
preserves the positivity w.r.t. $\Cone$ and  write\footnote{This symbol was introduced by
 Miura \cite{Miura}. Bratteli, Kishimoto and Robinson studied the
 commutative cases in \cite{BKR, KR}.}  
\begin{align*}
A\unrhd 0 \ \ \mbox{ w.r.t. $\Cone$.}
\end{align*} 
Note that 
\begin{align}
A\unrhd 0 \ \ \mbox{w.r.t. $\Cone$} \Longrightarrow \la x|Ay\ra\ge 0\ \ \forall x, y\in \Cone.\label{Useful}\ \ \ \ \diamondsuit
\end{align} 
}
\end{define} 

The following proposition is often useful.
\begin{Prop}{\rm \cite{Miura}}\label{Prop1} We have the following:
\begin{itemize}
\item[{\rm (i)}] If  $A\unrhd 0,  B\unrhd 0$ w.r.t. $\Cone$ and
	     $\alpha\ge 0, \beta \ge 0$, then $\alpha A+\beta B\unrhd 0$
	     w.r.t. $\Cone$.
\item[{\rm (ii)}] If $A  \unrhd 0$ and $B \unrhd 0$
	     w.r.t.
 $\Cone$,  then $AB \unrhd 0$ w.r.t. $\Cone$.
\end{itemize} 
\end{Prop}

Our first setting is as follows.
\begin{itemize}
\item[{\bf(A)}] There exists a complete orthonormal system (CONS)
	     $\{e_n\}_{n\in \BbbN}$ of
	   $\h$
such that $e_n\in  \Cone$ for all $n\in \BbbN$.
\end{itemize} 

The system's Hamiltonian  is denoted by $H$. $H$ is self-adjoint and
bounded from below.
To state the first quantum (i.e., noncommutative)  Griffiths  inequality, we need the
following conditions:
\begin{itemize}
\item[{\bf (H. 1)}] $\ex^{-\beta H} \unrhd 0$  w.r.t. $\Cone$ for all
	     $\beta \ge 0$.
\end{itemize}

For each $A\in \mathscr{B}(\h)$, the thermal expectation value of $A$ is
defined as 
\begin{align}
\la A\ra_{\beta}=\Tr\big[
A\, \ex^{-\beta H}
\big]\Big/Z_{\beta},\ \ \ Z_{\beta}
=\Tr\big[
\ex^{-\beta H}
\big].
\end{align} 
\begin{rem}
{\rm 
In this section, we always assume that $\ex^{-\beta H}$ is in the trace class
 for all $\beta >0$. $\diamondsuit$
}
\end{rem} 

Theorem \ref{AbstI} is a prototype of the Griffiths inequality.

\begin{Thm}\label{AbstI}
Assume {\bf (A)}  and {\bf (H. 1)}.
 If $A\unrhd 0$ w.r.t. $\Cone$, then $\la A\ra_{\beta}\ge 0$ for all
 $\beta \ge 0$.
\end{Thm} 
{\it Proof.}
By our assumptions and Proposition \ref{Prop1}, we have   $A \, \ex^{-\beta H}
\unrhd 0$ w.r.t. $\Cone$ for all $\beta \ge 0$. Thus,  applying Proposition \ref{Prop2}, 
we conclude Theorem \ref{AbstI}. $\Box$
\medskip\\

To discuss the case  where $\beta=\infty$, we assume that 
\begin{itemize}
\item[{\bf (A')}] $H$ has a unique ground state,  i.e., $\dim \ker(H-E)=1$,  where $E=\inf \mathrm{spec}(H)$.
\end{itemize} 
Under this condition, we can define the ground state expectation  value as 
\begin{align}
\la A\ra_{\infty}=\la \psi|A\psi\ra,\ \ A\in \mathscr{B}(\h),
\end{align} 
where $\psi$ is the unique ground state of $H$ such that $\|\psi\|=1$.

\begin{Thm}\label{AbstAbsI}
Assume {\bf (A')} and {\bf (H. 1)}.
 If $A\unrhd 0$ w.r.t. $\Cone$, then $\la A\ra_{\infty}\ge 0$.
\end{Thm} 
{\it Proof.} By Proposition \ref{GSP}, we can choose $\psi$ as  $\psi\ge 0$
w.r.t. $\Cone$.
Thus,  this theorem immediately follows from (\ref{Useful}). $\Box$

\begin{rem}
{\rm 
If we assume that $\ex^{-\beta H}$ improves the positivity
 w.r.t. $\Cone$ for all $\beta >0$, then the ground state of $H$ is
 automatically unique, see \ref{UniqG} for details. $\diamondsuit$
}
\end{rem} 

Theorem \ref{PPThm} is a generalization of Theorem \ref{AbstI}.

\begin{Thm}\label{PPThm}
Assume  {\bf (A)} and {\bf (H. 1)}.
Let $A(s)=\ex^{-s H}A\, \ex^{sH}$. If $A_j\unrhd 0$  w.r.t. $\Cone$ for
 all $j=1,\dots, n$, we then have 
\begin{align}
\Bigg\la \prod_{j=1}^{{n}\atop{\longrightarrow}}A_j(s_j)\Bigg\ra_{\beta}
 \ge 0 \label{ExPP2}
\end{align}
for all $0\le s_1\le s_2\le \cdots \le s_n<\beta$,  where $\displaystyle 
\prod_{j=1}^{{n}\atop{\longrightarrow}}O_j=O_1O_2\cdots O_n
$, the ordered product.
\end{Thm} 
{\it Proof.} 
Let $
\displaystyle
\mathcal{S}=\Bigg[\prod_{j=1}^{{n}\atop{\longrightarrow}} A_j(s_j)\Bigg]\,
\ex^{-\beta H}
$.
By our assumptions, we see that 
\begin{align}
\mathcal{S}=\underbrace{\ex^{-s_1 H}}_{\unrhd 0}\underbrace{A_1}_{\unrhd
 0} \underbrace{\ex^{-(s_2-s_1)H}}_{\unrhd 0}\cdots
 \underbrace{A_n}_{\unrhd 0} \underbrace{\ex^{-(\beta-s_n)H}}_{\unrhd 0}
\unrhd 0\ \ \ \mbox{w.r.t. $\mathfrak{P}$.}
\end{align}  
Thus, by Proposition \ref{Prop2}, we obtain (\ref{ExPP2}). $\Box$

\begin{Thm}\label{BetaI}
Assume  {\bf (A')} and {\bf (H. 1)}.
Then (\ref{ExPP2}) holds true at $\beta=\infty$.
\end{Thm}

\subsection{Second inequality }
We consider the extended Hilbert space $\he=\h\otimes \h$. Let $\Conee$
 be a self-dual cone in $\h\otimes \h$.
Instead of {\bf (A)}, we assume the following:

\begin{itemize}
\item[{\bf(B)}] There exists a CONS
	     $\{E_n\}_{n\in \BbbN}$ of
	   $\he$
such that $E_n\in  \Conee$ for all $n\in \BbbN$.
\end{itemize}

To state Theorem \ref{AbstII},  the following condition is assumed:
\begin{itemize}
\item[{\bf (H. 2)}] 
Let $H_{\mathrm{ext}}=H\otimes \one +\one \otimes H$.
Then there exists a unitary operator $\mathscr{U}$ 
 such that 
$\mathscr{U}^*\ex^{-\beta  H_{\mathrm{ext}}}\mathscr{U}\unrhd 0$ w.r.t. $\Conee$ for all
	     $\beta\ge 0$.
\end{itemize}

There are several ways to state the second quantum Griffiths inequality.
First,  we give the following formulation.

\begin{Thm} \label{AbstII}Assume {\bf (B)} and {\bf (H. 2)}.
Let $A, B, C, D\in \mathscr{B}(\h)$ and  $A(s)=\ex^{-sH} A\,  \ex^{s H}$.
Assume  the following:
\begin{itemize}
\item[{\rm (i)}] $ \mathscr{U}^*A\otimes C \mathscr{U}  \unrhd 0$
	     w.r.t. $\Conee$.
\item[{\rm (ii)}] $ \mathscr{U}^*(B\otimes D - D\otimes B)\mathscr{U}
	     \unrhd 0$ w.r.t. $\Conee$.
\end{itemize}
Then we have 
\begin{align}
 \big\la A(s) B(t)\big\ra_{\beta} \big\la C(s)D(t)\big\ra_{\beta} 
-
\big\la A(s)D(t)\big\ra_{\beta} \big\la C(s)B(t)\big\ra_{\beta}
\ge 0 \label{GriIIInq1}
\end{align}
for all  $0\le s \le t <  \beta$.
In addition, assume {\bf (A)} and {\bf (H. 1)}. 
If $A\unrhd 0, B\unrhd 0, C\unrhd 0$ and $D\unrhd 0$ w.r.t. $\Cone$,  we obtain
\begin{align}
\big\la A(s) B(t)\big\ra_{\beta}\ge 0,\ \  \big\la C(s)D(t)\big\ra_{\beta}\ge 0,\ \ 
\big\la A(s)D(t)\big\ra_{\beta} \ge 0,\ \  \big\la
 C(s)B(t)\big\ra_{\beta}\ge 0
\label{GriIIInq2}
\end{align} 
for all  $0\le s \le t <  \beta$.
\end{Thm}
{\it Proof.}
Let
\begin{align}
\la\!\la X\ra\!\ra_{\beta}= \Tr\big[X \, \ex^{-\beta
 H_{\mathrm{ext}}}\big]\Big/ Z_{\beta}^{2}.\label{DoubleExt}
\end{align} 
Then we can derive (\ref{GriIIInq1}) from the following: 
\begin{align}
\Big\la\!\!\Big\la
A(s)\otimes C(s)
\Big(
B(t)\otimes D(t)-D(t)\otimes B(t)
\Big)\Big\ra\!\!\Big\ra_{\beta}\ge 0.
\end{align} 
But this  follows immediately   from Proposition \ref{Prop2} and the fact that  
\begin{align}
&\mathscr{U}^*A(s)\otimes C(s)
\Big(
B(t)\otimes D(t)-D(t)\otimes B(t)
\Big)\, \ex^{-\beta \He}\mathscr{U}
\no
=&
\mathscr{U}^*
\ex^{-s H_{\mathrm{ext}}}
A\otimes C 
\, \ex^{-(t-s)H_{\mathrm{ext}}}
\big(
B\otimes D-D\otimes B
\big) \ex^{-(\beta-t)H_{\mathrm{ext}}}\mathscr{U}
\unrhd 0
\end{align} 
w.r.t. $\mathfrak{P}_{\mathrm{ext}}$ for all $0\le s \le t <  \beta$. 

By {\bf (H. 1)}, it follows that  $ \ex^{-s H}A\,  \ex^{-(t-s) H}\, B \ex^{-(\beta-t) H}  \unrhd 0
 $ w.r.t. $\Cone$ for all $\beta \ge 0$. Thus, by  Proposition
 \ref{Prop2}, we obtain that $\la A(s) B(t)\ra_{\beta} \ge 0$ for all $\beta \ge 0$.
  $\Box$
\medskip\\

\begin{Thm}
If we replace {\rm {\bf (A)}} and {\rm {\bf (B)} } by {\rm {\bf (A')}}
 in Theorem \ref{AbstII}, then (\ref{GriIIInq1}) and (\ref{GriIIInq2})
 hold true at $\beta =\infty$.
\end{Thm} 
{\it Proof.}  Since
$H$
has a unique ground state $\psi$, $H_{\mathrm{ext}}$ has a unique ground
state $\psi\otimes \psi$ as well. By {\bf (H. 2)} and Proposition
\ref{GSP},
 it follows that $\Phi=\mathscr{U}^* \psi\otimes \psi\ge 0$
 w.r.t. $\Cone_{\mathrm{ext}}$. Thus, by (\ref{Useful}), 
\begin{align}
&\Big\la\!\!\Big\la
 \ex^{-s H_{\mathrm{ext}}} A\otimes C
 \ex^{-(t-s)H_{\mathrm{ext}}}
(B\otimes D-D\otimes B) \, \ex^{tH_{\mathrm{ext}}}
\Big\ra\!\!\Big\ra_{\infty}\no
=& \ex^{2(t-s)E}
\Big\la
\underbrace{\Phi}_{\ge 0}\Big| \underbrace{\mathscr{U}^* A\otimes C\ex^{-(t-s)H_{\mathrm{ext}}}
(B\otimes D-D\otimes B) \mathscr{U}}_{\unrhd 0} \underbrace{\Phi}_{\ge 0}
\Big\ra\ge 0,
\end{align}
where $\la\!\la X\ra\!\ra_{\infty}=\la\psi\otimes \psi|X\psi\otimes \psi\ra$.
This completes the proof. $\Box$
\medskip\\

We introduce the Duhamel two-point function,
\begin{align}
(A, B)_{\beta}=Z_{\beta}^{-1} \int_0^1 \Tr \Big[
A\, \ex^{-x\beta H} B\,    \ex^{-(1-x)\beta H} 
\Big]d x,\ \  A, B\in \mathscr{B}(\h). 
\end{align} 

\begin{coro}
Assume {\bf (B)} and {\bf (H. 2)}.
Let $A, B \in \mathscr{B}(\h)$. 
Assume  the following:
\begin{itemize}
\item[{\rm (i)}] $ \mathscr{U}^*A\otimes \one \mathscr{U}  \unrhd 0$
	     w.r.t. $\Conee$.
\item[{\rm (ii)}] $ \mathscr{U}^*(B\otimes \one  - \one\otimes B)\mathscr{U}
	     \unrhd 0$ w.r.t. $\Conee$.
\end{itemize}
Then we have
\begin{align}
(A, B)_{\beta}&-\la A\ra_{\beta}\la B\ra_{\beta}\ge 0,\\
\la A B\ra_{\beta}&-\la A\ra_{\beta}\la B\ra_{\beta}\ge 0.
\end{align} 
In addition, assume {\bf (A)} and {\bf (H. 1)}. 
If $A\unrhd 0$ and $B\unrhd 0$ w.r.t. $\Cone$,   we obtain
\begin{align}
(A,  B)_{\beta}\ge 0,\ \ \la A B\ra_{\beta}\ge 0,\ \  \la A\ra_{\beta}\ge 0,\ \ \la B\ra_{\beta}\ge 0.
\end{align} 
\end{coro}

Our second formulation of the second quantum Griffiths inequality is as follows.

\begin{Thm}\label{AbstIII}
Assume {\bf (B)} and {\bf (H. 2)}.
Let $A, B, C, D\in \mathscr{B}(\h)$ and $A(s)=\ex^{-sH} A\,  \ex^{s H}$.
Assume  the following:
\begin{align}
\mathscr{U}^* \Big(
 A \otimes C -C \otimes A\Big)
\mathscr{U} \unrhd 0,\ \ \
 \mathscr{U}^*\Big(B \otimes D -D \otimes B
\Big) \mathscr{U} \unrhd 0\ \ \ \mbox{w.r.t. $\Conee$}.
\end{align} 
Then we have 
\begin{align}
 \big\la A(s) B(t)\big\ra_{\beta}\big\la C(s) D(t)\big\ra_{\beta}    
-
\big\la A(s) D(t) \big\ra_{\beta} \big\la C(s) B(t)\big\ra_{\beta}
\ge 0 \label{GriIIInq3}
\end{align}
for all $0\le s \le t <  \beta$. 
In addition, assume {\bf (A)} and {\bf (H. 1)}. 
If $A\unrhd 0, B\unrhd 0, C\unrhd 0 $ and $D\unrhd 0$  w.r.t. $\Cone$,  we obtain
\begin{align}
\big\la A(s) B(t)\big\ra_{\beta}\ge 0,\ \  \big\la C(s)D(t)\big\ra_{\beta}\ge 0,\ \ 
\big\la A(s)D(t)\big\ra_{\beta} \ge 0,\ \  \big\la
 C(s)B(t)\big\ra_{\beta}\ge 0
\label{GriIIInq4}
\end{align} 
for all   $0\le s \le t <  \beta$.
\end{Thm} 
{\it Proof.} 
Note that  we can conclude (\ref{GriIIInq3}) from the following: 
\begin{align}
\Big\la\!\!\Big\la
\Big(
A(s)\otimes C(s)-C(s)\otimes A(s)
\Big)
\Big(
B(t)\otimes D(t)-D(t)\otimes B(t)
\Big)\Big\ra\!\!\Big\ra_{\beta}\ge 0.
\end{align} 
To show this, we use  Proposition \ref{Prop2}  and the  fact that 
\begin{align}
&\mathscr{U}^*\Big(
A(s)\otimes C(s)-C(s)\otimes A(s)
\Big)
\Big(
B(t)\otimes D(t)-D(t)\otimes B(t)
\Big)\, \ex^{-\beta \He} \mathscr{U}\no
=&
\mathscr{U}^* \ex^{-sH_{\mathrm{ext}}}
\big(
A\otimes C -C\otimes A
\big)
\, \ex^{-(t-s)H_{\mathrm{ext}}}
\big(
B\otimes D-D\otimes B
\big) \ex^{-(\beta-t)H_{\mathrm{ext}}}\mathscr{U}
\unrhd 0
\end{align} 
w.r.t. $\mathfrak{P}_{\mathrm{ext}}$ for all $0\le s \le t <  \beta$. 
 $\Box$

\begin{Thm}
If we replace {\rm {\bf (A)}} and {\rm {\bf (B)} } by {\rm {\bf (A')}}
 in Theorem \ref{AbstIII}, then (\ref{GriIIInq3}) and (\ref{GriIIInq4})
 hold true at $\beta =\infty$.
\end{Thm} 

\begin{coro}\label{AbstCoroIII}
Assume {\bf (B)} and {\bf (H. 2)}.
Let $A, B\in \mathscr{B}(\h)$. 
Assume  the following:
\begin{align}
\mathscr{U}^* \Big(
 A \otimes \one -\one \otimes A\Big)
\mathscr{U} \unrhd 0,\ \ \
 \mathscr{U}^*\Big(B \otimes \one -\one \otimes B
\Big) \mathscr{U} \unrhd 0\ \ \ \mbox{w.r.t. $\Conee$}.
\end{align} 
Then we have 
\begin{align}
 \big\la A(s) B(t)\big\ra_{\beta}  
-
\big\la A \big\ra_{\beta} \big\la B \big\ra_{\beta}
\ge 0
\end{align}
for all $0\le s \le t <  \beta$.  In particular, we have 
\begin{align}
(A, B)_{\beta}-\la A \ra_{\beta} \la B\ra_{\beta}&\ge 0,\\
\la AB\ra_{\beta}-\la A\ra_{\beta}\la B\ra_{\beta}& \ge 0.
\end{align} 
In addition, assume {\bf (A)} and {\bf (H. 1)}. 
If $A\unrhd 0, B\unrhd 0$  w.r.t. $\Cone$, then we obtain
\begin{align}
(A, B)_{\beta}\ge 0,\ \ \la AB\ra_{\beta}\ge 0,\ \ 
\big\la A\big\ra_{\beta} \ge 0,\ \  \big\la B\big\ra_{\beta}\ge 0.
\end{align} 
\end{coro} 

\subsection{Further generalization}

Theorem \ref{AbstIII} can be generalized as follows.

\begin{Thm}\label{GeneAbstII}Assume {\bf (B)} and {\bf (H. 2)}.
Let $A_j, B_j\in \mathscr{B}(\h),\  j=1,\dots, n$ and $A(s)=\ex^{-sH} A\,  \ex^{s H}$.
Assume  the following:
\begin{align}
\mathscr{U}^* \Big(
 A_j \otimes B_j +\vepsilon_jB_j \otimes A_j \Big) 
 \mathscr{U} \unrhd 0\ \ \ \mbox{w.r.t. $\Conee$}, \ \ j=1,\dots, n,\label{Assump}
\end{align} 
where $\vepsilon_j=1$ or $-1$.
Then we have, for all $0  \le s_1 \le s_2\le \cdots \le s_{n} < \beta$, 
\begin{align}
\sum_{I\subseteq \{1,2,\dots, n\}} \vepsilon_I \la T_I\ra_{\beta}\la
 T_{I^c}\ra_{\beta} \ge 0, \label{GeneralCorrI}
\end{align}
where $I^c=\{1,2,\dots, n\}\backslash I$, $\vepsilon_I=\prod_{j\in I } \vepsilon_j$
  and 
\begin{align}
T_I=\prod_{j=1}^{{n}\atop{\longrightarrow}}T_{j}(s_{j}),
 \ \ \ 
T_j(s_j)=\begin{cases}
A_j(s_j) & j\in I\\
B_j(s_j) & j\in I^c
\end{cases}. 
\end{align} 
In addition, assume {\bf (A)} and {\bf (H. 1)}. 
If $A_{j}\unrhd 0, \ B_j\unrhd 0$  w.r.t. $\Cone$ for all $j=1,\dots, n$,   we obtain
\begin{align}
\big\la T_I\big\ra_{\beta}\ge 0 \label{GeneGri1}
\end{align} 
for all   $0 \le s_1 \le s_2\le \cdots \le s_{n} < \beta$ and $I\subseteq \{1,2,\dots, n\}$.
\end{Thm}

\begin{rem}\label{Equiv1}
{\rm 
Let $\la\!\la \cdot \ra\!\ra_{\beta}$ be defined by
 (\ref{DoubleExt}). Then we obtain (\ref{GeneralCorrI}) from  the
 following:
\begin{align}
\Bigg\la\!\!\!\Bigg\la
\prod_{j=1}^{{n}\atop{\longrightarrow}}
\Big[
A_j(s_j)\otimes B_j(s_j)+\vepsilon_jB_j(s_j)\otimes A_j(s_j)
\Big]
\Bigg\ra\!\!\!\Bigg\ra_{\beta} \ge 0. \label{GeneG}
\end{align} 
Thus (\ref{GeneG}) can be regarded as a generalization of the second
 quantum Griffiths inequality as well.
This expression will be useful in the later sections. $\diamondsuit$

}
\end{rem}

\begin{Thm}
If we replace {\rm {\bf (A)}} and {\rm {\bf (B)} } by {\rm {\bf (A')}}
 in Theorem \ref{GeneAbstII}, then (\ref{GeneralCorrI}) and (\ref{GeneGri1})
 hold true at $\beta =\infty$.
\end{Thm}

\begin{example}
{\rm
When $n=2$, by (\ref{GeneralCorrI}),   we have
\begin{align}
\vepsilon_1 \vepsilon_2 \la A_1 A_2\ra \la B_1 B_2 \ra+\vepsilon_1 \la
 A_1 B_2\ra \la B_1 A_2\ra+\vepsilon_2 \la B_1 A_2\ra \la A_1 B_2\ra
+\la B_1B_2\ra\la A_1A_2\ra \ge 0.
\end{align} 
Here $\la A_1A_2 \cdots A_n\ra$ is an abbreviation of  $\big\la A_1(s_1)
 A_2(s_2)\cdots  A_n(s_n)\big\ra_{\beta}$.
Thus,  if $\vepsilon_1=\vepsilon_2=-1$, we obtain Theorem \ref{AbstIII}. $\diamondsuit$
}
\end{example}

\begin{example}
{\rm
Consider the case where $n=3$ and $B_1=B_2=B_3=\one$. In this case,
 (\ref{GeneralCorrI}) is meaningful only if
 $\vepsilon_1\vepsilon_2\vepsilon_3=1$:
\begin{align}
\la A_1 A_2 A_3\ra+\vepsilon_1\la A_1\ra\la A_2A_3\ra
+\vepsilon_2\la A_2\ra\la A_1A_3\ra
+\vepsilon_3\la A_3\ra\la A_1A_2\ra
\ge 0.
\end{align}
Moreover,  suppose that    assumption (\ref{Assump}) is satisfied for
\begin{align}
(\vepsilon_1, \vepsilon_2, \vepsilon_3)=(1, -1, -1),\  (-1, 1, -1),\  (-1,
 -1, 1),
\end{align}  
then we obtain
\begin{align}
\la A_1 A_2 A_3\ra-\la A_1\ra\la A_2A_3\ra&\ge 0,\\
\la A_1 A_2 A_3\ra-\la A_2\ra\la A_1A_3\ra&\ge 0,\\
\la A_1 A_2 A_3\ra-\la A_3\ra\la A_1A_2\ra&\ge 0,
\end{align} 
which implies that 
\begin{align}
3\la A_1 A_2 A_3\ra-\la A_1\ra\la A_2A_3\ra
-\la A_2\ra\la A_1A_3\ra
-\la A_3\ra\la A_1A_2\ra
\ge 0.\ \ \ \ \diamondsuit
\end{align} 
}
\end{example} 

\begin{example}{\rm
Consider the case where  $n=4, \
 \vepsilon_1\vepsilon_2\vepsilon_3\vepsilon_4=1$,   and
 $B_1=B_2=B_3=B_4=\one$. In this case,  (\ref{GeneralCorrI}) implies  that 
\begin{align}
&\la A_1 A_2 A_3 A_4\ra+\vepsilon_3 \vepsilon_4\la A_1 A_2\ra\la A_3
 A_4\ra+
\vepsilon_2\vepsilon_4\la A_1 A_3\ra \la
 A_2 A_4\ra+
\vepsilon_2 \vepsilon_3\la A_1 A_4\ra \la A_2 A_3\ra\no
\vepsilon_4& \la A_1 A_2 A_3\ra \la A_4\ra+\vepsilon_3\la A_1 A_2 A_4\ra \la A_3\ra+\vepsilon_2\la A_1
 A_3 A_4\ra \la A_2\ra
+\vepsilon_1\la A_2 A_3 A_4\ra \la A_1\ra\no
&\ge 0.
\end{align}
Let $S=\big\{( \vepsilon_1, \vepsilon_2, \vepsilon_3, \vepsilon_4)\in \{\pm
 1\}^4\, |\, \vepsilon_1\vepsilon_2\vepsilon_3\vepsilon_4=1 \big\}$.
If  assumption (\ref{Assump}) holds true for all $(
 \vepsilon_1, \vepsilon_2, \vepsilon_3, \vepsilon_4)\in S
$,  we obtain
\begin{align}
3 \la A_1A_2A_3A_4\ra-\la A_1A_2\ra\la A_3A_4\ra
-\la A_1A_3\ra\la A_2A_4\ra
-\la A_1A_4\ra\la A_2A_3\ra
\ge 0.\ \ \ \diamondsuit
\end{align} 

}
\end{example}

The following theorem offers us a connection between Corollary
\ref{AbstCoroIII} and Theorem \ref{GeneAbstII} (similar arguments can
be found in \cite{Ginibre}):
\begin{Thm}\label{Equiv2}
Assume {\bf (B)} and {\bf (H. 2)}.
Let $A_j \in \mathscr{B}(\h),\  j=1,\dots, n$. 
Assume  that
\begin{align}
\mathscr{U}^* \Big(
 A_j \otimes \one  + \one  \otimes A_j \Big) 
 \mathscr{U} \unrhd 0,\ \ \
\mathscr{U}^* \Big(
A_j \otimes \one  - \one  \otimes A_j \Big) 
 \mathscr{U} \unrhd 0\ \ \
 \mbox{w.r.t. $\Conee$} \label{PAssum}
\end{align} 
for all $j=1, \dots, n$.
For each $I=\{i_1, \dots, i_k\}\subseteq \{1, \dots, n\}$, we set 
\begin{align}
A_I=\prod_{\ell=1}^{{k}\atop{\longrightarrow}}A_{i_{\ell}}.
\end{align} 
Then we obtain
\begin{align}
\la A_I A_K\ra_{\beta}-\la A_I\ra_{\beta}\la A_K\ra_{\beta}\ge 0
\end{align} 
for all $I, K\subseteq \{1, \dots, n\}$.
\end{Thm} 
{\it Proof.}
 For each $\vepsilon\in \{\pm 1\}$,  define
$
B_j^{(\vepsilon)}=\frac{1}{2}(A_j\otimes \one +\vepsilon \one \otimes A_j)
$. By (\ref{PAssum}), we have $B_j^{(\vepsilon)}\unrhd 0$
w.r.t. $\Cone_{\mathrm{ext}}$ for all $j=1, \dots, n$. Since 
$
A_j\otimes \one =B_j^{(+)}+B_j^{(-)}
$
and $\one \otimes A_j=B_j^{(+)}-B_j^{(-)}$, we see that 
\begin{align}
A_I\otimes \one -\one \otimes A_I
&=\prod_{\ell=1}^{{k}\atop{\longrightarrow}}
\Big[
B_{i_{\ell}}^{(+)}+B_{i_{\ell}}^{(-)}
\Big]
-
\prod_{\ell=1}^{{k}\atop{\longrightarrow}}
\Big[
B_{i_{\ell}}^{(+)}-B_{i_{\ell}}^{(-)}
\Big]\no
&=\sum_{\vepsilon_1,\dots, \vepsilon_k\in \{\pm 1\}} C_{\vepsilon_1,
 \dots, \vepsilon_k} B_{i_1}^{(\vepsilon_{1})}\cdots
 B_{i_k}^{(\vepsilon_k)},
\label{Difference}
\end{align} 
where $
C_{\vepsilon_1,
 \dots, \vepsilon_k} \ge 0
$
for all $
\vepsilon_1,\dots, \vepsilon_k\in \{\pm 1\}
$.
 Thus,  the RHS of (\ref{Difference})$\unrhd 0$ w.r.t. $\Conee$.
Similarly,  $A_K\otimes \one -\one \otimes A_K \unrhd 0$ w.r.t. $\Conee$.
By applying Corollary \ref{AbstCoroIII}, we obtain the result. $\Box$

\section{Reflection positivity}\label{GeneralRP}
\setcounter{equation}{0}

In Section \ref{Sec1}, we give a general framework of the
Griffiths inequality. In our proofs,   assumptions {\bf (A)}  and   {\bf (B)}
 are   basic inputs. Unfortunately,  these  assumptions  are not satisfied  in
 several models. To overcome this situation, we employ  the 
 concept of reflection positivity.
As we indicated in  \cite{Miyao2},    reflection positivity can be considered  an
 operator inequality associated  with a special self-dual cone.
This viewpoint makes it possible to visualize a common mathematical
structure among various quantum models.
Reflection positivity originates from  axiomatic quantum field
theory  \cite{OS}. Glimm, Jaffe, and Spencer first applied reflection
positivity to the rigorous study  of the phase transition \cite{GJS}. 
This idea was successfully  further developed by Dyson,  Fr\"ohlich, Israel,
Lieb,  Simon,  \cite{DLS,FILS,FSS} and many others.   Lieb also  discovered a crucial application of  
reflection positivity to many-electron systems, called the spin
reflection positivity \cite{Lieb}. Recently,  Jaffe and Pedrocchi  studied the
topological order by  reflection positivity \cite{JaffeP1, JaffeP2}.

For each $p\in \BbbN$, we denote the trace ideal by   $\mathscr{L}^p(\h)$, which is  defined as 
\begin{align}
\mathscr{L}^p(\h)=\big\{\xi\in \mathscr{B}(\h)\, \big|\, \Tr[|\xi|^p]<\infty\big\}.
\end{align} 
$\mathscr{L}^1(\h)$  is called the trace class, while  $\mathscr{L}^2(\h)$ is called the Hilbert--Schmidt class. 
$\mathscr{L}^2(\h)$ becomes a Hilbert space if  we define the inner product
 as  $\la \eta|\xi\ra_{\mathscr{L}^2}=\Tr[\eta^* \xi]$ for all $\eta,\xi\in \mathscr{L}^2(\h)$.

\begin{define}[Bounded operators]\label{MultDef1}{\rm
Let $A\in \mathscr{B}(\h)$.
\begin{itemize}
\item[{\rm (i)}] The left multiplication operator $\mathcal{L}(A)$ is
	     defined as  $\mathcal{L}(A)\xi=A\xi $ for all $\xi\in
	     \mathscr{L}^2(\h)$.
\item[{\rm (ii)}] The right multiplication operator $\mathcal{R}(A)$ is
	     defined as $\mathcal{R}(A)\xi=\xi A $ for all $\xi\in
	     \mathscr{L}^2(\h)$.  $\diamondsuit$
\end{itemize} 

}
\end{define} 
\begin{rem}
{\rm 
\begin{itemize}
\item[(i)] $\mathcal{L}(A), \mathcal{R}(A)\in \mathscr{B}(\mathscr{L}^2(\h))$, the set of all bounded
 operators on $\mathscr{L}^2(\h)$. 

\item[(ii)] $\mathcal{L}(A) \mathcal{L}(B)=\mathcal{L}(AB)$.
\item[(iii)] $\mathcal{R}(A)\mathcal{R}(B)=\mathcal{R}(BA)$. $\diamondsuit$
\end{itemize} 
}
\end{rem} 

 Let $\vartheta$ be an
antilinear involution on $\h$. Let $\Phi_{\vartheta}$ be an isometric
isomorphism  from $\mathscr{L}^2(\h)$ onto $\h\otimes \h$ defined by
\begin{align}
\Phi_{\vartheta}(|x\ra\la y|)=x\otimes \vartheta y\ \ \ \forall x,y\in \h.
\end{align} 
We have the relations
\begin{align}
\mathcal{L}(A) =\Phi_{\vartheta}^{-1} A\otimes \one\Phi_{\vartheta} ,\ \
 \ \mathcal{R}(\vartheta
 A^*\vartheta)=\Phi_{\vartheta}^{-1} \one \otimes A\Phi_{\vartheta} \label{Bare}
\end{align} 
for each $A\in \mathscr{B}(\h)$. We simply write these facts as 
\begin{align}
\h\otimes \h=\mathscr{L}^2(\h),\ \ A\otimes \one =\mathcal{L}(A),\ \ \one \otimes
 A=\mathcal{R}(\vartheta A^*\vartheta), \label{Ident}
\end{align} 
if no confusion arises.

Definition \ref{MultDef1} can be extended to unbounded operators by
(\ref{Bare}) as follows.
\begin{define}[Unbounded operators]{\rm
Let $A$ be  a densely defined closed operator on $\h$.
\begin{itemize}
\item[{\rm (i)}] The left multiplication operator $\mathcal{L}(A)$ is defined as 
$
\mathcal{L}(A) =\Phi_{\vartheta}^{-1} A\otimes \one\Phi_{\vartheta}
$.
\item[{\rm (ii)}]
The right multiplication operator $\mathcal{R}(A)$ is defined as 
$
\mathcal{R}(A) =\Phi_{\vartheta}^{-1} \one\otimes \vartheta A^* \vartheta \Phi_{\vartheta}
$. $\diamondsuit$
\end{itemize} 
}
\end{define} 
\begin{rem}{\rm
\begin{itemize}
\item[{\rm (i)}]Both $\mathcal{L}(A)$ and $\mathcal{R}(A)$ are closed
 operators on $\mathscr{L}^2(\h)$. 
\item[{\rm (ii)}]If $A$ is self-adjoint,    so are
 $\mathcal{L}(A)$ and $\mathcal{R}(A)$. 
\item[{\rm (iii)}] We will also use the conventional identification
	     (\ref{Ident}). $\diamondsuit$ 
\end{itemize} 
}
\end{rem}

\begin{define}
{\rm
A canonical cone in $\mathscr{L}^2(\h)$ is defined by 
\begin{align}
\mathscr{L}^2(\h)_+=\big\{ \xi\in \mathscr{L}^2(\h)\, \big|\, \xi\ge 0\ \mbox{as a linear
 operator in $\h$}\big\}.
\end{align} 
$\mathscr{L}^2(\h)_+$ is self-dual. $\diamondsuit$
}
\end{define}

 The following proposition is often useful.

\begin{Prop}
For each $A\in \mathscr{B}(\h)$, we have 
$
\mathcal{L}(A)\mathcal{R}(A^*) \unrhd 0
$ w.r.t. $\mathscr{L}^2(\h)_+$.
\end{Prop} 
{\it Proof.} For each $\xi\in \mathscr{L}^2(\h)_+$, we  can see that 
$\mathcal{L}(A)\mathcal{R}(A^*)\xi =A \xi A^* \ge 0$. $\Box$

\begin{define}
{\rm
We define
\begin{align}
\mathfrak{A}=\mathrm{Coni}\Big\{
\mathcal{L}(A) \mathcal{R}(A^*)\in \mathscr{B}(\mathscr{L}^2(\h))\, \Big|\, A\in \mathscr{B}(\h)
\Big\}^{\mbox{---$\mathrm{w}$}},
\end{align} 
where 
$\mathrm{Coni}(X)$ is the conical hull of   $X$ and 
 $S^{\mbox{--$\mathrm{w}$}}$ represents the closure of $S$ under a  weak
 topology in $\mathscr{B}(\mathscr{L}^{2}(\h))$.

If $A\in \mathfrak{A}$, then we write $A\Br 0$
 w.r.t. $\mathscr{L}^2(\h)_+$. 
$\diamondsuit$
}
\end{define} 
\begin{rem}\label{InqRem}{\rm
\begin{itemize}
\item[{\rm (i)}]$
A\Br 0 \Longrightarrow A\unrhd 0.
$\footnote{From this fact, we understand that  reflection positivity
	     is closely related to  the notion of positivity preservation
	     discussed in  Section \ref{Sec1}.}
\item[{\rm (ii)}] $A\Br 0, B\Br 0, a,b \ge 0 \Longrightarrow aA+bB \Br
	     0$.
\item[{\rm (iii)}] $A\Br  0, B\Br  0\Longrightarrow AB\Br  0$. \ $\diamondsuit$
\end{itemize}
}
\end{rem}

The following proposition is a guiding principle of reflection
positivity \cite{DLS, FILS, Miyao2}.
The point is that  assumptions {\bf (A)} and  {\bf (B)} are  unnecessary.
\begin{Prop}[Reflection positivity]\label{PPImplyP}
Assume that $A$ is a  trace class operator on  $\mathscr{L}^2(\h)$, i.e., $A\in
 \mathscr{L}^1(\mathscr{L}^2(\h))$. If $A \Br  0$
 w.r.t. $\mathscr{L}^2(\h)_+$, then  we have $\Tr_{\mathscr{L}^2}[A] \ge 0$.
\end{Prop} 
{\it Proof.} It suffices to consider the case where 
$A=\sum_{j=1}^N\mathcal{L}(a_j)\mathcal{R}(a_j^*),\ N\in \BbbN$.
In this case, we can easily  see that 
$\Tr_{\mathscr{L}^2}[A]=\sum_{j=1}^N|\Tr_{\mathfrak{H}}[a_j]|^2\ge 0$. $\Box$
\medskip\\

As before, the system's Hamiltonian $H$ is a self-adjoint
operator acting  in
$\mathscr{L}^2(\h)$ and 
bounded from below. In this section, we continue to  assume that $\ex^{-\beta
H}$ is a  trace class operator for all $\beta >0$.
Corresponding to {\bf (H. 1)}, we need the following condition: 
\begin{flushleft}
{\bf (H. 3)} $\ex^{-\beta H}\Br 0$ w.r.t. $\mathscr{L}^2(\h)_+$ for all
 $\beta \ge 0$.
\end{flushleft} 

Let  $\la \cdot \ra_{\beta}$ be  the thermal average.
Theorem \ref{Prototype2} is another prototype  of the Griffiths inequality.

\begin{Thm}\label{Prototype2}
Assume {\bf (H. 3)}. If $A\Br 0$ w.r.t. $\mathscr{L}^2(\h)_+$, then 
$\la A \ra_{\beta} \ge 0$ for all $\beta \ge 0$.
\end{Thm} 
{\it Proof.} From Remark \ref{InqRem} (iii), we have 
$A \, \ex^{-\beta H} \Br 0$ w.r.t. $\mathscr{L}^2(\h)_+$ for all $\beta
\ge 0$. Thus by Proposition \ref{PPImplyP}, we conclude the theorem. $\Box$
\medskip

Theorem \ref{Prototype2} can be generalized as follows.

\begin{Thm}\label{BasicPPP}
Assume {\bf (H. 3)}. 
If $A_j\Br 0$  w.r.t. $\mathscr{L}^2(\h)_+$ for
 all $j=1,\dots, n$, then we  have 
\begin{align}
\Bigg\la \prod_{j=1}^{{n}\atop{\longrightarrow}}A_j(s_j)\Bigg\ra_{\beta}
 \ge 0\label{TimeOP}
\end{align}
for all $0\le s_1\le s_2\le \cdots \le s_n<\beta$. 
\end{Thm} 
{\it Proof.} Since 
\begin{align}
\Bigg[\prod_{j=1}^{{n}\atop{\longrightarrow}}A_j(s_j)\Bigg]\,  \ex^{-\beta H}=
\underbrace{\ex^{-s_1 H}}_{\Br 0} \underbrace{ A_1}_{\Br 0}
 \underbrace{\ex^{-(s_2-s_1)H}}_{\Br 0} \cdots \underbrace{A_n}_{\Br 0}
 \underbrace{ \ex^{-(\beta-s_n)H}}_{\Br 0}
\Br 0
\end{align} 
w.r.t. $\mathscr{L}^2(\h)_+$, we obtain (\ref{TimeOP}) by Proposition \ref{PPImplyP}.
 $\Box$
 
\begin{Thm}\label{BasicPPPG}
Assume  {\rm {\bf (A')}}. Then Theorem \ref{BasicPPP} holds true at $\beta =\infty$.
\end{Thm} 
{\it Proof.} Considering Remark \ref{InqRem} (i), we know that  Theorem \ref{BasicPPPG} follows
from  Theorem \ref{BetaI}. $\Box$

\section{ Quantum Ising model}\label{Sec2}
\setcounter{equation}{0}

\subsection{Results}
Let $\Lambda$ be a finite subset of $\BbbR^d$.
The Hamiltonian of the quantum Ising model is  given by 
\begin{align}
H_{\Lambda}=-\sum_{x, y\in \Lambda} J_{xy} \sigma_x^{(3)}\sigma_y^{(3)}
-\sum_{x\in \Lambda}\mu_x \sigma_x^{(3)}
-\sum_{x\in
 \Lambda} \lambda_x \sigma_x^{(1)}.
\end{align} 
$\sigma^{(1)}, \sigma^{(2)}$,  and $\sigma^{(3)}$ are the Pauli matrices:
\begin{align}
\sigma^{(1)}=\left(\begin{array}{ll}
0 & 1\\
1 & 0
\end{array} 
\right),\ \ \ 
\sigma^{(2)}=\left(\begin{array}{ll}
0 & -i\\
i & 0
\end{array} 
\right),\ \ \ 
\sigma^{(3)}
=\left(\begin{array}{ll}
1 & 0\\
0 & -1
\end{array} 
\right).
\end{align} 
$H_{\Lambda}$ acts in the Hilbert space
$
\mathfrak{H}_{\Lambda}=\otimes_{x\in \Lambda} \BbbC^2.
$
$(J_{xy})_{x, y\in \BbbZ^d}$ is a family of coupling constants, and
$\mu_x, \lambda_x\in \BbbR$ are the magnetic fields.
In this section, we always assume the following:
\begin{itemize}
\item[{\bf (J)}] \hspace{2cm}$J_{xy}\ge 0,\ \ J_{xy}=J_{yx},\ \ J_{xx}=0. $
\end{itemize} 

The thermal average is defined by 
\begin{align}
\la A\ra_{\beta}= \Tr\big[A\, \ex^{-\beta
 H_{\Lambda}}\big]\Big/Z_{\beta},\ \ Z_{\beta}=\Tr\big[
\ex^{-\beta H_{\Lambda}}
\big].
\end{align}

Let
\begin{align}
\tau_x=\frac{1}{2}(\one+\sigma_x^{(1)}).
\end{align} 
Set 
\begin{align}
S_x^{(\vepsilon)}=
\begin{cases}
\tau_x & \mbox{ if $\vepsilon=1$}\\
\sigma_x^{(3)} & \mbox{ if $\vepsilon=3$}
\end{cases}.
\end{align} 
We define
\begin{align}
\mathfrak{A}=\mathrm{Coni}
\Big\{
S_{x_1}^{(\vepsilon_1)}\cdots S_{x_n}^{(\vepsilon_n)}\, \Big|\, x_1,\dots,
 x_n\in \Lambda,\ \vepsilon_1,\dots, \vepsilon_n\in \{1,3\}, n\in \BbbN
\Big\},
\end{align} 
where $\mathrm{Coni}(S)$ is the conical hull of $S$.

\begin{Thm}[First Griffiths inequality]\label{G1}
Assume {\bf (J)}.
Assume that $\mu_x\ge 0$ for all $x\in \Lambda$.
 For all $A_1, \dots, A_n\in \mathfrak{A}, \lambda_x\in \BbbR$
 and $0\le s_1\le \cdots \le s_n \le \beta$, we have 
\begin{align}
\Bigg\la \prod_{j=1}^{{n\atop{\longrightarrow}}}A_j(s_j)\Bigg\ra_{\beta}
 \ge 0.
\end{align} 
\end{Thm}

For each $A\subseteq \Lambda$, set
\begin{align}
\sigma_A^{(3)}=\prod_{x\in A} \sigma_x^{(3)},\ \ \ \tau_A=\prod_{x\in A} \tau_x.
\end{align} 
To state the second Griffiths inequality, we introduce the following notations:
\begin{align}
\la \!\la X\ra\!\ra_{\beta}&=\Tr_{\mathfrak{H}\otimes \mathfrak{H}}\Big[
X\, \ex^{-\beta H_{\mathrm{ext}}}
\Big]\Big/ Z_{\beta}^2,\\
 H_{\mathrm{ext}}&=H_{\Lambda}\otimes \one +\one \otimes H_{\Lambda}. \label{IHamiExt}
\end{align}

\begin{Thm}[Second Griffiths inequality]\label{G2}
Assume {\bf (J)}.
Assume that $\mu_x\ge 0, \lambda_x\ge 0$ for all $x\in \Lambda$.
For all $A, B, C, D\subseteq \Lambda$ and $\beta\ge 0$,  we have
\begin{align}
\Big\la\!\!\Big\la 
\Big(
\sigma_A^{(3)}(s)
\otimes \tau_C(s)
-\tau_C(s)\otimes \sigma_A^{(3)}(s)
\Big)
 \Big(
\sigma_B^{(3)}(t)
 \otimes \tau_D(t) -  \tau_D(t)\otimes \sigma_B^{(3)}(t)
\Big)
\Big\ra\!\!\Big\ra_{\beta}
\ge 0\label{QIdouble}
\end{align} 
for all $0 \le s \le t \le \beta$, where $
\sigma_A^{(3)}(t)=\ex^{-t H_{\Lambda}} \sigma_A^{(3)} \ex^{ t H_{\Lambda}}
$ and $\tau_B(t)=
\ex^{-t H_{\Lambda}}\,  \tau_B\,  \ex^{ t H_{\Lambda}}
$.  
\end{Thm} 

\begin{rem}
{\rm 
(\ref{QIdouble}) can be expressed  as follows:
\begin{align}
\Big\la \sigma_A^{(3)}(s) \sigma_B^{(3)}(t)\Big\ra_{\beta}\Big\la
 \tau_C(s) \tau_D(t)\Big\ra_{\beta} -  \Big\la\sigma_A^{(3)}(s)
 \tau_D(t)\Big\ra_{\beta}\Big\la \tau_C(s)
 \sigma_B^{(3)}(t)\Big\ra_{\beta}
\ge 0.\ \ \  \ \diamondsuit \label{QIII}
\end{align} 
}
\end{rem} 

From this theorem (or (\ref{QIII})), we can derive the well-known formula.
\begin{coro}Under the same assumptions as Theorem \ref{G2}, we have 
\begin{align}
\Big\la \sigma_A^{(3)} \sigma_B^{(3)}\Big\ra_{\beta}-
\Big\la \sigma_A^{(3)} \Big\ra_{\beta}\Big\la  \sigma_B^{(3)}\Big\ra_{\beta}\ge 0,\ \ \ 
\big\la \tau_A \tau_B\big\ra_{\beta}-
\big\la \tau_A \big\ra_{\beta}\big\la  \tau_B\big\ra_{\beta}\ge 0.
\end{align} 
\end{coro}

The following theorem is an extension of Theorem \ref{G2}.

\begin{Thm}\label{IsingGGriII}
Assume {\bf (J)}.
Assume that $\mu_x\ge 0, \lambda_x\ge 0$ for all $x\in \Lambda$.
Let $A_1, \dots, A_{n}, B_1, \dots, B_{n}\subseteq \Lambda$.
Then, for all $0 \le t_1\le t_2\le
 \cdots \le t_{n} \le \beta$,  we have
\begin{align}
\Bigg\la\! \!\!\Bigg\la
\prod_{j=1}^{{n}\atop{\longrightarrow}}
\Big(
\sigma_{A_j}^{(3)}(t_j)\otimes \tau_{B_j}(t_j)
-\tau_{B_j}(t_j)\otimes \sigma_{A_j}^{(3)}(t_j)
\Big)
\Bigg\ra\!\!\!\Bigg\ra_{\beta}\ge 0.
\end{align} 
\end{Thm} 

By Theorem \ref{GeneAbstII}, we obtain the following corollary.

\begin{coro}\label{IsingGGriIICoro}
Assume {\bf (J)}.
Assume that $\mu_x\ge 0, \lambda_x\ge 0$ for all $x\in \Lambda$.
Let $A_1, \dots, A_{n}, B_1, \dots, B_{n}\subseteq \Lambda$.
For each $I=\{i_1, \dots, i_k\}\subseteq \{1,\dots, n\}$ with 
$i_1 <i_2<\cdots <i_k$,
 we define 
\begin{align}
S_I(\mb{t})=\prod_{j=1}^{{k}\atop{\longrightarrow}}
 S_j(t_{j}),\ \ \ S_j(t_j)=\begin{cases}
\sigma_{A_j}^{(3)}(t_j)\ \  &\mbox{if $j\in I$}\\
\tau_{B_j}(t_j)\ \  & \mbox{if $j\in I^c$}
\end{cases}.
\end{align} 
Then we have, for all $0 \le t_1\le t_2\le
 \cdots \le t_{n} \le \beta$, 
\begin{align}
\sum_{I\subseteq \{1,2,\dots, n\}}
(-1)^{|I|}
\Big\la S_I(\mb{t})\Big\ra_{\beta}
\Big\la S_{I^c}(\mb{t})\Big\ra_{\beta} \ge 0.
\end{align} 
In addition, we have 
\begin{align}
\Big\la S_I(\mb{t})\Big\ra_{\beta} \ge 0
\end{align} 
 for all $0 \le t_1\le t_2\le
 \cdots \le t_{n} \le \beta$ and $I\subseteq \Lambda$.
\end{coro} 

\begin{example}\label{IsingEx}
{\rm We have the following:
\begin{itemize}
\item[(i)] $\la \sigma_A^{(3)}\ra_{\beta}$ is monotonically increasing
	   in $J_{xy}$ and $\mu_x$.
\item[(ii)] $\la \tau_A\ra_{\beta}$ is monotonically decreasing in
	    $J_{xy}$ and $\mu_x$.
\item[(iii)] $\la \sigma_A^{(3)}\ra_{\beta}$ is monotonically decreasing
	   in  $\lambda_x$.
\item[(iv)] $\la \tau_A\ra_{\beta}$ is monotonically increasing in
	    $\lambda_x$.
\end{itemize} 
We will prove  this example in Section \ref{IsingExPr}. $\diamondsuit$ 

}
\end{example} 

\begin{rem}
{\rm
\begin{itemize}
\item[{\rm (i)}]
Our results can be extended to  a more general Hamiltonian of the form
\begin{align}
H_{\Lambda}=-\sum_{A\subseteq \Lambda} J_A \sigma_A^{(3)}-\sum_{A\subseteq \Lambda} K_A \tau_A
\end{align} 
with $J_A \ge 0$ and $K_A\ge 0$. 
\item[{\rm (ii)}] Assume that $\mu_x>0$ or $\lambda_x>0$ for all $x\in \Lambda$. Then since the ground state of $H_{\Lambda}$ is unique for
	     all $\Lambda$\footnote{This fact can be proven by the
	     Perron--Frobenius--Faris theorem \cite{Faris}.}, our results
	     are valid at $\beta =\infty$.
The results at $\beta =\infty$ are used in the study of  quantum
	     phase transitions  \cite{CI, DLP}. 
 $\diamondsuit$
\end{itemize} 
}
\end{rem} 
\newpage

\subsection{Proof of Theorem \ref{G1}}\label{Sub1}

Let $\Omega=\{-1, +1\}$ be the set of possible  values of a spin.
Given $\Lambda$, $\Omega^{\Lambda}$ is the set of spin  configurations in
$\Lambda$. Set
\begin{align}
|+1\ra=
\begin{pmatrix}
1\\
0
\end{pmatrix}
,\ \ \ 
|-1\ra=
\begin{pmatrix}
0\\
1
\end{pmatrix}.
\end{align} 

For each $\bomega=\{\omega_x\}_{x\in \Lambda}\in \Omega^{\Lambda}$, 
we define 
\begin{align}
|\bomega\ra=\bigotimes_{x\in \Lambda} |\omega_x\ra.\label{CONS1}
\end{align} 
Then $\big\{|\bomega\ra\, |\, \bomega\in \Omega^{\Lambda}\big\}$ is a CONS of
$\mathfrak{H}_{\Lambda}$.
\begin{define}{\rm 
A standard self-dual cone in $\mathfrak{H}_{\Lambda}$ is defined by 
\begin{align}
\mathfrak{H}_{\Lambda, +}=\Bigg\{
\Psi\in\mathfrak{H}_{\Lambda}\, \Bigg|\, \Psi=\sum_{\bomega\in
 \Omega^{\Lambda}}
C_{\bomega} |\bomega\ra,\ \ C_{\bomega}\ge 0\ \forall \bomega\in \Omega^{\Lambda}
\Bigg\}.\ \ \ \ \diamondsuit
\end{align}  
}
\end{define} 
\begin{rem}
{\rm 
$|\bomega\ra\in \h_{\Lambda, +}$ for all $\bomega\in \Omega^{\Lambda}$. $\diamondsuit$
}
\end{rem} 

Let $U$ be a unitary operator\footnote{This unitary operator is 
well-known \cite{KT, Miyao1}.} on $\mathfrak{H}_{\Lambda}$ given by 
\begin{align}
U=\bigotimes_{x\in \Lambda}u,\ \ \ \ 
u=\frac{1}{\sqrt{2}}\left(\begin{array}{ll}
1 & 1\\
-1 & 1
\end{array} 
\right). \label{Unitary}
\end{align} 
Since $u^* \sigma^{(3)} u =\sigma^{(1)}$ and $u^* \sigma^{(1)} u=-\sigma^{(3)}$, we have 
\begin{align}
U^* \sigma_x^{(3)} U=\sigma_x^{(1)},\ \ \ U^* \sigma_x^{(1)} U=-\sigma_x^{(3)}
\end{align} 
for all $x\in \Lambda$.
Thus, 
\begin{align}
\hat{H}_{\Lambda}=U^* H_{\Lambda}U
=-\sum_{x, y\in \Lambda} J_{xy} \sigma_x^{(1)}\sigma_y^{(1)}
-\sum_{x\in \Lambda}\mu_x
\sigma_x^{(1)}+\sum_{x\in
 \Lambda} \lambda_x \sigma_x^{(3)}.
\end{align}

\begin{Prop}\label{ElementsPP}
We have the following:
\begin{itemize}
\item[{\rm (i)}] $\sigma_x^{(1)} \unrhd 0$ w.r.t. $\mathfrak{H}_{\Lambda,
	     +}$ for all $x\in \Lambda$.
\item[{\rm (ii)}] $\frac{1}{2}(\one-\sigma_x^{(3)}) \unrhd 0$ w.r.t. $\mathfrak{H}_{\Lambda,
	     +}$ for all $x\in \Lambda$.
\item[{\rm (iii)}] $
\ex^{-\beta \hat{H}_{\Lambda}} \unrhd 0
$ w.r.t. $\mathfrak{H}_{\Lambda,
	     +}$ for all $\beta \ge 0$.
\end{itemize} 
\end{Prop} 
{\it Proof.} (i), (ii) Let $r$ be a map on $\Omega$ defined by $r(-1)=+1$ and
$r(+1)=-1$.  Clearly,  $\sigma^{(1)} |\omega\ra=|r(\omega)\ra$ holds.
Then $\sigma_x^{(1)} |\bomega\ra= |r_x(\bomega)\ra$,  where
$(r_x(\bomega))_y= r(\omega_x)$ if $y=x$, $(r_x(\bomega))_y=\omega_y$ if
$y\neq x$. Thus,  for all $\bomega\in \Omega^{\Lambda}$, it holds that
$\sigma_x^{(1)} |\bomega\ra \in \mathfrak{H}_{\Lambda, +}.$ Thus, we conclude
(i).  (ii) is obvious.

(iii) Let
\begin{align}
\hat{T}= \sum_{x, y\in \Lambda} J_{xy} \sigma_x^{(1)} \sigma_y^{(1)},
\ \ \ 
\hat{V}_{\mu}=-\sum_{x\in \Lambda}\mu_x
 \sigma_x^{(1)},\ \ \ \hat{V}_{\lambda}=\sum_{x\in \Lambda }\lambda_x\sigma_x^{(3)}.
\end{align} 
Set $
\hat{V}= \hat{V}_{\mu}+\hat{V}_{\lambda}
$.
Then we have $
\hat{H}_{\Lambda}=-\hat{T}+\hat{V}
$.
By (i), we have $\hat{T}\unrhd 0$ w.r.t. $\mathfrak{H}_{\Lambda, +}$.
On the other hand, 
since $-\hat{V}_{\mu}\unrhd 0$ w.r.t. $\mathfrak{H}_{\Lambda,
+}$\footnote{We used the assumption $\mu_x\ge 0$ here.}, we
have  $\ex^{-\beta \hat{V}_{\mu}} \unrhd 0$
w.r.t. $\mathfrak{H}_{\Lambda, +}$ for all $\beta \ge 0$ by Proposition \ref{SumPP}.
In addition, we have 
\begin{align}
\ex^{-\beta \hat{V}_{\lambda}} |\bomega\ra=\underbrace{\exp\Bigg\{
-\beta \sum_{x\in \Lambda} \lambda_x \omega_x
\Bigg\}}_{\ge 0}
|\bomega\ra \in \mathfrak{H}_{\Lambda, +},
\end{align} 
which  implies $\ex^{-\beta \hat{V}_{\lambda}} \unrhd 0$
w.r.t. $\mathfrak{H}_{\Lambda, +}$. 
By Proposition \ref{TK},  we have  $\ex^{-\beta \hat{V}} \unrhd 0$
w.r.t. $\mathfrak{H}_{\Lambda, +}$ for all $\beta \ge 0$.
Now we can apply Proposition \ref{BasicPP} with $ A=-\hat{V}, B=\hat{T}$.
$\Box$

\begin{flushleft}
{\it Proof of Theorem \ref{G1}}
\end{flushleft} 
For each $x\in\Lambda$,   by Proposition
\ref{ElementsPP} (i) and (ii), we have
\begin{align}
\hat{\sigma}_x^{(3)}=U^* \sigma_x^{(3)}U=\sigma_x^{(1)} \unrhd 0,\ \ \ 
\hat{\tau}_x=U^* \tau_x U =\frac{1}{2}(\one-\sigma_x^{(3)}) \unrhd 0
\end{align} 
w.r.t. $\mathfrak{H}_{\Lambda, +}$.
Thus,  $U^* S_{x_1}^{(\vepsilon_1)} \cdots S_{x_n}^{(\vepsilon_n)}U \unrhd
0$ w.r.t. $\mathfrak{H}_{\Lambda, +}$,  implying  that  $\hat{A}=U^* A U
\unrhd 0$ w.r.t. $\mathfrak{H}_{\Lambda, +}$ for all $A\in \mathfrak{A}$.
By applying   Theorem \ref{PPThm}, we conclude Theorem \ref{G1}. $\Box$

\subsection{Proof of Theorems \ref{G2} and \ref{IsingGGriII}}\label{Sub2}

\subsubsection{Preliminaries}
Let $\mathfrak{K}=\BbbC^2\otimes \BbbC^2=\BbbC^4$.
Then $\{|\omega, \omega'\ra\, |\, \omega, \omega'\in \Omega\}$ is  a
CONS of $\mathfrak{K}$, where $|\omega, \omega'\ra=|\omega\ra\otimes
|\omega'\ra$. 
We label $\{|\omega, \omega'\ra\, |\, \omega, \omega'\in \Omega\}$  as 
\begin{align}
|e^1\ra=|+1, +1\ra,\ \ |e^2\ra=|-1, -1\ra,\ \ |e^3\ra=|+1, -1\ra,\ \
 |e^4\ra=|-1, +1\ra.
\end{align} 
Thus,  each $|\vphi\ra=\sum_{j=1}^4 c_j |e_j\ra\in \mathfrak{K}$  can be identified with 
 $(c_1, c_2, c_3, c_4)^T\in \BbbC^4$.
We introduce linear operators on $\mathfrak{K}$  as 
\begin{align}
\psi&=\frac{1}{\sqrt{2}} \big(\sigma^{(3)}\otimes \one +\one \otimes
 \sigma^{(3)}\big),\label{Var1}\\
\phi&=\frac{1}{\sqrt{2}} \big(\sigma^{(3)}\otimes \one -\one \otimes
 \sigma^{(3)}\big),\label{Var2}\\
\eta&=\frac{1}{\sqrt{2}} \big(\sigma^{(1)}\otimes \one +\one \otimes
 \sigma^{(1)}\big),\\
\xi&=\frac{1}{\sqrt{2}} \big(\sigma^{(1)}\otimes \one -\one \otimes
 \sigma^{(1)}\big)
.\label{Var3}
\end{align}  
In general, each operator $X$ in $\mathfrak{K}$ can be expressed as a
$4\times 4$ matrix:
$X=(X_{ij})_{i,j=1,2,3,4}$ with $X_{ij}=\la e_i|X e_j\ra$. In particular, we have
\begin{align}
\psi=\sqrt{2}
\left(\begin{array}{ll}
\sigma^{(3)} & 0\\
0 & 0
\end{array} 
\right),\ \ 
\phi=\sqrt{2}
\left(\begin{array}{ll}
0 & 0\\
0 & \sigma^{(3)}
\end{array} 
\right),\no
\eta=
\sqrt{2}\left(\begin{array}{ll}
\sigma^{(1)} & \alpha\\
\alpha  & \sigma^{(1)}
\end{array} 
\right), \ \ 
\xi=
\sqrt{2}\left(\begin{array}{ll}
0 & \gamma\\
\gamma  & 0
\end{array} 
\right),
\label{44Rep}
\end{align}
where 
\begin{align} 
\alpha
=\frac{1}{\sqrt{2}}\left(\begin{array}{ll}
1 & 1\\
1 & 1
\end{array} 
\right),\ \ \ \ \gamma=\frac{1}{\sqrt{2}}
\left(\begin{array}{ll}
-1 & 1\\
1 & -1
\end{array} 
\right)
.
\end{align} 
Let $u$ be the unitary operator given by (\ref{Unitary}) and let
\begin{align}
\vartheta=\left(\begin{array}{ll}
u & 0\\
0 & u
\end{array} 
\right). \label{theta}
\end{align} 
For each operator $X$ on $\mathfrak{K}$, we write $\tilde{X}=\vartheta^*
X\vartheta$. By (\ref{44Rep}), we obtain 
\begin{align}
\tilde{\psi}&=\sqrt{2}
\left(\begin{array}{ll}
\sigma^{(1)} & 0\\
0 & 0
\end{array} 
\right),\ \ 
\tilde{\phi}=\sqrt{2}
\left(\begin{array}{ll}
0 & 0\\
0 & \sigma^{(1)}
\end{array} 
\right),\no
\tilde{\eta}&=
\sqrt{2}\left(\begin{array}{ll}
-\sigma^{(3)} & \hat{\alpha}\\
\hat{\alpha}  & -\sigma^{(3)}
\end{array} 
\right), \ \ 
\tilde{\xi}=\sqrt{2}
\left(\begin{array}{ll}
0 & \hat{\gamma}\\
\hat{\gamma}  & 0
\end{array} 
\right)
\label{MatrixRep}
\end{align} 
where $\hat{\alpha}=u^* \alpha u=(\one_2-\sigma^{(3)})/\sqrt{2}$ and 
$\hat{\gamma}=-(\one_2+\sigma^{(3)})/\sqrt{2}$.

\begin{define}
{\rm
Let 
\begin{align}
\mathfrak{K}_+= \Bigg\{|\vphi\ra\in \mathfrak{K}\, \Bigg|\, |\vphi\ra=\sum_{j=1}^4
c_j |e_j\ra,\ \ c_j\ge 0, j=1,2,3,4
\Bigg\}.
\end{align}
Clearly,  $\mathfrak{K}_+$ is a self-dual cone in $\mathfrak{K}$.
 $\diamondsuit$ 
}
\end{define}

\begin{Prop}\label{OneP}
We have the following:
\begin{itemize}
\item[{\rm (i)}] $\tilde{\psi}\unrhd 0$ w.r.t. $\mathfrak{K}_+$.
\item[{\rm (ii)}] $\tilde{\phi}\unrhd 0$ w.r.t. $\mathfrak{K}_+$.
\item[{\rm (iii)}] $-\tilde{\xi}\unrhd 0$  w.r.t. $\mathfrak{K}_+$.
\item[{\rm (iv)}] $\one_4+\frac{1}{\sqrt{2}} \tilde{\eta} \unrhd 0$
	     w.r.t. $\mathfrak{K}_+$.
\item[{\rm (v)}]$\exp(\beta \tilde{\psi}) \unrhd 0$ w.r.t. $\mathfrak{K}_+$
for all $\beta\ge  0$.
\item[{\rm (vi)}] $
\exp(\beta \tilde{\eta}) \unrhd 0$ w.r.t. $\mathfrak{K}_+$
for all $\beta\ge  0$.
\end{itemize} 
\end{Prop} 
{\it Proof.}  Note that by Proposition \ref{Prop3}, a linear operator $X$ on $\mathfrak{K}$
satisfies $X\unrhd 0$ w.r.t. $\mathfrak{K}_+$ if and only if $X_{ij}=\la
e_i|X e_j\ra \ge 0$ for all $i, j=1,2,3,4$. Thus,  (i), (ii),  (iii),  and (iv)
immediately  follow from (\ref{MatrixRep}).

(v) By (i) and Proposition \ref{SumPP}, we can see that 
$
\exp(\beta \tilde{\psi}) \unrhd 0 $ w.r.t. $\mathfrak{K}_+$ for all
$\beta \ge 0$.

To show (vi), we write
$\tilde{\eta}=\tilde{\eta}_d+\tilde{\eta}_o$, where
\begin{align}
\tilde{\eta}_d
=\sqrt{2}\left(\begin{array}{ll}
- \sigma^{(3)} & 0\\
0 &-  \sigma^{(3)}
\end{array} 
\right),\ \ 
\tilde{\eta}_o=
\sqrt{2}\left(\begin{array}{ll}
0 & \hat{\alpha}\\
\hat{\alpha} & 0
\end{array} 
\right).
\end{align} 
Suppose that 
\begin{itemize}
\item[{\rm (a)}] $\exp(\beta \tilde{\eta}_d) \unrhd 0$
	     w.r.t. $\mathfrak{K}_+$ for all $\beta \ge 0$,
\item[{\rm (b)}] $\exp(\beta \tilde{\eta}_o) \unrhd 0$
	     w.r.t. $\mathfrak{K}_+$ for all $\beta \ge 0$.
\end{itemize} 
Then  we can immediately  conclude (vi) by Proposition \ref{TK}. Hence,  it
suffices to prove (a) and (b).

To show (a), observe that 
\begin{align}
\exp(\beta \tilde{\eta}_d)=
\left(\begin{array}{ll}
\ex^{-\sqrt{2} \beta \sigma^{(3)}} & 0 \\
0  & \ex^{-\sqrt{2}\beta \sigma^{(3)}}
\end{array} 
\right).
\end{align} 
Since all matrix elements of $\exp(-\sqrt{2}\beta \sigma^{(3)})$ are positive, we
conclude that  (a) is true by Proposition \ref{Prop3}.

Since $\tilde{\eta}_o\unrhd 0$ w.r.t. $\mathfrak{K}_+$, we find  that 
$
\exp(\beta \tilde{\eta}_o)
 \unrhd 0 
$  w.r.t. $\mathfrak{K}_+$ for all $\beta\ge 0$ by Proposition \ref{SumPP}. 
Thus, we conclude (b). $\Box$

\subsubsection{Completion of proof of Theorems \ref{G2} and \ref{IsingGGriII} }
Let $\He$ be given by (\ref{IHamiExt}).
$\He$ acts in  the extended Hilbert space
$\mathfrak{K}_{\Lambda}=\mathfrak{H}_{\Lambda} \otimes
\mathfrak{H}_{\Lambda}$.
For each $x\in \Lambda$, let 
\begin{align}
\psi_x &=\frac{1}{\sqrt{2}}\big(\sigma_x^{(3)}\otimes \one+ \one \otimes
 \sigma_x^{(3)}\big),\\
\phi_x &=\frac{1}{\sqrt{2}}\big(\sigma_x^{(3)}\otimes \one- \one \otimes
 \sigma_x^{(3)}\big),\\
\eta_x&= \frac{1}{\sqrt{2}}\big(\sigma_x^{(1)}\otimes \one +\one \otimes
 \sigma_x^{(1)}\big),\\
\xi_x&= \frac{1}{\sqrt{2}}\big(\sigma_x^{(1)}\otimes \one -\one \otimes
 \sigma_x^{(1)}\big).
\end{align} 
$\He$ can be expressed as 
\begin{align}
\He =-\sum_{x, y\in \Lambda}J_{xy}(\psi_x\psi_y+\phi_x\phi_y)
-\sqrt{2}\sum_{x\in \Lambda}\mu_x\psi_x
-\sqrt{2}\sum_{x\in
 \Lambda}\lambda_x \eta_x.
\end{align} 
We employ the following identification\footnote{Indeed, we have 
$
\mathfrak{K}_{\Lambda}=\Big(
\bigotimes_{x\in \Lambda} \BbbC^2
\Big)\otimes
\Big(
\bigotimes_{x\in \Lambda} \BbbC^2
\Big) \cong \bigotimes_{x\in \Lambda} (\BbbC^2\otimes \BbbC^2)
\cong \bigotimes_{x\in \Lambda}\mathfrak{K}.
$
 } of $\mathfrak{K}_{\Lambda}$:
\begin{align}
\mathfrak{K}_{\Lambda}=\bigotimes_{x\in \Lambda}\mathfrak{K},
\end{align} 
where $\mathfrak{K}=\BbbC^4$.
Thus, $\psi_x, \phi_x$,  and $\eta_x$ can be expressed as 
\begin{align}
\psi_x=\otimes_{y\in \Lambda}(\psi)^{\delta_{xy}},\ \
 \phi_x=\otimes_{y\in \Lambda} (\phi)^{\delta_{xy}},\ \
 \eta_x=\otimes_{y\in \Lambda} (\eta)^{\delta_{xy}}. \label{Tensor}
\end{align} 
Here $(X)^{\delta_{xy}}=\one$ if $x\neq y$, $(X)^{\delta_{xy}}=X$ if $x=y$.
 Let $\vartheta$ be given by
(\ref{theta}). 
Set $\Theta=\otimes_{x\in \Lambda}\vartheta$.
For each linear operator $X$ on $\mathfrak{K}_{\Lambda}$,
set $\tilde{X}=\Theta^* X \Theta$. Then we obtain
\begin{align}
\The=-\sum_{x, y\in \Lambda}J_{xy}
(\tilde{\psi}_x\tilde{\psi}_y+\tilde{\phi}_x\tilde{\phi}_y)
-\sqrt{2}\sum_{x\in \Lambda}\mu_x\tilde{\psi}_x
-\sqrt{2}\sum_{x\in
 \Lambda}\lambda_x \tilde{\eta}_x,
\end{align} 
where $\tilde{\psi}_x, \tilde{\phi}_x, \tilde{\eta}_x, \tilde{\xi}_x$ are defined
through (\ref{MatrixRep}) and (\ref{Tensor}).

\begin{define}\label{bigCone}
{\rm
We define a self-dual cone in $\mathfrak{K}_{\Lambda}$ by 
\begin{align}
\CK:=\bigg\{
|\Psi\ra \in \mathfrak{K}_{\Lambda}\ \bigg|\ 
|\Psi\ra=\sum_{\mathbf{n}\in \{1,2,3,4\}^{\Lambda}}
C_{\mathbf{n}} |e_{\mathbf{n}}\ra,\ C_{\mathbf{n}} \ge 0\  \forall
 \mathbf{n}\in \{1,2,3,4\}^{\Lambda}
\bigg\},
\end{align}
where
 $|e_{\mathbf{n}}\ra=\otimes_{x\in \Lambda} |e_{n_x}\ra$  for each
 $\mathbf{n}=\{n_x\}_{x\in \Lambda}\in \{1,2,3,4\}^{\Lambda}$. $\diamondsuit$ 
}
\end{define} 
\begin{rem}
{\rm 
$|e_{\mathbf{n}}\ra\in \mathfrak{K}_{\Lambda, +}$ for all $\mathbf{n}
 \in \{1,2,3,4\}^{\Lambda}$. $\diamondsuit$
}
\end{rem} 

\begin{Prop}\label{ManyP}
We have the following:
\begin{itemize}
\item[{\rm (i)}] $\tilde{\psi}_x\unrhd 0$ w.r.t. $\CK$ for all $x\in
	     \Lambda$.
\item[{\rm (ii)}] $\tilde{\phi}_x\unrhd 0$ w.r.t. $\CK$ for all $x\in
	     \Lambda$.
\item[{\rm (iii)}] $-\tilde{\xi}_x\unrhd 0$ w.r.t. $\CK$ for all $x\in
	     \Lambda$.
\item[{\rm (iv)}] $\one+\frac{1}{\sqrt{2}}\tilde{\eta}_x \unrhd 0$  $\CK$ for all $x\in
	     \Lambda$.
\item[{\rm (v)}]$\exp(\beta \tilde{\psi}_x) \unrhd 0$ w.r.t. $\CK$
	     for all $x\in \Lambda$ and $\beta \ge 0$.
\item[{\rm (vi)}] $\exp(\beta \tilde{\eta}_x) \unrhd 0$ w.r.t. $\CK$
	     for all $x\in \Lambda$ and $\beta \ge 0$.
\end{itemize} 
\end{Prop} 
{\it Proof.} By Proposition \ref{Prop3},  a linear operator $X$ in $\mathfrak{K}_{\Lambda}$ satisfies
$X\unrhd 0$ w.r.t. $\CK$ if and only if 
$\la e_{\mb{n}}|X e_{\mb{m}}\ra \ge 0$ for all $\mb{m}, \mb{n}\in
\{1,2,3,4\}^{\Lambda}$. Thus,  the assertions immediately follow from 
Proposition \ref{OneP}. $\Box$

\begin{coro}\label{HPP}
$
\exp(-\beta \The) \unrhd 0
$
w.r.t. $\CK$ for all $\beta \ge 0$.
\end{coro} 
{\it Proof.}
Set $\The=-\tilde{T}+\tilde{V}$, where
\begin{align}
\tilde{T}&=\sum_{x, y\in \Lambda}J_{xy}
(\tilde{\psi}_x\tilde{\psi}_y+\tilde{\phi}_x\tilde{\phi}_y),\ \ \ 
\tilde{V}=
\tilde{V}_{\mu}+\tilde{V}_{\lambda}
\end{align} 
with $\tilde{V}_{\mu}=-\sqrt{2}\sum_{x\in \Lambda} \mu_x\tilde{\psi}_x$
 and $\tilde{V}_{\lambda}=-\sqrt{2}\sum_{x\in \Lambda} \lambda_x \tilde{\eta}_x$.
By Proposition \ref{ManyP} (i) and (ii), it holds that $\tilde{T}\unrhd 0$ w.r.t. $\CK$.
On the other hand, we can see that  by Proposition \ref{ManyP} (v) and (vi), 
\begin{align}
\ex^{-\beta V_{\mu}}=\prod_{x\in \Lambda} \underbrace{\ex^{\sqrt{2}\beta \mu_x
 \tilde{\psi}_x}}_{\unrhd 0} \unrhd 0,\ \ \ 
\ex^{-\beta V_{\lambda}}=\prod_{x\in \Lambda} \underbrace{\ex^{\sqrt{2}\beta \lambda_x
 \tilde{\eta}_x}}_{\unrhd 0} \unrhd 0
\end{align} 
 w.r.t. $\CK$ for all
 $\beta \ge 0$.\footnote{Here we have used the assumptions $\mu_x\ge 0$ and
 $\lambda_x\ge 0$.}
Thus,  by Proposition \ref{TK}, we obtain  $\ex^{-\beta \tilde{V}} \unrhd 0$ w.r.t. $\CK$ for all $\beta \ge 0$.
 By applying Proposition \ref{BasicPP}, we conclude
the desired assertion. $\Box$

\begin{coro}\label{SPP}
For all $A, B, C, D\subseteq \Lambda$, we have the following:
\begin{itemize}
\item[{\rm (i)}]
$
\Theta^* 
\sigma_A^{(3)}\otimes \tau_C\Theta\unrhd 0\ \ \mbox{w.r.t. $\CK$}.
$
\item[{\rm (ii)}]
$
\Theta^* \big(
\sigma_B^{(3)}\otimes \tau_D -\tau_D\otimes \sigma_B^{(3)}
\big)\Theta\unrhd 0\ \ \mbox{w.r.t. $\CK$}.
$
\end{itemize} 
\end{coro} 
{\it Proof.} (i)
Let $\tilde{\ell}_x=\frac{1}{2}(\one+\frac{1}{\sqrt{2}}\tilde{\eta}_x)$
and $\tilde{m}_x=-\frac{1}{2\sqrt{2}}\tilde{\xi}_x$. By Proposition \ref{ManyP} (iii) and
(iv), it holds that $\tilde{\ell}_x\unrhd 0$ and $\tilde{m}_x\unrhd 0$
w.r.t. $\CK$.
 Thus, we have 
\begin{align}
\Theta^* \sigma_A^{(3)} \otimes \tau_C \Theta=2^{-|A|/2} \prod_{x\in
 A}(\tilde{\psi}_x+\tilde{\phi}_x) \prod_{x\in C}(\tilde{\ell}_x+\tilde{m}_x)
\unrhd 0
\end{align} 
w.r.t. $\CK$ by Proposition \ref{ManyP}. 

(ii) We have
\begin{align}
&\Theta^* \big(\sigma_B^{(3)}\otimes \tau_D -\tau_D\otimes
 \sigma_B^{(3)}\big)\Theta\no
=&2^{-|B|/2} \prod_{x\in B}\prod_{y\in D} (\tilde{\psi}_x+\tilde{\phi}_x)
 (\tilde{\ell}_y+\tilde{m}_y)
-2^{-|B|/2}\prod_{x\in B}\prod_{y\in D} (\tilde{\psi}_x-\tilde{\phi}_x)
 (\tilde{\ell}_y-\tilde{m}_y)
\\
=&\sum_{X_1, X_2 \subset B} \sum_{Y_1, Y_2\subset D}K_{X_1X_2 Y_1
 Y_2}\tilde{\psi}_{X_1}\tilde{\phi}_{X_2} \tilde{\ell}_{Y_1} \tilde{m}_{Y_2}
\end{align} 
with $K_{X_1X_2Y_1Y_2}\ge 0, \tilde{\psi}_{X_1}
=\prod_{x\in
X_1}\tilde{\psi}_x, \ \tilde{\phi}_{X_2}=\prod_{x\in
X_2}\tilde{\phi}_x, \tilde{\ell}_{Y_1}=\prod_{x\in Y_1} \tilde{\ell}_x,\
\tilde{m}_{Y_2}=\prod_{x\in Y_2} \tilde{m}_x$. Thus,
 $\Theta^*
(
\sigma_B^{(3)}\otimes \tau_D -\tau_D\otimes
 \sigma_B^{(3)}
)\Theta \unrhd 0
$
w.r.t. $\CK$.  $\Box$

\begin{flushleft}
{\it Proof of Theorems \ref{G2} and \ref{IsingGGriII}}
\end{flushleft} 
By Corollaries \ref{HPP}, \ref{SPP} and  Theorem
\ref{Prop2}, we have
\begin{align}
&\Big\la\!\!\Big\la 
\Big(
\sigma_A^{(3)}(s)
\otimes \tau_C(s)
-\tau_C(s)\otimes \sigma_A^{(3)}(s)
\Big)
 \Big(
\sigma_B^{(3)}(t)
 \otimes \tau_D(t) -  \tau_D(t)\otimes \sigma_B^{(3)}(t)
\Big)
\Big\ra\!\!\Big\ra_{\beta}\no
&=Z_{\beta}^{-2}\Tr\Bigg[
\underbrace{\ex^{- s\tilde{H}_{\mathrm{ext}}}}_{\unrhd 0}
\underbrace{
\Theta^*
\Big(
\sigma_A^{(3)}
\otimes \tau_C
-\tau_C\otimes \sigma_A^{(3)}
\Big)
\Theta
}_{\unrhd 0}
\underbrace{
\ex^{-(t-s)\tilde{H}_{\mathrm{ext}}}
}_{\unrhd 0}\no
 &\ \ \ \ \ \times
\underbrace{
\Theta^* \Big(
\sigma_B^{(3)}
 \otimes \tau_D -  \tau_D\otimes \sigma_B^{(3)}
\Big)
\Theta
}_{\unrhd 0}
\underbrace{
\ex^{-(\beta-t)\tilde{H}_{\mathrm{ext}}}
}_{\unrhd 0}
\Bigg]\ge 0.
\end{align} 
  This concludes Theorem \ref{G2}.

Similarly, we can show  Theorem \ref{IsingGGriII} by  Corollary \ref{SPP}  and
Theorem \ref{Prop2}.
$\Box$

\subsection{Proof of Example \ref{IsingEx}}\label{IsingExPr}
We only prove (i) and (ii), since (iii) and (iv) can be proved in a
similar manner.

Recall the Duhamel formula
\begin{align}
\ex^{-t(A+B)}=\sum_{n\ge 0} \int_{0\le t_1 \le t_2 \le \cdots \le t_n
 \le \beta}
(-B(t_1))\cdots (-B(t_n))\, \ex^{-tA}dt_1\cdots dt_n\label{Duhamel}
\end{align} 
for any bounded self-adjoint operators $A$ and $B$  with $B(t)=\ex^{-tA}
B\ex^{tA}$. Using this, we have
\begin{align}
\frac{\partial}{\partial J_{xy}} \ex^{-\beta H}&=\sum_{n\ge 1} D_n,\\
D_n&=n(J_{xy})^{n-1} \int_{0\le t_1 \le \cdots \le t_n \le \beta}
T_{xy}[t_1]\cdots T_{xy}[t_n] \, \ex^{-\beta H^{'}}dt_1\cdots dt_n,
\end{align} 
where
$T_{xy}=\sigma_x^{(3)}\sigma_y^{(3)},\
H^{'}=H_{\Lambda}+J_{xy}\sigma_x^{(3)}\sigma_y^{(3)}$ and $T_{xy}[t]=\ex^{-tH^{'}}
T_{xy} \ex^{tH^{'}}$.
Note that in a similar manner to Sections \ref{Sub1} and \ref{Sub2}, we
have the following:
\begin{itemize}
\item[(a)] $\ex^{-\beta \hat{H}^{'}} \unrhd 0$
	   w.r.t. $\mathfrak{H}_{\Lambda, +}$ for all $\beta \ge 0$,
	   where
$\hat{H}^{'}=U^* H^{'}U$.
\item[(b)] $\ex^{-\beta \tilde{H}_{\mathrm{ext}}^{'}} \unrhd 0$
	   w.r.t. $\mathfrak{K}_{\Lambda, +}$ for all $\beta \ge 0$,
	   where $H^{'}_{\mathrm{ext}}=H^{'}\otimes \one +\one \otimes
	   H^{'}$.
\end{itemize} 
Hence, by setting $\mathsf{M}_n=T_{xy}[t_1]\cdots T_{xy}[t_n]$,  we obtain
\begin{align}
&\frac{\partial}{\partial J_{xy}} \la \sigma_A^{(3)} \ra_{\beta}\no
=&\sum_{n\ge 1} n (J_{xy})^{n-1} 
 \int_{0\le t_1 \le \cdots \le  \beta}
\Big\{
\big\la \sigma_A^{(3)} \mathsf{M}_n
\big\ra_{H^{'}, \beta}
-\big\la \sigma_A^{(3)} \big\ra_{H^{'}, \beta}
\big\la \mathsf{M}_n \big\ra_{H^{'}, \beta}
\Big\}dt_1\cdots dt_n\no
=& \sum_{n\ge 1}\frac{n (J_{xy})^{n-1}}{2}
\int_{0\le t_1 \le \cdots \le \beta}
\Big\la\!\!\Big\la
\big(
\sigma_A^{(3)}\otimes \one -\one \otimes \sigma_A^{(3)}
\big)
\big(
\mathsf{M}_n\otimes \one -\one \otimes \mathsf{M}_n
\big)
\Big\ra\!\!\Big\ra_{H^{'}, \beta}dt_1\cdots dt_n\no
&\ge 0,
\end{align} 
where $\la \cdot \ra_{H^{'}, \beta}$ and $\la\!\la\cdot\ra\!\ra_{H^{'},
\beta}$
 are the thermal averages associated with $H^{'}$ and
 $H^{'}_{\mathrm{ext}}$.
(Here we used the facts that 
$
\Theta^* (\sigma_A^{(3)}\otimes \one -\one \otimes \sigma_A^{(3)})
\Theta\unrhd 0
$
and $\Theta^*(\mathsf{M}_n\otimes \one -\one \otimes \mathsf{M}_n)
\, \ex^{-\beta H^{'}_{\mathrm{ext}}}\Theta \unrhd 0$
w.r.t. $\mathfrak{K}_{\Lambda, +}$,  which follow from Corollary \ref{SPP}.)
Thus, we have proved (i). Similarly, by applying the fact that   $\Theta^* (\tau_A\otimes \one
-\one \otimes \tau_A)\Theta \unlhd 0$ w.r.t. $\mathfrak{K}_{\Lambda, +}$,
which follows from  Corollary \ref{SPP},
 we have 
\begin{align}
&\frac{\partial}{\partial J_{xy}} \la \tau_A \ra_{\beta}\no
=& \sum_{n\ge 1}\frac{n (J_{xy})^{n-1}}{2}
\int_{0\le t_1 \le \cdots \le \beta}
\Big\la\!\!\Big\la
\big(
\tau_A\otimes \one -\one \otimes \tau_A
\big)
\big(
\mathsf{M}_n\otimes \one -\one \otimes \mathsf{M}_n
\big)
\Big\ra\!\!\Big\ra_{H^{'}, \beta}dt_1\cdots dt_n\no
&\le 0.
\end{align} 
Thus,  we have proved  (ii). $\Box$

\section{Quantum rotor model}\label{SecQM}
\setcounter{equation}{0}
\subsection{Results}
Let $\Lambda$ be a finite subset of $\BbbR^2$.
The quantum rotor model on $\Lambda$ is defined by 
\begin{align}
H=\sum_{x\in \Lambda}
\frac{U_x}{2}
\Big(- i \frac{\partial}{\partial
 \theta_x}\Big)^2
- \sum_{x, y\in \Lambda}t_{xy}\cos(\theta_x-\theta_y).
\end{align} 
The Hilbert space is $\mathfrak{H}=\otimes_{x\in \Lambda}L^2(\mathbb{T})$ with
$\mathbb{T}=[-\pi, \pi]$.
$U_x> 0$  being the strength of the on site repulsion and $t_{xy}\ge 0$ being  the hopping
strength.
$H$ is a self-adjoint operator acting in the Hilbert space 
$\mathfrak{H}$.\footnote{The precise definition of $-i 
\frac{\partial}{\partial \theta}$ is given by 
\begin{align*}
\D\Big(-i
\frac{\partial}{\partial \theta}\Big)
&= \{f\in C^1(\mathbb{T})\, |\, f(-\pi)=f(\pi)\},\\
-i
\frac{\partial}{\partial \theta} f&=-i  f^{'}\, \ \ \forall f\in
 \D\Big(- i
\frac{\partial}{\partial \theta}\Big).
\end{align*} 
Then $
-i
\frac{\partial}{\partial \theta}
$
 is essentially self-adjoint. We still denote its closure by the same symbol. 
}
We refer  readers who want to learn the physical background to \cite{BFLLS,Sachdev}.

\begin{rem}{\rm
In this study, we simply write $M_f$, the multiplication operator by the function $f$,  as $f(\theta)$ if no confusion
 occurs. $\diamondsuit$
}
\end{rem} 
Let $T_x=\ex^{i  \theta_x}$. For each $A=\{m_x\}_{x\in \Lambda}\in \BbbZ^{\Lambda}$, we set
\begin{align}
T^A=\prod_{x\in \Lambda}(T_x)^{m_x}.
\end{align} 
Let
\begin{align}
\mathfrak{A}=\mathrm{Coni}\big\{ T^A\, |\, A\in \BbbZ^{\Lambda}\big\}^{\mbox{---$\mathrm{w}$}}.
\end{align}  

The thermal expectation value $\la \cdot \ra_{\beta}$ is defined by
\begin{align}
\la A \ra_{\beta}=\Tr\big[A\, \ex^{-\beta H}\big]\Big/ Z_{\beta},\ \ \
 Z_{\beta}=\Tr\big[
\ex^{-\beta H}
\big]
\end{align} 
for all $A\in \mathscr{B}(\mathfrak{H})$.

\begin{Thm}[First Griffiths inequality]
Let $A_1, \dots, A_n\in \mathfrak{A}$.
\label{GriffithsI}
For all  $0\le s_1\le s_2\le \cdots \le
 s_n < \beta$, we have
\begin{align}
\Bigg\la 
\prod_{j=1}^{{n}\atop{\longrightarrow}} A_j(s_j)
\Bigg\ra_{\beta}\ge 0.
\end{align} 
\end{Thm}

To state second Griffiths inequality,   some conditions are required.
We introduce an extended Hilbert space $\mathfrak{H}_{\mathrm{ext}}$ by
$\mathfrak{H}_{\mathrm{ext}}=\mathfrak{H}\otimes \mathfrak{H}$.
For each $X\in \mathscr{B}(\mathfrak{H}_{\mathrm{ext}})$, we set
\begin{align}
\la \!\la X\ra\!\ra_{\beta}&=\Tr_{\mathfrak{H}_{\mathrm{ext}}}\big[
X\, \ex^{-\beta H_{\mathrm{ext}}}
\big]\Big/ Z_{\beta}^2,\\
H_{\mathrm{ext}}&=H\otimes \one +\one \otimes H.
\end{align} 
Let $C_x=\cos\theta_x$ and 
\begin{align}
C_x(s)=\ex^{-s H} C_x\ex^{sH}.
\end{align} 

\begin{Thm}[Second Griffiths inequality]\label{QGII1}
For all $x_1, \dots, x_n\in \Lambda$, $0\le s_1 \le s_2 \le \cdots
 \le s_n < \beta$ and $ \vepsilon_1,\dots, \vepsilon_n\in \{\pm 1\}$,
 we have 
\begin{align}
\Bigg\la\!\!\!\Bigg\la \prod_{j=1}^{{n}\atop{\longrightarrow}}
\Big[
C_{x_j}(s_j)\otimes \one +\vepsilon_j\one  \otimes C_{x_j}(s_j)
\Big]
\Bigg\ra\!\!\!\Bigg\ra_{\beta}\ge 0.
\end{align} 
\end{Thm} 

From Theorem \ref{QGII1}, we immediately obtain  Corollary
\ref{QGII2},  which has a form similar to (\ref{IsingQGII}).
(This is  why we call Theorem \ref{QGII1}   the second
Griffiths inequality, see Remark \ref{Equiv1} and Theorem \ref{Equiv2}
for general arguments.)

\begin{coro}\label{QGII2}
For each $A=\{m_x\}_{x\in \Lambda}\in \BbbN^{\Lambda}$, set
\begin{align}
C^A=\prod_{x\in \Lambda}(C_x)^{m_x}.
\end{align} 
For all $A, B\in \BbbN^{\Lambda}$, we obtain
\begin{align}
\big\la C^A C^B\big\ra_{\beta} \ge \big\la C^A \big\ra_{\beta} \big\la C^B\big\ra_{\beta}.
\end{align} 
\end{coro} 

Let
\begin{align}
n_x=-i \frac{\partial}{\partial\theta_x}.
\end{align} 
Set $n_x(s)=\ex^{-s H} n_x \ex^{s H}$. We have the following.
\begin{Thm}\label{QGNII2}
For all $x_1, \dots, x_n\in \Lambda,\, 0\le s_1\le s_2\le
 \cdots \le s_n<\beta$ and $\vepsilon_1, \dots, \vepsilon_n\in
 \{\pm 1\}$, we have 
\begin{align}
\Bigg\la\!\!\!\Bigg\la
\prod_{j=1}^{{n}\atop{\longrightarrow}}
\Big[
n_{x_j}(s_j)\otimes \one +\vepsilon_j \one \otimes n_{x_j}(s_j)
\Big]
\Big[
n_{x_j}(s_j)\otimes \one +\overline{\vepsilon}_j \one \otimes n_{x_j}(s_j)
\Big]
\Bigg\ra\!\!\!\Bigg\ra_{\beta}\ge 0,
\end{align} 
where $\overline{\vepsilon}_j=-\vepsilon_j$.
\end{Thm}

We can construct several extensions of Theorems \ref{QGII1} and
\ref{QGNII2}.
Theorem \ref{QGII3} illustrates this fact.
Let 
\begin{align}
\alpha_x^{(1)}(s)&=C_x(s)\otimes \one +\one \otimes C_x(s),\\
\alpha_x^{(2)}(s)&=C_x(s)\otimes \one -\one \otimes C_x(s),\\
\alpha_x^{(3)}(s)&=\Big[
n_{x}(s)\otimes \one + \one \otimes n_{x}(s)
\Big]
\Big[
n_{x}(s)\otimes \one - \one \otimes n_{x}(s)
\Big].
\end{align}

\begin{Thm}\label{QGII3}
For all $x_1, \dots, x_n\in \Lambda,\ \mu_1,\dots, \mu_n\in \{1,2,3\}$
 and $0\le s_1\le s_2\le \cdots \le s_n\le \beta$, we have 
\begin{align}
\Bigg\la\!\!\!\Bigg\la
\prod_{j=1}^{{n}\atop{\longrightarrow}} \alpha_{x_j}^{(\mu_j)}(s_j)
\Bigg\ra\!\!\!\Bigg\ra_{\beta}\ge 0.
\end{align} 
\end{Thm} 

\begin{example}\label{MonoCorr}{\rm 
For all $A\subseteq \Lambda$ and $ x,y, z\in \Lambda$, we have the following:
\begin{itemize}
\item[{\rm (i)}]
  $\la C^A\ra_{\beta}$ is monotonically increasing in
$t_{xy}$.
\item[{\rm (ii)}] $\la n_z^2 \ra_{\beta}$ is monotonically increasing  in $t_{xy}$.
\item[{\rm (iii)}] $
\displaystyle
\frac{\partial}{\partial U_x} \la C^A\ra_{\beta}\Big|_{U_x=0}\le 0
$.
\end{itemize} 
We will provide a proof of this example in Section \ref{ProofExam}. $\diamondsuit$
}
\end{example}

\begin{rem}
{\rm
\begin{itemize}
\item[(i)] Our results can be extended to  a more general Hamiltonian of
	   the form
\begin{align}
H=\sum_{x\in \Lambda}
\frac{U_x}{2}
\Big(- i \frac{\partial}{\partial
 \theta_x}\Big)^2
- \sum_{A \in \BbbN^{\Lambda}}J_A C^A 
\end{align} 
with $J_A\ge 0$, where the sum converges under   a   uniform topology. 
\item[(ii)] Since the ground state of $H$ is unique, our results are
	    valid at $\beta=\infty$. The results at $\beta=\infty$
are essential for the study of   quantum phase transitions
 \cite{KP}. $\diamondsuit$
\end{itemize} 
}
\end{rem} 
\subsection{Proof of Theorem \ref{GriffithsI}}\label{ProofPP}

Let $\mathcal{F}$ be the Fourier transformation \footnote{
To be precise, $\mathcal{F}$ is a unitary operator given by 
\begin{align}
(\mathcal{F}f)(\mathbf{n})=(2\pi)^{-|\Lambda|/2}\int_{\mathbb{T}^{\Lambda}}
f({\boldsymbol \theta})\, \ex^{-i {\boldsymbol \theta}\cdot \mathbf{n}}
d{\boldsymbol \theta}\ \ \forall f\in \mathfrak{H}.
\end{align} 
} on $\h$ and let
$\hat{H}=\mathcal{F} H\mathcal{F}^{-1}$. We have
\begin{align}
\hat{H}=\sum_{x\in \Lambda}\frac{U_x}{2}\hat{n}_x^2
+ \frac{1}{2}\sum_{x, y\in \Lambda}(-t_{xy})(\hat{T}_x \hat{T}_{y}^*+\hat{T}_{x}^*\hat{T}_y).
\end{align}  
$\hat{H}$ acts in the Hilbert space $\hat{\mathfrak{H}}=\mathcal{F}\mathfrak{H}=\otimes_{x\in
\Lambda}\ell^2(\BbbZ)$. 
$\hat{n}_x$ and $\hat{T}_x$ are defined by 
$
\hat{n}_x=\mathcal{F} n_x\mathcal{F}^{-1}
$ and $\hat{T}_x=\mathcal{F} T_x\mathcal{F}^{-1}$.

For  each $n\in \BbbZ$, set $e_n(m)=\delta_{mn}\in \ell^2(\BbbZ)$.
$\{e_n\, |\, n\in \BbbZ\}$ is a CONS in $\ell^2(\BbbZ)$.
For each $\mb{n}=\{n_x\}_{x\in \Lambda}\in \BbbZ^{\Lambda}$, let
$
e_{\mathbf{n}}=\otimes_{x\in \Lambda} e_{n_x}.
$ 
Clearly, $\{e_{\mathbf{n}}\, |\, \mathbf{n}\in \BbbZ^{\Lambda}\}$ is a CONS of $\hat{\mathfrak{H}}$ as well.
Remarkably,  for each $\mathbf{n}=\{n_x\}_{x\in
\Lambda}\in \BbbZ^{\Lambda}$,  
\begin{align}
\hat{n}_xe_{\mathbf{n}}=n_x e_{\mathbf{n}},\ \ \ \hat{T}_x e_{\mathbf{n}}=
e_{\mb{n}+{\boldsymbol \delta_{x}}}, \label{DefET}
\end{align} 
where ${\boldsymbol \delta_x}=\{\delta_{xy}\}_{y\in \Lambda}\in \BbbZ^{\Lambda}$.
In other words, $\hat{n}_x$ is the number operator and $\hat{T}_x$ is the
creation  operator at site $x$.

\begin{define}
{\rm 
Let
\begin{align}
\hat{\mathfrak{H}}_{+}=\Bigg\{
F=\sum_{\mathbf{n}\in \BbbZ^{\Lambda}} F(\mathbf{n}) e_{\mathbf{n}}\in \hat{\mathfrak{H}}\, \Big|\, 
F(\mathbf{n}) \ge 0\ \ \forall \mathbf{n}\in \BbbZ^{\Lambda}
\Bigg\}.\label{StandardCone}
\end{align}
Note  that $\hat{\mathfrak{H}}_{+}$ is a  self-dual cone in
 $\hat{\mathfrak{H}}$. 
Clearly,  $e_{\mb{n}}\in \hat{\mathfrak{H}}_+$ for all $\mb{n}\in
 \BbbZ^{\Lambda}$.
$\diamondsuit$

}
\end{define}

\begin{Prop}\label{PP}
We have  the following:
\begin{itemize}
\item[{\rm (i)}] $\hat{T}_x\unrhd 0$ w.r.t. $\hat{\mathfrak{H}}_{+}$ for all
	     $x\in \Lambda$.
\item[{\rm (ii)}]$\ex^{-\beta \hat{H}}\unrhd 0$
	     w.r.t. $\hat{\mathfrak{H}}_{+}$ for all $\beta \ge 0$.
 \end{itemize} 
\end{Prop} 
{\it Proof.} (i)  Note that  $\hat{\mathfrak{H}}_{+}=\mathrm{Coni}\{e_{\mathbf{n}}\, |\, \mathbf{n}\in \BbbZ^{\Lambda}\}^{-}$,
 where $\mathrm{Coni}(S)^{-}$ is the closure of $\mathrm{Coni}(S)$. Thus,  it suffices to show that $\hat{T}_x
 e_{\mathbf{n}}\ge 0$ w.r.t. $\hat{\mathfrak{H}}_+$ for all
 $\mathbf{n}\in \BbbZ^{\Lambda}$. This is 
 trivial according to  (\ref{DefET}).

(ii)  Let 
\begin{align}
-\hat{\mathbf{K}}=\frac{1}{2}\sum_{x, y\in \Lambda}
 t_{xy}(\hat{T}_x\hat{T}_{y}^*+\hat{T}_{x}^*\hat{T}_y),\ \ \
 \hat{\mathbf{U}}=\sum_{x\in \Lambda} \frac{U_x}{2}\hat{n}_x^2.\label{DefUK}
\end{align} 
By (i), we can see that $-\hat{\mathbf{K}}\unrhd 0$ w.r.t. $\hat{\mathfrak{H}}_{
+}$. 
On the other hand, since 
\begin{align}
\ex^{-\beta \hat{\mathbf{U}}}e_{\mathbf{n}}
 =\underbrace{\exp\Bigg\{-\beta \sum_{x\in \Lambda} \frac{U_x}{2}
 n_x^2\Bigg\}}_{\ge 0} e_{\mathbf{n}}\ \ \mbox{for all
 $\mathbf{n}=\{n_x\}\in \BbbZ^{\Lambda}$},
\end{align} 
we have $\ex^{-\beta \hat{\mathbf{U}}} \unrhd 0$
w.r.t. $\hat{\mathfrak{H}}_{ +}$. Thus,  by Proposition \ref{BasicPP},
 we conclude (ii). $\Box$

\subsubsection{ Completion of proof of Theorem \ref{GriffithsI}  } 
By Proposition \ref{PP} (i), we have $A\unrhd 0$
w.r.t. $\hat{\mathfrak{H}}_+$ for all $\mathfrak{A}$.
Applying Theorem \ref{PPThm}, we
 prove Theorem \ref{GriffithsI}. 
$\Box$

\subsection{Proof of Theorems \ref{QGII1}, \ref{QGNII2}, and \ref{QGII3} and Corollary \ref{QGII2}}
First, note  the following identification:
\begin{align}
\mathfrak{H}_{\mathrm{ext}}=L^2
\big(\mathbb{T}^{\Lambda}\times
 \mathbb{T}^{\Lambda}, d{\boldsymbol \theta} d{\boldsymbol \theta}^{'}
\big). \label{HIden}
\end{align} 
Under the identification (\ref{HIden}), we see that 
\begin{align}
H_{\mathrm{ext}}=&H\otimes \one +\one \otimes H\no
=&\sum_{x\in \Lambda}
\frac{U_x}{2}
\Bigg\{\Big(- i \frac{\partial}{\partial
 \theta_x}\Big)^2+
\Big(- i \frac{\partial}{\partial
 \theta_x^{'}}\Big)^2
\Bigg\}\no
&- \sum_{x, y\in \Lambda}
t_{xy}\Big\{
\cos(\theta_x-\theta_y)
+
\cos(\theta_x^{'}-\theta_y^{'})\Big\}.
\end{align} 
Next,  we introduce  a new coordinate system $\{\phi_x, \phi_x'\}$ with
\begin{align}
\phi_x=\frac{1}{2}(\theta_x^{'}-\theta_x),\ \ \ 
\phi^{'}_x=\frac{1}{2}(\theta_x^{'}+\theta_x).
\end{align} 
Then we easily see that 
\begin{align}
\mathfrak{H}_{\mathrm{ext}}=L^2
\big(\mathbb{T}^{\Lambda}\times
 \mathbb{T}^{\Lambda}, d{\boldsymbol \phi} d{\boldsymbol \phi}^{'}
\big). \label{HIden2}
\end{align} 
Using the identity
\begin{align}
\cos\theta+\cos \theta^{'}=2\cos\frac{\theta^{'}+\theta}{2} \cos \frac{\theta^{'}-\theta}{2},
\end{align} 
we obtain
\begin{align}
H_{\mathrm{ext}}
=\sum_{x\in \Lambda}\frac{U_x}{4}(\nu_x^2+\nu_x^{'2})
-2\sum_{x, y\in \Lambda} t_{xy} \cos(\phi_x-\phi_y)
 \cos(\phi_x^{'}-\phi_y^{'}),
\end{align} 
where 
\begin{align}
\nu_x=-i \frac{\partial}{\partial \phi_x},\ \ \ \nu_x^{'}=-i
 \frac{\partial}{\partial \phi_x^{'}}.
\end{align} 
Let
$\mathfrak{X}=L^2(\mathbb{T}^{\Lambda}, d\bphi)$. Then by
(\ref{HIden2}), we  obtain the following  identification:
\begin{align}
\mathfrak{H}_{\mathrm{ext}}=L^2(\mathbb{T}^{\Lambda}, d\bphi)\otimes
 L^2(\mathbb{T}^{\Lambda}, d\bphi)=
\mathfrak{X}\otimes \mathfrak{X}.\label{HIden3}
\end{align} 
 Moreover, we obtain the following proposition.

\begin{Prop}
We have $H_{\mathrm{ext}}=\mathsf{T}-\mathsf{V}$, where
\begin{align}
\mathsf{T}&=\sum_{x\in \Lambda}\frac{U_x}{4}\big(\nu_x^2\otimes \one+\one
 \otimes \nu_x^{2}\big),\\
\mathsf{V}&=2\sum_{x, y\in \Lambda} t_{xy} \cos(\phi_x-\phi_y)\otimes  \cos(\phi_x-\phi_y).
\end{align} 
\end{Prop} 

Let $\vartheta$ be the antilinear isomorphism defined by 
\begin{align}
(\vartheta f)(\bphi)=\overline{f}(\bphi)\ \ \mbox{a.e., \ \ $f\in
 L^2(\mathbb{T}^{\Lambda}, d\bphi)$}.
\end{align} 
By (\ref{Ident}) and (\ref{HIden3}), we have the identification
$
\mathfrak{H}_{\mathrm{ext}}=\mathscr{L}^2(\mathfrak{X})
$
 by $\vartheta$.
Moreover, by (\ref{Ident}), we have the following proposition:

\begin{Prop}
We have $H_{\mathrm{ext}}=\mathsf{T}-\mathsf{V}$, where
\begin{align}
\mathsf{T}&=\sum_{x\in \Lambda}\frac{U_x}{4}\big\{\mathcal{L}(\nu_x^2)+\mathcal{R}(\nu_x^{2})\big\},\\
\mathsf{V}&=2\sum_{x, y\in \Lambda} t_{xy}\mathcal{L}\Big[\cos(\phi_x-\phi_y)\Big]\mathcal{R}\Big[  \cos(\phi_x-\phi_y)\Big].
\end{align} 
\end{Prop}

By Corollary \ref{GeneralPP2}, we immediately obtain the following:
\begin{coro}\label{SemiPP}
We have 
$\exp(-\beta H_{\mathrm{ext}}) \Br 0$
 w.r.t. $\mathscr{L}^2(\mathfrak{X})_+$ for all $\beta \ge 0$.
\end{coro}

\subsubsection{Completion of proof of Theorem \ref{QGII1} and Corollary \ref{QGII2}}

\begin{Prop}\label{PPMate}
We have the following:
\begin{itemize}
\item[{\rm (i)}]
$\displaystyle 
\cos \theta_x\otimes \one+\one \otimes \cos \theta_x
=2\mathcal{L}(\cos \phi_x )\mathcal{R}( \cos \phi_x) \Br 0
$ w.r.t. $\mathscr{L}^2(\mathfrak{X})_+$.
\item[{\rm (ii)}]
$\displaystyle 
\cos \theta_x\otimes \one-\one \otimes \cos \theta_x
=2\mathcal{L}(\sin \phi_x )\mathcal{R}( \sin \phi_x) \Br 0
$ w.r.t. $\mathscr{L}^2(\mathfrak{X})_+$.
\end{itemize} 
\end{Prop} 
{\it Proof.} (i), (ii) We apply Ginibre's idea \cite{Ginibre}:
\begin{align}
\cos a+\cos b&=2\cos \frac{b+a}{2} \cos\frac{b-a}{2},\\
\cos a-\cos b&=2\sin \frac{b+a}{2} \sin\frac{b-a}{2}.   \ \ \ \ \Box
\end{align}

Put 
\begin{align}
2V_{x}^{(\vepsilon)}=C_{x}\otimes \one +\vepsilon\one \otimes C_{x},\ \
     \vepsilon=\pm 1.
\end{align} 
Then by Proposition \ref{PPMate}, we have 
$V_{x}^{(\vepsilon)}\Br 0$ w.r.t. $\mathscr{L}^2(\mathfrak{X})_+$ for
all $x\in \Lambda$ and $\vepsilon\in \{\pm 1\}$. Since $\exp(-\beta H_{\mathrm{ext}}) \Br 0$
w.r.t. $\mathscr{L}^2(\mathfrak{X})_+$ for
all $\beta \ge 0$ by Corollary \ref{SemiPP}, we can apply Theorem \ref{BasicPPP}.
Thus,  we conclude Theorem \ref{QGII1}.

For each $\mathcal{A}\subseteq \Lambda$, define
$[\mathcal{A}]=\{m_x\}_{x\in \Lambda}\in
\{0, 1\}^{\Lambda}$ by $m_x=1$ if $x\in \mathcal{A}$ and $m_x=0$ otherwise.
For simplicity, we will consider the case where $A=[\mathcal{A}]$ and $B=[\mathcal{B}]$.
To prove Corollary  \ref{QGII2}, we note
\begin{align}
C_{x}\otimes \one =V_x^{(+1)}+V_x^{(-1)},\ \ \ \one \otimes
 C_x=V_x^{(+1)}-V_x^{(-1)}.
\end{align} 
Observe that 
\begin{align}
&2\big\la C^A C^B\big\ra_{\beta}-2\la C^A \ra_{\beta} \la C^B\ra_{\beta}\no
=&
\Big\la\!\!\Big\la
\Big(
C^A\otimes \one -\one \otimes C^A
\Big)
\Big(
C^B\otimes \one -\one \otimes C^B
\Big)
\Big\ra\!\!\Big\ra_{\beta}\no
=&\sum_{\mathcal{X}\subseteq \mathcal{A}}\sum_{\mathcal{Y}\subseteq \mathcal{B}}
 \underbrace{\Big[1-(-1)^{|\mathcal{X}|}\Big]\Big[1-(-1)^{|\mathcal{Y}|}
\Big]}_{\ge 0} \underbrace{ \Big\la\!\!\Big\la
V_{\mathcal{A}\backslash \mathcal{X}}^{(+1)}V_{\mathcal{X}}^{(-1)} V_{\mathcal{B}\backslash \mathcal{Y}}^{(+1)}V_{\mathcal{Y}}^{(-1)}
\Big\ra\!\!\Big\ra_{\beta}}_{\ge 0{\rm\   by\  Theorem\  \ref{QGII1}}}
\ge 0,
\end{align}
where $V_{\mathcal{A}}^{(\pm 1)}=\prod_{x\in \mathcal{A}}V_x^{(\pm 1)}$. Hence,  we conclude
Corollary \ref{QGII2}. $\Box$

\subsubsection{Completion of proof of Theorem \ref{QGNII2}}

\begin{Prop}\label{NumPP}
For all $x\in \Lambda, \beta \ge 0$ and $\vepsilon\in \{\pm 1\}$, we have 
\begin{align}
(n_x\otimes \one + \one \otimes n_x) 
(n_x\otimes \one - \one \otimes n_x) 
\Br 0\ \ \ \mbox{w.r.t. $\mathscr{L}^2(\mathfrak{X})_+$}.
\end{align} 
\end{Prop} 
{\it Proof.}
Note that since $\vartheta \nu_x^* \vartheta=-\nu_x$, we have 
\begin{align}
n_x\otimes \one +\one \otimes n_x=\one \otimes \nu_x=-\mathcal{R}(\nu_x),
\end{align} 
and  
\begin{align}
n_x\otimes \one -\one \otimes n_x=-\nu_x \otimes \one=-\mathcal{L}(\nu_x).
\end{align} 
Thus,  we have 
$
(n_x\otimes \one +\one \otimes n_x) 
(n_x\otimes \one - \one \otimes n_x) 
=\mathcal{L}(\nu_x)\mathcal{R}(\nu_x)\Br 0
$
w.r.t. $\mathscr{L}^2(\mathfrak{X})_+$.
 $\Box$
\medskip\\

By Proposition \ref{NumPP}, we see that 
\begin{align}
&\prod_{j=1}^{{n}\atop{\longrightarrow}}
\Big[
n_{x_j}(s_j)\otimes \one + \one \otimes n_{x_j}(s_j)
\Big]
\Big[
n_{x_j}(s_j)\otimes \one - \one \otimes n_{x_j}(s_j)
\Big]\, \ex^{-\beta H_{\mathrm{ext}}}\no
=&\underbrace{\ex^{-sH_{\mathrm{ext}}}
}_{\Br 0}
\underbrace{\Big[
n_{x_1}\otimes \one + \one \otimes n_{x_1}
\Big]
\Big[
n_{x_1}\otimes \one - \one \otimes n_{x_1}
\Big]}_{\Br 0}
\,\underbrace{ \ex^{-(s_2-s_1)H_{\mathrm{ext}}}
}_{\Br 0}\times 
\cdots\no
&\cdots\times 
\underbrace{\Big[
n_{x_n}\otimes \one + \one \otimes n_{x_n}
\Big]
\Big[
n_{x_n}\otimes \one - \one \otimes n_{x_n}
\Big]
}_{\Br 0}
\, \underbrace{\ex^{-(\beta -s_n)H_{\mathrm{ext}}}}_{\Br 0}
\Br 0 \ \ \ \mbox{w.r.t. $\mathscr{L}^2(\mathfrak{X})_+$}.
\end{align} 
Therefore,  Theorem \ref{QGNII2} follows from Proposition  \ref{PPImplyP}. $\Box$

\subsubsection{Completion of proof of Theorem \ref{QGII3}}
By Propositions \ref{PPMate} and \ref{NumPP}, we know 
$\displaystyle 
\Bigg[\prod_{j=1}^{{n}\atop{\longrightarrow}} \alpha_{x_j}^{(\mu_j)}(s_j)\Bigg]\,
\ex^{-\beta H_{\mathrm{ext}}}
\Br 0
$
w.r.t. $\mathscr{L}^2(\mathfrak{X})_+$.
Thus,  Theorem \ref{QGII3} immediately follows from 
 Proposition  \ref{PPImplyP}. $\Box$

\subsection{Proof of Example \ref{MonoCorr}}\label{ProofExam}
The proof of Example \ref{MonoCorr} is similar to that of Example
\ref{IsingEx},
so we  only provide a sketch.
By the Duhamel formula (\ref{Duhamel}), we obtain
\begin{align}
&\frac{\partial}{\partial t_{xy}}\big\la C^A\ra_{\beta}\no
=&\sum_{n\ge 1} \frac{n(t_{xy})^{n-1}}{2} \int_{0\le t_1\le \cdots \le
 t_n\le \beta}\Big\la\!\!\Big\la
\big(
C^A\otimes \one-\one \otimes C^A
\big)
\big(
\mathsf{K}_n\otimes\one -\one \otimes \mathsf{K}_n
\big)\Big\ra\!\!\Big\ra_{H^{'}_{\mathrm{ext}}, \beta} dt_1\cdots dt_n,\label{RotorDu}
\end{align} 
where
$\mathsf{K}_n=\ex^{-t_1 H^{'}} \cos(\theta_x-\theta_y)\, \ex^{t_1H^{'}}\cdots
 \ex^{-t_nH^{'}} \cos(\theta_x-\theta_y)\, \ex^{t_nH^{'}}$
with $H^{'}=H+t_{xy}\cos(\theta_x-\theta_y)$
 and $\He^{'}=H^{'}\otimes \one +\one \otimes H^{'}$.
Since $\ex^{-tH^{'}_{\mathrm{ext}}} \Br 0,\ C^A\otimes\one -\one \otimes C^A\Br 0$ and 
$\mathsf{K}_n\otimes \one -\one \otimes \mathsf{K}_n\Br 0$
w.r.t. $\mathscr{L}^2(\mathfrak{X})_+$, we know that the RHS
of (\ref{RotorDu}) is positive. Thus,  we obtain (i).
Similarly,   we have
\begin{align}
&\frac{\partial}{\partial t_{xy}}\big\la n_z^2\ra_{\beta}\no
=&\sum_{n\ge 1} \frac{n(t_{xy})^{n-1}}{2} \int_{0\le t_1\le \cdots \le
 t_n\le \beta}\Big\la\!\!\Big\la
\underbrace{
\big(
n_z^2\otimes \one-\one \otimes n_z^2
\big)
}_{\Br 0\ \mathrm{ by\ Proposition\  \ref{NumPP}}}
\big(
\mathsf{K}_n\otimes\one -\one \otimes \mathsf{K}_n
\big)\Big\ra\!\!\Big\ra_{H^{'}_{\mathrm{ext}},  \beta} dt_1\cdots dt_n\no
\ge& 0.
\end{align} 
Hence,  we arrive at (ii).

(iii) Let $H^{''}=H-\frac{U_x}{2}n_x^2$.
By Proposition \ref{NumPP},  Corollary \ref{SemiPP}, and the Duhamel
formula (\ref{Duhamel}), we obtain
\begin{align}
&\frac{\partial}{\partial U_x} \big\la C^A\big\ra_{\beta}\Big|_{U_x=0}\no
 =&-\frac{\beta}{2}\int_0^{\beta} \Big\la\!\!\Big\la 
 \underbrace{\big(C^A\otimes \one -\one \otimes C^A
\big)
}_{\Br 0} \, 
 \underbrace{\ex^{-t \He^{''}}
 }_{\Br 0}
 \underbrace{
\big(
n_x^2\otimes \one -\one \otimes n_x^2
 \big)
}_{\Br 0}
\ex^{t \He^{''}}
\Big\ra\!\!\Big\ra_{\He^{''},\beta} dt\no
\le& 0,
\end{align}
where $\He^{''}=H^{''}\otimes \one +\one \otimes H^{''}$. 
This completes the proof. $\Box$

\section{ Bose--Hubbard model}\label{Sec3}

\setcounter{equation}{0}
\subsection{Results}
Let $\Lambda$ be a finite subset of $\BbbR^d$.
The Bose--Hubbard model on $\Lambda$ is defined  by 
\begin{align}
H=\sum_{x, y\in \Lambda}(-t_{xy})a_{x}^* a_{y}+\sum_{x\in \Lambda}U_x
n_{x}(n_{x}-\one)-\sum_{x\in \Lambda}\lambda_x(a_x^* +a_x)-\mu N_{\mathrm{b}}.
 \end{align} 
$H$ acts in the bosonic Fock space $\mathfrak{B}=\oplus_{n=0}^{\infty}
\otimes^n_{\mathrm{s}} \ell^2(\Lambda)$, where
$\otimes^n_{\mathrm{s}}\ell^2(\Lambda)$ is the $n$-fold symmetric tensor
product of $\ell^2(\Lambda)$ with $\otimes^0_{\mathrm{s}}
\ell^2(\Lambda)=\BbbC$.
$a_x$ is the bosonic  annihilation operator satisfying the canonical
commutation relations (CCRs):
\begin{align}
[a_x, a_y^*]=\delta_{xy},\ \ \ [a_x, a_y]=0.
\end{align} 
$n_x=a_x^* a_x$ is the number operator at site $x\in \Lambda$ and
$N_{\mathrm{b}}=\sum_{x\in \Lambda} n_x$ is the total number operator.

We assume the following:
\begin{itemize}
\item[{\bf (A. 1)}] $t_{xy} \ge 0,\ \ U_x> 0,\ \ \lambda_x\ge 0$ for all
	     $x,y\in \Lambda$.
\item[{\bf (A. 2)}] $t_{xy}=t_{yx}$ for all $x,y\in \Lambda$ and
	     $t_{xx}=0$ for all $x\in \Lambda$.
\item[{\bf (A. 3)}] $\mu\in \BbbR$.
\end{itemize} 
Under these conditions, we  see that $\ex^{-\beta H}$ is in the  trace
class  for all $\beta >0$.
The thermal expectation  value is defined as
\begin{align}
\la X\ra_{\beta}=\Tr\big[X\, \ex^{-\beta H}\big]\Big/Z_{\beta},\ \ \
 Z_{\beta}=\Tr\big[\ex^{-\beta H}\big].
\end{align} 
For each densely defined linear operator $X$, $X^{\#}$ ($\#=+$ or $-$
) means 
\begin{align} 
X^{\#}=
\begin{cases}
X & \mbox{ if $\#=-$}\\
X^* & \mbox{ if $\#=+$}.
\end{cases} 
\end{align} 
Set $\BbbN_0=\{0\} \cup \BbbN$.
For each $\mb{m}=\{m_x\}_{x\in \Lambda}\in \BbbN_0^{\Lambda}$ 
 and ${\boldsymbol
\#}=\{\#_x\}_{x\in \Lambda}\in \{\pm \}^{\Lambda}$,  define
\begin{align}
I(\mb{m}; {\boldsymbol
\#} )=\prod_{x\in \Lambda} \big(a_x^{\#_x}\big)^{m_x}
\end{align} 
with $
\big(a_x^{\#_x}\big)^{0}=\one
$.
Now we define 
\begin{align}
\mathfrak{A}=\mathrm{Coni}
\Big\{
I(\mb{m}; {\boldsymbol
\#} )\ \Big|\
\mb{m}\in \BbbN_0^{\Lambda},\ {\boldsymbol
\#}\in \{\pm \}^{\Lambda}
\Big\}.
\end{align} 
Note that for all $A\in \mathfrak{A}$, $A\,  \ex^{-\beta H}$ is in  the trace
class for all $\beta >0$. Thus,  $\la A \ra_{\beta}$ is finite.

\begin{Thm}[First Griffiths inequality]\label{BHGriI2}
Let $A_1, \dots, A_n\in \mathfrak{A}$. For all $0\le s_1\le s_2\le \cdots \le s_n < \beta$, we have 
\begin{align}
\Bigg\la
 \prod_{j=1}^{{n}\atop{\longrightarrow}}A_j(s_j)
\Bigg\ra_{\beta} \ge 0,
\end{align} 
where $A(s)=\ex^{-s H} A \ex^{sH}$. 
\end{Thm} 

To state the second quantum Griffiths inequality, we introduce the
following notation:
\begin{align}
\la\!\la Y\ra\!\ra_{\beta}:=\Tr_{\mathfrak{B}\otimes \mathfrak{B}} \Big[
Y \, \ex^{-\beta H_{\mathrm{ext}}}
\Big]\Big/Z_{\beta}^2,\ \ H_{\mathrm{ext}}=H\otimes \one +\one \otimes H.
\end{align}  

\begin{Thm}\label{BHGriII}
Let $x_1,\dots,  x_n, y_1,\dots, y_n\in \Lambda$.
For each   $0 \le s_1\le t_1 \le s_2\le t_2\le \cdots\le
 s_{n}\le t_n < \beta,\ \#_1,\dots, \#_n\in \{\pm\}$ and $\vepsilon_1,\dots, \vepsilon_n\in \{\pm 1\}$,
 we have 
\begin{align}
\Bigg\la\!\!\!\Bigg\la \prod_{j=1}^{{n\atop{\longrightarrow}}}
\Big[
a_{x_j}^{\#_j}(s_j)\otimes \one+\vepsilon_j\one \otimes a_{x_j}^{\#_j}(s_j)
\Big]
\Big[
a_{y_j}^{\overline{\#}_j}(t_j)\otimes \one+\vepsilon_j\one \otimes a_{y_j}^{\overline{\#}_j}(t_j)
\Big]
\Bigg\ra\!\!\!\Bigg\ra_{\beta} \ge 0,
\end{align} 
where 
 $\overline{\#}=-\#$ \footnote{To be precise, $\overline{+}=-$ and
 $\overline{-}=+$.} and $a_x^{\#}(s)=\ex^{-s H} a_x^{\#}\ex^{sH}$.
\end{Thm} 

\begin{example}
{\rm 
Consider the case where  $n=1$, $\vepsilon_1=-1$,  and
 $\#_1=+$. Then we have 
\begin{align}
\la a_x^*(s) a_y(t)\ra_{\beta}-\la a_x^*\ra_{\beta}\la a_y\ra_{\beta}\ge 0
\end{align} 
for all $x,y\in \Lambda$ and $0\le s \le t < \beta$.
From this, we have 
\begin{align}
(a_x^*, a_y)_{\beta}-\la a_x^*\ra_{\beta}\la a_y\ra_{\beta}\ge 0.
\end{align} 
In addition, by Theorem \ref{BHGriI2}, it follows  that 
\begin{align}
(a_x^*, a_y)_{\beta}\ge 0,\ \ \la a_x^*\ra_{\beta}\ge 0,\ \ \la
 a_y\ra_{\beta}\ge 0.\ \ \ \ \ \diamondsuit
\end{align} 
}
\end{example}

We can generalize  Theorem \ref{BHGriII}.  
To state our result, we need to introduce the following:
\begin{align}
\alpha_{+1, x}&=a_x\otimes \one+\one \otimes a_x,\\
\alpha_{-1, x}&= - i\big(a_x\otimes \one-\one \otimes a_x\big),
\end{align} 
where  $i=\sqrt{-1}$.

\begin{Thm}[Second Griffiths inequality]\label{BHGriIII}
Let $x_1,\dots, x_n\in \Lambda$. For all $ \#_1, \dots, \#_n\in \{\pm
 \},\ \vepsilon_1, \dots, \vepsilon_n\in \{\pm 1\}$ and $0\le s_1\le s_2\le \cdots \le s_n < \beta$, we have 
\begin{align}
\Bigg\la\!\!\!\Bigg\la
 \prod_{j=1}^{{n}\atop{\longrightarrow}}\alpha_{\vepsilon_j, x_j}^{\#_j}(s_j)
\Bigg\ra\!\!\!\Bigg\ra_{\beta} \ge 0,
\end{align} 
where $\alpha_{\vepsilon, x}^{\#}(s)=\ex^{-s \He} \alpha_{\vepsilon,
 x}^{\#} \ex^{s\He}$.
\end{Thm} 
\begin{rem}
{\rm
If $\lambda_x> 0$ for all $x\in \Lambda$, then we can prove that the ground state of $H$
 is unique\footnote{This fact follows from an application of the
 Perron--Frobenius--Faris theorem\cite{Faris}.}. In this case, our results are valid at $\beta =\infty$. $\diamondsuit$
}
\end{rem}

\begin{example}
{\rm 
Consider the case where $n=3,\ \#_1=+,  \#_2=\#_3=-$,  and
 $\vepsilon_1\vepsilon_2\vepsilon_3=1$.
We have 
\begin{align}
\la a_1^* a_2 a_3\ra
-\la a_1^*\ra \la a_2 a_3\ra
-\la a_2\ra\la a_1^* a_3\ra
+\la a_3\ra \la a_1^* a_2\ra\ge 0 \label{BHEx1}
\end{align} 
for $
(\vepsilon_1, \vepsilon_2,
 \vepsilon_3)=(-1, -1, +1)
$, and 
\begin{align}
\la a_1^* a_2 a_3\ra
-\la a_1^*\ra \la a_2 a_3\ra
+\la a_2\ra\la a_1^* a_3\ra
-\la a_3\ra \la a_1^* a_2\ra\ge 0 \label{BHEx2}
\end{align} 
for $
(\vepsilon_1, \vepsilon_2,
 \vepsilon_3)=(-1, +1, -1)
$, where we use the  abbreviation $a_j^{\#}=a^{\#}_{x_j}(s_j)$. On the other hand,
 we have
\begin{align}
\la a_1^* a_2 a_3\ra
+\la a_1^*\ra \la a_2 a_3\ra
-\la a_2\ra\la a_1^* a_3\ra
-\la a_3\ra \la a_1^* a_2\ra\le 0
\end{align} 
for $
(\vepsilon_1, \vepsilon_2,
 \vepsilon_3)=(+1, -1, -1)
$. Combining (\ref{BHEx1}) and (\ref{BHEx2}), we get
\begin{align}
\la a_1^* a_2a_3\ra-\la a_1^*\ra \la a_2a_3\ra \ge 0.\ \ \ \ \diamondsuit
\end{align} 

}
\end{example} 

If $U_x\equiv 0$, then we obtain a stronger result as follows. 

\begin{Thm}\label{U0}
Assume that $U_x=0$ for all $x\in \Lambda$. Assume that the matrix
$(-t_{xy}-\mu \delta_{xy})_{x, y}$ is positive-definite.\footnote{
This assumption is needed in order to guarantee that $\ex^{-\beta H}$ is
 a  trace class operator.}
Let $x_1, \dots, x_n\in \Lambda$. For all $\#_1, \dots, \#_n\in
 \{\pm\},\ 
\vepsilon_1,\dots, \vepsilon_n\in \{\pm 1\}$ and 
  $0\le s_1\le \cdots\le s_n  <\beta$, we have 
\begin{align}
\Bigg\la\!\!\!\Bigg\la
 \prod_{j=1}^{{n}\atop{\longrightarrow}}
\bigg[
a_{x_j}^{\#_j}(s_j)\otimes \one +\vepsilon_j \one \otimes a_{x_j}^{\#_j}(s_j)
\bigg]
\Bigg\ra\!\!\!\Bigg\ra_{\beta} \ge 0.
\end{align} 
\end{Thm} 

\begin{coro}\label{U02}
Under the same assumptions as  Theorem \ref{U0},
 we have 
\begin{align}
\la A_1A_2\ra_{\beta}-\la A_1\ra_{\beta}\la A_2\ra_{\beta} \ge 0
\end{align} 
for  all $A_1, A_2\in \mathfrak{A}$.
\end{coro} 

\begin{example}
{\rm 
Let $A\in \mathfrak{A}$. Under the same assumptions as in Theorem \ref{U0},
 we have the following:
\begin{itemize}
\item[(i)] $\la A\ra_{\beta}$ is monotonically increasing in $t_{xy}$.
\item[(ii)] $\la A\ra_{\beta}$ is monotonically increasing in $\lambda_x$.
\end{itemize} 
The proofs of these properties are similar to those of  Examples \ref{IsingEx} and \ref{MonoCorr}. $\diamondsuit$
}
\end{example} 

\subsection{Proof of Theorem  \ref{BHGriI2}}

In this section, we will  often discuss unbounded operators. Thus,  we have to
extend definitions of our operator inequalities as follows:

\begin{define}
{\rm 
Let $A$ be  a densely defined linear operator in $\mathfrak{H}$. 
If $Ax \ge 0$ w.r.t. $\Cone$ for all $x\in \Cone\cap \D(A)$,
then we also write $A\unrhd 0$ w.r.t. $\Cone$.  Note that  
\begin{align}
\la x|Ay\ra \ge  0\ \ \ \mbox{for all $x\in \Cone$ and $y\in \Cone\cap
 \D(A)$.\ \ \ \ $\diamondsuit$}
\end{align} 
}
\end{define}

For each $\mb{N}=\{N_x\}_{x\in \Lambda}\in \BbbN_0^{\Lambda}$, we set
\begin{align}
|\mb{N}\ra=\Bigg(\prod_{x\in \Lambda} N_x!\Bigg)^{-1/2} \prod_{x\in
 \Lambda}
(a_x^*)^{N_x} \Omega,
\end{align} 
where $\Omega$ is the Fock vacuum. Then $\big\{|\mb{N}\ra\, |\, \mb{N}\in
\BbbN_0^{\Lambda}\big\}$ is a CONS of $\mathfrak{B}$.

\begin{define}{\rm 
A standard self-dual cone in $\mathfrak{B}$ is defined by 
\begin{align}
\mathfrak{B}_+=\Bigg\{
\psi\in \mathfrak{B}\ \Bigg|\ \psi=\sum_{\mb{N}\in \BbbN_0^{\Lambda}}
 \psi_{\mb{N}} |\mb{N}\ra,\ \ \psi_{\mb{N}}\ge 0\ \forall \mb{N}\in \BbbN_0^{\Lambda}
\Bigg\}.
\end{align}
$\mathfrak{B}_+$ was introduced by Fr\"ohlich \cite{JFroehlich1}, see also
 \cite{Miyao4}. $\diamondsuit$
}
\end{define} 
\begin{rem}
{\rm 
$|\mb{N}\ra\in \mathfrak{B}_+$ for all $\mb{N}\in \BbbN_0^{\Lambda}$. $\diamondsuit$
}
\end{rem} 

The following lemma is useful in this section.
\begin{lemm}\label{UbPP}
Let $A$ be a densely defined  linear operator on $\mathfrak{B}$.
Let $P_{\ell}$ be the orthogonal projection onto
 $\oplus_{n=0}^{\ell}\otimes^n_{\mathrm{s}} \ell^2(\Lambda)$.
Assume the following:
\begin{itemize}
\item[{\rm (i)}] $|\mb{N}\ra \in \D(A)$ for all $\mb{N}\in
	     \BbbN_0^{\Lambda}$.
\item[{\rm (ii)}] $AP_{\ell}\vphi \to A\vphi$ as $\ell \to \infty$ for all
	     $\vphi\in  \D(A)$.
\end{itemize} 
Then the following are equivalent.
\begin{itemize}
\item[{\rm (a)}] $A\unrhd 0$ w.r.t. $\mathfrak{B}_+$.
\item[{\rm (b)}] $\la \mb{M}|A|\mb{N}\ra \ge 0$ for all $\mb{M},
	     \mb{N}\in \BbbN_0^{\Lambda}$\footnote{$\la \psi
	     |X|\phi\ra:=\la \psi|X\phi\ra$.
}.
\end{itemize} 
\end{lemm} 
{\it Proof.} (a) $\Longrightarrow $(b): This is immediate.

(b) $\Longrightarrow$ (a): Let $A_{\ell}=P_{\ell} A P_{\ell}$. Then,
for all $\vphi\in \mathfrak{B}_+$ and $\psi\in \D(A)\cap \mathfrak{B}_+$, we see that 
\begin{align}
\la \vphi| A_{\ell} \psi\ra= \sum_{|\mb{M}| \le \ell, |\mb{N}| \le \ell}
 \underbrace{\vphi_{\mb{M}}}_{\ge 0} \underbrace{\psi_{\mb{N}}}_{\ge 0}
 \underbrace{\la \mb{M}|A|\mb{N}\ra}_{\ge 0} \ge 0,
\end{align} 
where $|\mb{N}|=\sum_{x\in \Lambda}N_x$. Taking $\ell\to \infty$, we
obtain $\la \vphi|A\psi\ra \ge 0$, which implies $A\psi\ge 0$
w.r.t. $\mathfrak{B}_+$. $\Box$

\begin{Prop}\label{BosonPP}
We have  $a_x\unrhd 0,\ a_x^*\unrhd 0$ w.r.t. $\mathfrak{B}_+$ for all
	     $x\in \Lambda$.

\end{Prop} 
{\it Proof.}
It is not difficult   to  verify  that $a_x$ and $a_x^*$ satisfy the
assumptions of Lemma \ref{UbPP}.
Moreover,   we see that $
\la \mb{M}|a_x|\mb{N}\ra\ge 0
$
and $\la \mb{M}|a_x^*|\mb{N}\ra\ge 0$ for all $\mb{M}, \mb{N}\in
\BbbN_0^{\Lambda}$. Thus,  we obtain the desired assertion by Lemma \ref{UbPP}. 
$\Box$

\begin{coro}\label{APP}
For all $A\in \mathfrak{A}$, it holds that  $A\unrhd 0$ w.r.t. $\mathfrak{B}_+$.
\end{coro} 

\begin{Prop}\label{BHExpPP}
We have
$\ex^{-\beta  H} \unrhd 0$ w.r.t. $\mathfrak{B}_+$ for all
	     $\beta \ge 0$.
\end{Prop} 
{\it Proof.}   
Let $P_{\ell}$ be the orthogonal projection defined
in Lemma \ref{UbPP}. Let $H_{\ell}=P_{\ell} H P_{\ell}$.
Since $H_{\ell}$ converges to $H$ in the strong resolvent sense as
$\ell\to \infty$, it suffices to show that 
\begin{align}
\ex^{-\beta H_{\ell}} \unrhd 0\ \ \ \mbox{w.r.t. $\mathfrak{B}_+$ for all
 $\beta \ge0$ and $\ell \in \BbbN$.}
\end{align} 
To this end, we set
\begin{align}
T=\sum_{x, y\in \Lambda}t_{xy}a_{x}^* a_{y}+\sum_{x\in
 \Lambda}\lambda_x(a_x+a_x^*)+\mu N_{\mathrm{b}},\ \ \ 
U=\sum_{x\in \Lambda}U_x
n_{x}(n_{x}-\one). \label{DefT}
\end{align} 
Let $T_{\ell}=P_{\ell} TP_{\ell},\
U_{\ell}=P_{\ell} U P_{\ell}$ . Then $T_{\ell}$ and $U_{\ell}$
are bounded for each $\ell \in \BbbN$.  We   observe that 
$
\la \mb{M}|T_{\ell}|\mb{N}\ra\ge 0.
$
Thus,  by Proposition \ref{Prop3}, $T_{\ell} \unrhd 0$ w.r.t. $\mathfrak{B}_+$ holds
 for all $\ell\in \BbbN$.
 On the other hand, 
\begin{align}
\la \mb{M}|\ex^{-\beta U_{\ell}}|\mb{N}\ra
=
\begin{cases}
\exp\Big\{
-\beta \sum_{x\in \Lambda}N_x(N_x-1)\Big\}
\delta_{\mathbf{ MN}} & \mbox{if
 $|\mb{M}|\le \ell$  and $|\mb{N}|\le \ell$}\\
\delta_{\mathbf{ MN}} & \mbox{if
 $|\mb{M}|> \ell$  or  $|\mb{N}|> \ell$}
\end{cases}. 
\end{align} 
This means $\la \mb{M}|\ex^{-\beta U_{\ell}}|\mb{N}\ra \ge 0$. Thus, 
applying Proposition \ref{Prop3}, we conclude $\ex^{-\beta U_{\ell }} \unrhd
0$
w.r.t. $\mathfrak{B}_+$ for all $\beta \ge 0$ and $\ell\in \BbbN$.
Hence,  by Proposition \ref{BasicPP}, we conclude $\ex^{-\beta H_{\ell}}
\unrhd 0$ w.r.t. $\mathfrak{B}_+$ for all $\beta \ge 0$ and $\ell\in \BbbN$.
$\Box$

\begin{coro}\label{TimeBHPP}
Let $x_1,\dots, x_n\in \Lambda$. For all $ \#_1, \dots, \#_n\in \{\pm
 \}$ and $0 \le  s_1\le s_2\le \cdots \le s_n < \beta$, we have 
\begin{align}
\ex^{-s_1 H} a_{x_1}^{\#_1} \ex^{-(s_2-s_1)H}a_{x_2}^{\#_2}\ex^{-(s_3-s_2)H}\cdots 
\ex^{-(s_{n}-s_{n-1})H} a_{x_n}^{\#_n}
\ex^{-(\beta-s_n)H}\unrhd 0
\end{align} 
w.r.t. $\mathfrak{B}_+$.
\end{coro} 

\subsubsection{Completion of proof of Theorem \ref{BHGriI2}}

By Corollary \ref{APP} and Proposition \ref{BHExpPP}, we have 
\begin{align}
&\Bigg[\prod_{j=1}^{{n}\atop{\longrightarrow}} A_j(s_j)\Bigg]\ex^{-\beta
 H}\no
=&\underbrace{\ex^{-sH}}_{\unrhd 0}
 \underbrace{A_1}_{\unrhd 0}
\underbrace{\ex^{-(s_2-s_1)H}}_{\unrhd 0}\cdots 
\underbrace{A_n}_{\unrhd 0}
\underbrace{\ex^{-(\beta-s_n)H}}_{\unrhd 0}\unrhd 0\ \ \ \mbox{w.r.t. $\mathfrak{B}_+$}.
\end{align} 
 Thus,  by Proposition \ref{Prop2}, we conclude Theorem \ref{BHGriI2}. $\Box$

\subsection{Proof of Theorems \ref{BHGriII} and \ref{BHGriIII}}
Let $\mathfrak{B}_{\mathrm{ext}}=\mathfrak{B}\otimes \mathfrak{B}$.
We introduce a new  representation of the CCRs as follows.
Let 
\begin{align}
\xi_x=\frac{1}{\sqrt{2}}\big(a_x\otimes \one +\one \otimes a_x\big),\ \
 \ \eta_x=-\frac{1}{\sqrt{2}}\big(
a_x\otimes \one-\one \otimes a_x
\big).
\end{align} 
$\xi_x$ and $\eta_x$ act in $\mathfrak{B}_{\mathrm{ext}}$ and  are closable. We denote their closures by the same symbols.
Then $\{\xi_x,\eta_x\}$ satisfies the following  CCRs:
\begin{align}
&[\xi_x, \xi_y]=0, \ \ \ \ [\eta_x, \eta_y]=0,\ \ \ [\xi_x, \eta_y]=0,\\
&[\xi_x, \xi_y^*]=\delta_{xy},\ [\eta_x, \eta_y^*]=\delta_{xy},\ [\xi_x,
 \eta_y^*]=0.
\end{align} 
Using $\xi_x$ and $\eta_x$, we can rewrite $H$ as 
\begin{align}
H_{\mathrm{ext}}=-\mathbb{T}+\mathbb{U},
\end{align} 
where
\begin{align}
\mathbb{T}=& \sum_{x, y\in \Lambda}t_{xy}(\xi_x^* \xi_y+\eta_x^*\eta_y)
+\sqrt{2}\sum_{x\in \Lambda} \lambda_x(\xi_x+\xi_x^*)
\end{align} 
and
\begin{align}
\mathbb{U}=& \mathbb{U}_d+\mathbb{U}_o,\\
\mathbb{U}_d=&\sum_{x\in \Lambda}\frac{U_x}{2} 
\Big(
\xi_x^*\xi_x \xi_x^*\xi_x
+4\xi_x^*\xi_x\eta_x^*\eta_x+\eta_x^*\eta_x \eta_x^*\eta_x
\Big)\no
&-\sum_{x\in \Lambda}\Big(\frac{1}{2}U_x+\mu\Big)(\xi_x^*
 \xi_x+\eta_x^*\eta_x),\no
\mathbb{U}_o=&\sum_{x\in \Lambda}\frac{U_x}{2} \Big(
\xi_x^* \xi_x^*\eta_x\eta_x
+
\xi_x \xi_x\eta_x^*\eta_x^*
\Big).
\end{align} 

Let $\Omega_{\mathrm{ext}}=\Omega\otimes \Omega\in
\mathfrak{B}_{\mathrm{ext}}$. For each $\mb{M}, \mb{N}\in
\BbbN_0^{\Lambda}$,  we define
\begin{align}
|\mb{M}, \mb{N}\ra\!\ra=
\Bigg(
\prod_{x\in \Lambda}M_x! N_x!
\Bigg)^{-1/2} \prod_{x\in \Lambda} (\xi_x^*)^{M_x} (\eta_x^*)^{N_x} \Omega_{\mathrm{ext}}.
\end{align} 
Clearly,  $
\big\{
|\mb{M}, \mb{N}\ra\!\ra\, |\, \mb{M}, \mb{N}\in \BbbN_0^{\Lambda}
\}
$ is a CONS of $\mathfrak{B}_{\mathrm{ext}}$.

\begin{define}
{\rm 
We define a self-dual cone in $\mathfrak{B}_{\mathrm{ext}}$ by 
\begin{align}
\mathfrak{B}_{\mathrm{ext}, +}
=\Bigg\{
\Psi\in \mathfrak{B}_{\mathrm{ext}}
\ \Bigg|\ \Psi=\sum_{\mb{M}, \mb{N}\in \BbbN_0^{\Lambda}}
\Psi_{\mb{M}, \mb{N}}|\mb{M}, \mb{N}\ra\!\ra,\ \Psi_{\mb{M}, \mb{N}}\ge
 0\ 
\forall \mb{M}, \mb{N}\in \BbbN_0^{\Lambda}
\Bigg\}.\ \ \ \diamondsuit
\end{align} 
}
\end{define} 

We can prove the following in a  manner similar to  that used for 
Proposition \ref{BosonPP}.
\begin{Prop}\label{ExtAniPP}
We have  $\xi_x^{\#}\unrhd 0,\  \eta^{\#}_x \unrhd 0$
 w.r.t. $\mathfrak{B}_{\mathrm{ext}, +}$ for all $x\in
 \Lambda$ and $ \#\in \{\pm\}$.
\end{Prop} 

\begin{Prop}\label{BHExtHPP}
Let 
\begin{align}
\mathscr{U}=\exp\Bigg\{
-i \frac{\pi}{2}\sum_{x\in \Lambda}\eta_x^*\eta_x
\Bigg\}.
\end{align}
Then we have 
$
\mathscr{U}^* \ex^{-\beta H_{\mathrm{ext}}} \mathscr{U}\unrhd 0
$ w.r.t. $\mathfrak{B}_{\mathrm{ext}, +}$ for all $\beta \ge 0$.
\end{Prop} 
{\it Proof.}
Let $\hat{H}_{\mathrm{ext}}=\mathscr{U}^* H_{\mathrm{ext}} \mathscr{U}$.
It is important to note that 
\begin{align}
\mathscr{U}^* \mathbb{U}_o \mathscr{U}=- \mathbb{U}_o.
\end{align} 
Thus,  we have 
\begin{align}
\hat{H}_{\mathrm{ext}}=-\mathbb{K}+\mathbb{U}_d,
\end{align} 
where $\mathbb{K}=\mathbb{T}+\mathbb{U}_o$.
Let $\mathcal{P}_{\ell}$ be the orthogonal projection onto the closed
subspace spanned by $\big\{
|\mb{M}, \mb{N}\ra\!\ra\, |\, \mb{M}, \mb{N} \in \BbbN_0^{\Lambda},\
|\mb{M}|+|\mb{N}|\le \ell
\big\}$.
Let $\hat{H}_{\mathrm{ext}, \ell}=\mathcal{P}_{\ell}
\hat{H}_{\mathrm{ext}}\mathcal{P}_{\ell}$. Since $\hat{H}_{\mathrm{ext},
\ell}$ converges to $\hat{H}_{\mathrm{ext}}$ in the strong resolvent
sense as $\ell\to \infty$, it suffices to show that 
\begin{align}
\exp\big(-\beta \hat{H}_{\mathrm{ext}, \ell}\big) \unrhd 0\ \ \ 
 \mbox{w.r.t. $\mathfrak{B}_{\mathrm{ext}, +}$ for all $\beta\ge 0$ and
$\ell\in \BbbN$.} \label{BHExtPP}
\end{align}   The proof of this is almost parallel to that of
Proposition \ref{BHExpPP}. For reader's convenience, we provide
a sketch of it.  Let $\mathbb{K}_{\ell}=\mathcal{P}_{\ell} \mathbb{K}
\mathcal{P}_{\ell}$ and $\mathbb{U}_{d, \ell}=\mathcal{P}_{\ell}
\mathbb{U}_d \mathcal{P}_{\ell}$. 
 First, we show that $\mathbb{K}_{\ell} \unrhd 0$
 w.r.t. $\mathfrak{B}_{\mathrm{ext}, +}$ for all $\ell\in \BbbN$.
Next we show that $\exp\big(-\beta \mathbb{U}_{d, \ell}\big) \unrhd 0$
w.r.t. $\mathfrak{B}_{\mathrm{ext}, +}$ for all $\beta\ge 0$ and
$\ell\in \BbbN$. Then  by Proposition \ref{BasicPP}, we conclude
(\ref{BHExtPP}). $\Box$

\begin{Prop}\label{alphahat}
Set $
\hat{\alpha}_{\vepsilon, x}^{\#}=\mathscr{U}^* \alpha_{\vepsilon ,x}^{\#}\mathscr{U}
$. Then we have 
$
\hat{\alpha}_{\vepsilon, x}^{\#} \unrhd 0
$ w.r.t.  $\mathfrak{B}_{\mathrm{ext}, +}$ 
for all $x\in \Lambda,\vepsilon\in \{\pm 1\}$ and $\#\in \{-, +\}$.
\end{Prop} 
{\it Proof.}
By Proposition \ref{ExtAniPP}, we have 
\begin{align}
\hat{\alpha}_{+1. x}^{\#}&=\sqrt{2} \xi_x^{\#}\unrhd 0\ \ \
 \mbox{w.r.t. $\mathfrak{B}_{\mathrm{ext}, +}$},\\
\hat{\alpha}_{-1. x}^{\#}&=\sqrt{2} \eta_x^{\#}\unrhd 0\ \ \
 \mbox{w.r.t. $\mathfrak{B}_{\mathrm{ext}, +}$.\ \ \ $\Box$}
\end{align}

\subsubsection{
Completion of  proofs  of Theorems \ref{BHGriII} and \ref{BHGriIII}}
We only prove Theorem \ref{BHGriIII}, since Theorem \ref{BHGriII} is a
corollary of it.
Let $\hat{H}_{\mathrm{ext}}=\mathscr{U}^* H_{\mathrm{ext}} \mathscr{U}$.
Then we have 
\begin{align}
&\mathscr{U}^*
 \prod_{j=1}^{{n}\atop{\longrightarrow}}\alpha_{\vepsilon_j,
 x_j}^{\#_j}(s_j)\mathscr{U}\ex^{\beta \hat{H}_{\mathrm{ext}}}\no
=&
\underbrace{\ex^{-s_1 \hat{H}_{\mathrm{ext}}}}_{\unrhd 0}
\underbrace{ \hat{\alpha}_{\vepsilon_1, x_1}^{\#_1}}_{\unrhd 0}
\underbrace{\ex^{-(s_2-s_1) \hat{H}_{\mathrm{ext}}}}_{\unrhd 0}
\underbrace{\hat{\alpha}_{\vepsilon_2, x_2}^{\#_2}}_{\unrhd 0}
\cdots
\underbrace{\hat{\alpha}_{\vepsilon_n,
 x_n}^{\#_n}}_{\unrhd 0}
\underbrace{\ex^{-(\beta-s_n)\hat{H}_{\mathrm{ext}}}}_{\unrhd 0}
\unrhd 0\ \ \ \mbox{w.r.t. $\mathfrak{B}_{\mathrm{ext}, +}$} \label{LongEq}
\end{align} 
by Propositions \ref{BHExtHPP} and \ref{alphahat}.
Thus,  by Proposition \ref{Prop2}, we
obtain Theorem  \ref{BHGriIII}.
$\Box$

\subsection{Proof of Theorem \ref{U0} and Corollary \ref{U02}}
If $U_x=0$, then  we have $H=-T$, where $T$ is given by (\ref{DefT}). Thus,  instead of Proposition \ref{BHExtHPP}, we
have the following:
\begin{Prop}\label{U03}
We have $\ex^{-\beta \He}\unrhd 0$ w.r.t. $\mathfrak{B}_{\mathrm{ext},
 +}$
 for all $\beta \ge 0$.
\end{Prop} 
Note that the unitary operator $\mathscr{U}$ is unnecessary to prove 
Proposition \ref{U03}.
Hence, instead of (\ref{LongEq}), we obtain
\begin{align}
&\prod_{j=1}^{{n}\atop{\longrightarrow}}
\Big[
a_{x_j}^{\#_j}(s_j)\otimes \one +\vepsilon_j \one \otimes a_{x_j}^{\#_j}(s_j)
\Big]\, \ex^{-\beta \He}\no
=& \ex^{-s_1\He} 
\Big[
a_{x_1}^{\#_1}\otimes \one +\vepsilon_1 \one \otimes a_{x_1}^{\#_1}
\Big]
\ex^{-(s_2-s_1)\He}
\cdots 
\Big[
a_{x_n}^{\#_n}\otimes \one +\vepsilon_n \one \otimes a_{x_n}^{\#_n}
\Big]\, \ex^{-(\beta-s_n)\He}\no
\unrhd &0\ \ \ \mbox{w.r.t. $\mathfrak{B}_{\mathrm{ext}, +}$}.
\end{align} 
This completes the proof of Theorem \ref{U0}.
By applying Theorem \ref{Equiv2}, we prove Corollary \ref{U02}. $\Box$

\section{ Hubbard model}\label{Sec4}
\setcounter{equation}{0}
\subsection{Results}

\subsubsection{The finite temperature case}
Let $G=(\Lambda, E)$ be a graph with vertex set $\Lambda$ and edge
collection $E$. An edge
with end-points $x$ and $y$ will be denoted by $\{x, y\}$.
We  assume that $\{x, x\}\notin E$ for all  $x\in
\Lambda$, i.e., any loops are excluded.
In this section, we assume the following:
\begin{itemize}
\item[{\bf (G. 1)}] $|\Lambda|$ is even.
\item[{\bf (G. 2)}] $G$ is bipartite, i.e., $\Lambda$ admits a partition into two classes such
that every edge has its ends in different classes.
\end{itemize}

The Hubbard model on $G$ is given by 
\begin{align}
H=\sum_{\{x, y\}\in E}\sum_{\sigma\in \{\uparrow,
 \downarrow\}}(-t_{xy})c_{x\sigma}^* c_{y\sigma}+U\sum_{x\in \Lambda}
(n_{x\uparrow}-\tfrac{1}{2})(n_{x\downarrow}-\tfrac{1}{2}).
\end{align} 
$H$ acts  in the Hilbert space $\mathfrak{H}=\Fock\otimes \Fock$. $\Fock$
 is the fermionic Fock space defined by
 $\Fock=\oplus_{n\ge 0} \wedge^n
\ell^2(\Lambda)$, where $\wedge^n \ell^2(\Lambda)$ is  the $n$-fold
antisymmetric tensor product of $\ell^2(\Lambda)$ with $\wedge^0
\ell^2(\Lambda)=\BbbC$.
$c_{x\sigma}$ is the electron annihilation operator that satisfies the
canonical anticommutation relations (CARs):
\begin{align}
\{c_{x\sigma}, c_{x'\sigma'}^*\}=\delta_{xx'}\delta_{\sigma\sigma'},\ \ 
\{c_{x\sigma}, c_{x'\sigma'}\}=0.
\end{align}  
$n_{x\sigma}=c_{x\sigma}^* c_{x\sigma}$ is the number operator at vertex
$x\in \Lambda$. 
$t_{xy}\in \BbbR$ is the quantum mechanical amplitude of  an electron
hopping  from $y$ to $x$.  We assume that 
\begin{itemize}
\item[{\bf (T)}] $t_{xy}=t_{yx}\neq 0$ for all $\{x,y\}\in E$.
\end{itemize} 
$U$ is the strength of the Coulomb repulsion\footnote{
All results in this section can be extended  to a more general Coulomb
interaction of the form $\sum_{x, y\in
\Lambda}U_{xy}(n_{x\uparrow}-\frac{1}{2})(n_{y\downarrow}-\frac{1}{2})$,
 where $U_{xy}$ is real and positive semidefinite. 
} such that
\begin{itemize}
\item[{\bf (U)}] $U\ge0$.
\end{itemize}

Since $G$ is bipartite, $\Lambda$ can be divided into two disjoint sets
$\Lambda_e$ and $\Lambda_o$.
We  set $\mu(x)=0$ if $x\in \Lambda_e$, $\mu(x)=1$ if $x\in \Lambda_o$.
For each $x\in \Lambda$, define
\begin{align}
b_x=(-1)^{\mu(x)}c_{x\uparrow}^* \gamma_{\uparrow} c_{x\downarrow},
\end{align} 
where $\gamma_{\uparrow}=(-\one)^{N_{\uparrow}}$ with
$N_{\sigma}=\sum_{x\in \Lambda} n_{x\sigma}$.
Let 
\begin{align}
\al = \mathrm{Coni}\Big\{
b_{x_1}^{\#_1} b_{x_2}^{\#_2} \cdots b_{x_n}^{\#_n}\ \Big|\ 
x_1, \dots, x_n\in \Lambda,\ \#_1, \dots, \#_n \in \{+, -\}, \ n\in \BbbN
\Big\}.
\end{align}

We use the thermal average associated with the grand canonical Gibbs
state at  inverse temperature $\beta$:
\begin{align}
\la X\ra_{\beta}=\Tr\big[X\, \ex^{-\beta H}\big] \Big/ \Xi_{\beta},\ \
 \Xi_{\beta}=\Tr\big[
\ex^{-\beta H}
\big].
\end{align} 
For each $\beta>0$, we can verify  that
$\la n_x\ra_{\beta}=1$, where $n_x=n_{x\uparrow}+n_{x\downarrow}$.
This means that the system at half-filling will be considered.

\begin{Thm}[First Griffiths inequality]\label{HubbardGI2}
Let $A_1, \dots, A_n\in \mathfrak{A}$.
For all  $0 \le  s_1 \le  s_2 \le  \cdots \le s_n  \le \beta$, we have 
\begin{align}
\Bigg\la
 \prod_{j=1}^{{n}\atop{\longrightarrow}}
A_j(s_j)
\Bigg\ra_{\beta} \ge 0,
\end{align} 
where $A(s)=\ex^{-s H} A \, \ex^{s H}$.
\end{Thm} 
\begin{example}{\rm
For each $x_1, \dots, x_n\in \Lambda$, $\#_1, \dots, \#_n\in \{+,
 -\}$ and $0 \le  s_1 \le  s_2 \le  \cdots \le s_n  \le \beta$,
  we have
\begin{align}
\Big\la 
b_{x_1}^{\#_1}(s_1)b_{x_2}^{\#_2}(s_n)\cdots b_{x_n}^{\#_n}(s_n)
\Big\ra_{\beta}
\ge 0,
\end{align} 
where $b_x^{\#}(s)=\ex^{-s H} b_x^{\#} \ex^{s H}$.
}
\end{example} 

To state the second quantum Griffiths inequality, we introduce the
following notation:
\begin{align}
\la\!\la Y\ra\!\ra_{\beta}&=
\Tr_{\mathfrak{H} \otimes \mathfrak{H}}\big[Y\, \ex^{-\beta
 H_{\mathrm{ext}}}\big]\Big/ \Xi_{\beta}^2,\ \ \ H_{\mathrm{ext}}=H\otimes \one +\one
 \otimes H.
\end{align}

\begin{Thm}[Second Griffiths inequality]\label{HubbardG2}
For each $x\in \Lambda, \vepsilon\in \{\pm 1\}, \#\in \{\pm\}, \sigma\in
 \{\uparrow, \downarrow\}$ and $s\ge
 0$, 
we introduce 
\begin{align}
\alpha_{x\sigma; \vepsilon}^{\#}(s)=c_{x\sigma}^{\#}(s)\otimes \one+ \vepsilon \gamma \otimes c_{x\sigma}^{\#}(s),
\end{align} 
where  $\gamma=
 (-\one)^{\Ne}$  with $\Ne=N_{\uparrow}+N_{\downarrow}$ and $c_{x\sigma}^{\#}(s)=\ex^{-sH} c_{x\sigma}^{\#}\, \ex^{sH}$.
Let $x_1,\dots,  x_n \in \Lambda$.
For each   $0 \le s_1 \le s_2\le \cdots\le
 s_{n} \le \beta$, $\#_1, \dots, \#_n\in \{+, -\}$ and $\vepsilon_1,
 \dots, \vepsilon_n\in \{\pm 1\}$,
 we have 
\begin{align}
\Bigg\la\!\!\!\Bigg\la \prod_{j=1}^{{n\atop{\longrightarrow}}}
\bigg[(-1)^{\mu(x_j)}
\alpha_{x_j\uparrow; \vepsilon_j}^{\#_j}(s_j)
\gamma_{\uparrow}\otimes \gamma_{\uparrow}
\alpha_{x_j\downarrow; \vepsilon_j}^{\overline{\#}_j}(s_j)
\bigg]
\Bigg\ra\!\!\!\Bigg\ra_{\beta} \ge 0 \label{Hubbard1}
\end{align} 
and 
\begin{align}
\Bigg\la\!\!\!\Bigg\la \prod_{j=1}^{{n\atop{\longrightarrow}}}
\bigg[(-1)^{\mu(x_j)}
\alpha_{x_j\uparrow; \vepsilon_j}^{\#_j}(s_j)
\gamma_{\uparrow}\otimes \gamma_{\uparrow}
\alpha_{x_j\downarrow; -\vepsilon_j}^{\overline{\#}_j}(s_j)
\bigg]
\Bigg\ra\!\!\!\Bigg\ra_{\beta} \ge 0.\label{Hubbard2}
\end{align}  
\end{Thm}

\begin{coro}\label{HubbardG3}
Let $x_1,\dots,  x_n \in \Lambda$.
For each   $0 \le s_1 \le s_2\le \cdots\le
 s_{n} \le \beta$ and $\#_1, \dots, \#_n\in \{+, -\}$,
 we have 
\begin{align}
\Bigg\la\!\!\!\Bigg\la \prod_{j=1}^{{n\atop{\longrightarrow}}}
\bigg[
b_{x_j}^{\#_j}(s_j)\otimes \gamma_{\uparrow}
- \gamma_{\uparrow}\otimes b_{x_j}^{\#_j}(s_j)
\bigg]
\Bigg\ra\!\!\!\Bigg\ra_{\beta}\ge 0.
\end{align} 
\end{coro}

\begin{coro}\label{HubbardG4}
Let $x_1,\dots,  x_{2n} \in \Lambda$.
For each   $0 \le s_1 \le s_2\le \cdots\le
 s_{2n} \le \beta$ and $\#_1, \dots, \#_{2n}\in \{+, -\}$,
 we have 
\begin{align}
&\Bigg\la\!\!\!\Bigg\la \prod_{j=1}^{{2n\atop{\longrightarrow}}}
\bigg[
b_{x_j}^{\#_j}(s_j)\otimes \gamma_{\uparrow}
- \gamma_{\uparrow}\otimes b_{x_j}^{\#_j}(s_j)
\bigg]
\Bigg\ra\!\!\!\Bigg\ra_{\beta} \no
\ge&
\Bigg\la\!\!\!\Bigg\la \prod_{j=1}^{{2n\atop{\longrightarrow}}}
(-1)^{\mu(x_j)}
\bigg[
c_{x_j\uparrow}^{\overline{\#}_j}(s_j)\gamma_{\downarrow}
\otimes \gamma_{\uparrow}c_{x_j\downarrow }^{\#_j}(s_j)
-\gamma_{\downarrow}c_{x_j\downarrow}^{\#_j}(s_j)
\otimes c_{x_j\uparrow}^{\overline{\#}_j}(s_j) \gamma_{\uparrow}
\bigg]
\Bigg\ra\!\!\!\Bigg\ra_{\beta} \no
\ge& 0.\label{GGG}
\end{align} 
\end{coro}

\begin{example}
 {\rm
Consider the case where $n=2$. We then have
\begin{align}
&(-1)^{\mu(x)+\mu(y)}
\Big(
\big\la c_{x\downarrow}^* c_{x\uparrow} c_{y\uparrow}^* c_{y\downarrow}
\big\ra_{\beta}
-
\big\la
c_{x\downarrow}^* c_{x\uparrow}
\big\ra_{\beta}
\big\la c_{y\uparrow}^* c_{y\downarrow}
\ra_{\beta}
\Big)\no
\ge &
(-1)^{\mu(x)+\mu(y)}
\Big(
\big\la
 c_{x\uparrow}  c_{y\uparrow}^*
\big\ra_{\beta}
 \big\la
 c_{x\downarrow}^*  c_{y\downarrow}
\big\ra_{\beta}
+
\big\la
c_{x\uparrow} c_{y\downarrow}
\big\ra_{\beta}
\big\la c_{x\downarrow}^* c_{y\uparrow}^*
\ra_{\beta}
\Big)\no
\ge &0.
\end{align} 
Since $
\big\la
c_{x\downarrow}^* c_{x\uparrow}
\big\ra_{\beta}
=0
=\big\la
c_{x\uparrow} c_{y\downarrow}
\big\ra_{\beta}
$
by  the symmetries of the system, we arrive at 
\begin{align}
(-1)^{\mu(x)+\mu(y)}
\big\la c_{x\downarrow}^* c_{x\uparrow} c_{y\uparrow}^* c_{y\downarrow}
\ra_{\beta}
\ge 
(-1)^{\mu(x)+\mu(y)}
\big\la
 c_{x\uparrow}  c_{y\uparrow}^*
\big\ra_{\beta}
 \big\la
 c_{x\downarrow}^*  c_{y\downarrow}
\big\ra_{\beta}
\ge 0.
\end{align} 
If $x, y\in \Lambda_e$ or $x, y\in \Lambda_o$, then
 $(-1)^{\mu(x)+\mu(y)}=1$, 
so that we obtain a standard-type correlation inequality.
$\diamondsuit$
 }
\end{example} 

\begin{coro}\label{HubbardG5}
Let $\overline{n}_{x\uparrow}=\one -n_{x\uparrow}$. Let $x_1, \dots,
 x_n\in \Lambda$. We have 
\begin{align}
&\Bigg\la\!\!\!\Bigg\la
 \prod_{j=1}^{{n\atop{\longrightarrow}}}
\bigg[
\overline{n}_{x_j\uparrow}n_{x_j\downarrow}\otimes \one +\one \otimes 
\overline{n}_{x_j\uparrow}n_{x_j\downarrow}
\bigg]
\Bigg\ra\!\!\!\Bigg\ra_{\beta}\no
\ge &\Bigg\la\!\!\!\Bigg\la
 \prod_{j=1}^{{n\atop{\longrightarrow}}}
\bigg[
\overline{n}_{x_j\uparrow}\otimes n_{x_j\downarrow}
 +n_{x_j\downarrow} \otimes 
\overline{n}_{x_j\uparrow}
\bigg]
\Bigg\ra\!\!\!\Bigg\ra_{\beta}\no
\ge & 0. \label{EQN}
\end{align} 
\end{coro} 

\begin{example}
{\rm 
In  the case   where $n=2$, we have 
\begin{align}
&\Big\la
\overline{n}_{x\uparrow}\overline{n}_{y\uparrow}
n_{x\downarrow}n_{y\downarrow}
\Big\ra_{\beta}
+
\Big\la
\overline{n}_{x\uparrow}
n_{x\downarrow}
\Big\ra_{\beta}
\Big\la
\overline{n}_{y\uparrow}
n_{y\downarrow}
\Big\ra_{\beta}
\no
&-
\Big\la
\overline{n}_{x\uparrow}
\overline{n}_{y\uparrow}
\Big\ra_{\beta}
\Big\la
n_{x\downarrow}
n_{y\downarrow}
\Big\ra_{\beta}
-
\Big\la
\overline{n}_{x\uparrow}
n_{y\downarrow}
\Big\ra_{\beta}
\Big\la
\overline{n}_{y\uparrow}
n_{x\downarrow}
\Big\ra_{\beta}
\ge 0.\ \ \ \diamondsuit
\end{align} 
}
\end{example} 

\begin{rem}
{\rm
Our results can be extended to 
 a general class of electron--phonon(or photon) Hamiltonians, including
the Holstein--Hubbard model  and  the SSH model. 
$\diamondsuit$
}
\end{rem}

\subsubsection{The zero-temperature case}
Our results can be extended to the case where $\beta=\infty$.
Unfortunately,   the general theorems in Section \ref{GeneralRP} cannot
be  directly
applied  to this model. To clarify  the main points of modification,
we  state results without proofs.

We assume an additional condition.
\begin{itemize}
\item[{\bf (G. 3)}] $G$ is connected, i.e.,  any of its vertices are
	     linked by a path in $G$.
\end{itemize} 
We consider a half-filled system. Thus,  our  Hilbert space is
restricted to 
\begin{align}
\mathfrak{E}=\mathfrak{H}\cap \ker(\Ne-|\Lambda|).
\end{align}  
 Let $S^{(z)}=\frac{1}{2}(N_{\uparrow}-N_{\downarrow})$.
Since $S^{(z)}$ commutes with $H$, we have the following decomposition:
\begin{align}
\mathfrak{E}=\bigoplus_{M=-|\Lambda|/2}^{|\Lambda|/2}\mathfrak{E}_M,\ \
 \ \mathfrak{E}_M=\mathfrak{E}\cap \ker(S^{(z)}-M).
\end{align} 
$\mathfrak{E}_M$ is called the $M$-subspace.
For each $M\in \mathrm{spec}(S^{(z)})$, set $H_M=H\restriction
 \mathfrak{E}_M$.
The following theorem is important.

\begin{Thm}{\rm \cite{Lieb, Miyao2}}
 For each $M\in \{-|\Lambda|/2, -(|\Lambda|-2)/2,\dots,
 |\Lambda|/2\}$, $H_M$ has a unique ground state.
\end{Thm}

We denote  the normalized ground state of $H_M$  by $\psi_M$.
We define the ground state expectation value  by 
\begin{align}
\la X\ra_{\infty, M}=\la \psi_M|X\psi_M\ra.
\end{align}

\begin{Thm}\label{HubbardGI2-2}
Let $A_1,\dots, A_n\in \mathfrak{A}$.  For all  $0 \le  s_1 \le  s_2 \le  \cdots \le s_n$, we have 
\begin{align}
\Bigg\la
 \prod_{j=1}^{{n}\atop{\longrightarrow}}
A_j(s_j)
\Bigg\ra_{\infty, M} \ge 0.
\end{align} 
\end{Thm} 

We introduce the
following notation:
\begin{align}
\la\!\la Y\ra\!\ra_{\infty, M}&=
\Big\la\psi_M\otimes \psi_M\Big|
Y \psi_M\otimes \psi_M\Big\ra.
\end{align} 

\begin{Thm}\label{HubbardG2-2}
Let $x_1,\dots,  x_n \in \Lambda$.
For each   $0 \le s_1 \le s_2\le \cdots\le
 s_{n} $, $\#_1, \dots, \#_n\in \{+, -\}$ and $\vepsilon_1, \dots, \vepsilon_n\in \{\pm 1\}$,
 we have 
\begin{align}
\Bigg\la\!\!\!\Bigg\la \prod_{j=1}^{{n\atop{\longrightarrow}}}
\bigg[(-1)^{\mu(x_j)}
\alpha_{x_j\uparrow; \vepsilon_j}^{\#_j}(s_j)
\gamma_{\uparrow}\otimes \gamma_{\uparrow}
\alpha_{x_j\downarrow; \vepsilon_j}^{\overline{\#}_j}(s_j)
\bigg]
\Bigg\ra\!\!\!\Bigg\ra_{\infty, M} \ge 0 \label{Hubbard1-2}
\end{align} 
and 
\begin{align}
\Bigg\la\!\!\!\Bigg\la \prod_{j=1}^{{n\atop{\longrightarrow}}}
\bigg[(-1)^{\mu(x_j)}
\alpha_{x_j\uparrow; \vepsilon_j}^{\#_j}(s_j)
\gamma_{\uparrow}\otimes \gamma_{\uparrow}
\alpha_{x_j\downarrow; -\vepsilon_j}^{\overline{\#}_j}(s_j)
\bigg]
\Bigg\ra\!\!\!\Bigg\ra_{\infty, M} \ge 0.\label{Hubbard2-2}
\end{align} 
\end{Thm}

\subsection{Proof of Theorem  \ref{HubbardGI2}}

The hole--particle transformation $\mathcal{U}$ is a unitary operator
such that 
\begin{align}
\mathcal{U} c_{x\uparrow} \mathcal{U}^*=(-1)^{\mu(x)} c_{x\uparrow}^*,\ \
 \ 
\mathcal{U}c_{x\downarrow} \mathcal{U}^*=c_{x\downarrow}.
\end{align} 
Let $\hat{H}=\mathcal{U} H\mathcal{U}^* $. Then we obtain the attractive
Hubbard model:
\begin{align}
\hat{H}=\sum_{\{x, y\}\in E}\sum_{\sigma\in \{\uparrow,
 \downarrow\}}(-t_{xy})
c_{x\sigma}^* c_{y\sigma}
-U \sum_{x\in \Lambda} (n_{x\uparrow}-\tfrac{1}{2}) (n_{x\downarrow}-\tfrac{1}{2}).
\end{align}

Let $c_x$ be the annihilation operator on $\Fock$.  We
note that 
\begin{align}
c_{x\uparrow}=c_x\otimes \one,\ \ \ c_{x\downarrow}=(-\one)^{\mathsf{N}}\otimes c_x,
\end{align} 
where $\mathsf{N}=\sum_{x\in \Lambda} c_x^*c_x$.
Then we obtain  
\begin{align}
\hat{H}=\mathsf{T}\otimes \one +\one \otimes \mathsf{T}
-U\sum_{x\in \Lambda} (
\mathsf{n}_x-\tfrac{1}{2}
)
\otimes 
(\mathsf{n}_x-\tfrac{1}{2}), \label{HTensor}
\end{align} 
where $
\mathsf{n}_x=c_x^*c_x
$
and 
\begin{align}
\mathsf{T}= \sum_{\{x, y\}\in E} (-t_{xy}) c_x^* c_y.
\end{align}

Let $\vartheta_1$ be an antilinear involution on $\Fock$ defined by 
\begin{align}
\vartheta_1 c_{x_1}^* \cdots c_{x_{n}}^* \Omega= c_{x_1}^* \cdots
 c_{x_n}^* \Omega, \ \ x_1, \dots, x_{n}\in \Lambda,
\end{align} 
where $\Omega$ is the Fock vacuum in $\Fock$. 
By (\ref{Ident}),
 we have the following identification: 
\begin{align}
\mathfrak{H}=\mathscr{L}^2(\Fock).
\end{align} 
 Moreover, by (\ref{Ident}) and (\ref{HTensor}), we obtain the following:
\begin{Prop}\label{HubbardHami}
We have
\begin{align}
\hat{H}=\mathcal{L}(\mathsf{T})+\mathcal{R}(\mathsf{T})-U \sum_{x\in \Lambda}
 \mathcal{L}(\mathsf{n}_x-\tfrac{1}{2}) \mathcal{R}(\mathsf{n}_x-\tfrac{1}{2}),
\end{align} 
\end{Prop}

\begin{Prop}\label{HBasicPP}
We have  the following:
\begin{itemize}
\item[{\rm (i)}]$\hat{b}_x:=\mathcal{U} b_x\mathcal{U}^* \Br 0$ w.r.t. $\mathscr{L}^2(\Fock)_+$ for all $x\in
	     \Lambda$.
\item[{\rm (ii)}] $\ex^{-\beta \hat{H}} \Br 0$ w.r.t. $\mathscr{L}^2(\Fock)_+$
	     for all $\beta \ge 0$.
\end{itemize} 
\end{Prop} 
{\it Proof.} (i)  This immediately follows from the identification
$
\hat{b}_x=\mathcal{L}(c_x) \mathcal{R}(c_x^*)
$.

(ii) 
By Proposition \ref{HubbardHami} and Corollary \ref{GeneralPP2}, we obtain (ii)
 $\Box$

\begin{coro}\label{HAlgPP}
For all $A\in \mathfrak{A}$, we have $\mathcal{U}A\mathcal{U}^*\Br 0$
 w.r.t. $\mathscr{L}^2(\Fock)_+$.
\end{coro}

\subsubsection{
 Completion of proof of Theorem \ref{HubbardGI2}
}

By 
Theorem \ref{BasicPPP} and Corollary \ref{HAlgPP}, we obtain Theorem \ref{HubbardGI2}.
 $\Box$

\subsection{Proof of Theorem \ref{HubbardG2}}

Let $\mathfrak{H}_{\mathrm{ext}}=\h\otimes \h$. Let
\begin{align}
\phi_{x\sigma}=\frac{1}{\sqrt{2}}\big(
c_{x\sigma}\otimes \one+\gamma \otimes c_{x\sigma}
\big),\ \ \
\psi_{x\sigma}=\frac{1}{\sqrt{2}}\big(
c_{x\sigma}\otimes \one- \gamma \otimes c_{x\sigma}
\big). \label{HubbardCAR1}
\end{align} 
$\phi_{x\sigma}$ and $\psi_{x\sigma}$ act in
$\mathfrak{H}_{\mathrm{ext}}$
 as well.
These operators satisfy the  following CARs:
\begin{align}
\{\phi_{x\sigma}, \phi_{y\sigma'}^*\}&=\delta_{xy}\delta_{\sigma\sigma'},
 \ \ \{\phi_{x\sigma}, \phi_{y\sigma'}\}=0,\\
\{\psi_{x\sigma}, \psi_{y\sigma'}^*\}&=\delta_{xy}\delta_{\sigma\sigma'},
 \ \ \{\psi_{x\sigma}, \psi_{y\sigma'}\}=0,\\
\{\phi_{x\sigma}, \psi_{y\sigma'}^*\}&=0, \ \ \ \ \ \ \ \ 
 \ \ \{\phi_{x\sigma}, \psi_{y\sigma'}\}=0. \label{HubbardCAR}
\end{align} 

Let $\{\phi_x, \psi_x\, |\, x\in \Lambda\}$ be new annihilation
operators on $\mathfrak{X}= \Fock\otimes \Fock$ such that 
\begin{align}
\{\phi_{x}, \phi_{y}^*\}&=\delta_{xy},
 \ \ \{\phi_{x}, \phi_{y}\}=0, \label{cars3}\\
\{\psi_{x}, \psi_{y}^*\}&=\delta_{xy},
 \ \ \{\psi_{x}, \psi_{y}\}=0,\\
\{\phi_{x}, \psi_{y}^*\}&=0,\ \ \
 \ \ \{\phi_{x}, \psi_{y}\}=0, \label{cars4}
\end{align} 
and $\phi_x\Omega_{\mathfrak{X}}=0=\psi_x\Omega_{\mathfrak{X}}$, where $\Omega_{\mathfrak{X}}$ is the Fock vacuum in $\mathfrak{X}$.
Then we have the following identifications: 
\begin{align}
\phi_{x\uparrow}=\phi_x\otimes \one,\ \ \
 \phi_{x\downarrow}=(-\one)^{\mathcal{N}} \otimes \phi_x,\ \
 \psi_{x\uparrow}=\psi_x\otimes \one,\ \
 \psi_{x\downarrow}=(-\one)^{\mathcal{N}} \otimes \psi_x,
\end{align} 
where $\mathcal{N}=\sum_{x\in
\Lambda}(\phi_x^*\phi_x+\psi^*_x\psi_x)$.
Let
\begin{align}
\mathscr{U}=\mathcal{U}\otimes \mathcal{U}.
\end{align} 
Set
\begin{align}
\hat{H}_{\mathrm{ext}}= \mathscr{U} \He
 \mathscr{U}^*+\frac{1}{2}U |\Lambda|.
\end{align} 
Then $\hat{H}_{\mathrm{ext}}$ can be expressed as 
\begin{align}
\hat{H}_{\mathrm{ext}}=\mathbb{T}\otimes \one +\one \otimes \mathbb{T}-\mathbb{V},\label{HubbardTensor}
\end{align} 
where
\begin{align}
\mathbb{T}&= \sum_{\{x, y\}\in E}(-t_{xy})(\phi_x^* \phi_y+\psi_x^* \psi_y)+\frac{U}{2}\mathcal{N},\\
\mathbb{V}&=\frac{U}{2} \sum_{x\in \Lambda}(\mathcal{N}_x\otimes \mathcal{N}_x
+\mathcal{M}_x\otimes \mathcal{M}_x
),\\
\mathcal{N}_x&=\phi_x^* \phi_x+\psi_x^* \psi_x,\ \ 
\mathcal{M}_x=\phi_x^* \psi_x+\psi_x^* \phi_x.
\end{align}

Let $\vartheta_2 $ be an antilinear involution on $\mathfrak{X}$  defined by 
\begin{align}
\vartheta_2 \phi_x\vartheta_2=\phi_x,\ \ \vartheta_2 \psi_x \vartheta_2=\psi_x,\
 \ \vartheta_2 \Omega_{\mathfrak{X}}=\Omega_{\mathfrak{X}}. \label{ThetaDef}
\end{align} 

By (\ref{Ident}), we have the identification
\begin{align}
\mathfrak{H}_{ \mathrm{ext}}=\mathscr{L}^2(\mathfrak{X}). \label{ID2}
\end{align} 
In addition, we have the following expression:
\begin{Prop}
We have
$
\hat{H}_{\mathrm{ext}}=\mathcal{L}(\mathbb{T})+\mathcal{R}(\mathbb{T})-\mathbb{V},
$
where
\begin{align}
\mathbb{V}=\frac{U}{2} \sum_{x\in \Lambda}\Big\{\mathcal{L}(\mathcal{N}_x)\mathcal{R}( \mathcal{N}_x)
+\mathcal{L}(\mathcal{M}_x)\mathcal{R}( \mathcal{M}_x)
\Big\}.
\end{align} 
\end{Prop} 

By Corollary \ref{GeneralPP2}, we obtain the following:
\begin{coro}\label{HubbardPP}
For all $\beta\ge 0$, we have  $\exp(-\beta
 \hat{H}_{\mathrm{ext}})\Br 0$ w.r.t. $\mathscr{L}^2(\mathfrak{X})_+$.
\end{coro}

The below  proposition  immediately follows  from the definitions (note that $\mathscr{U} \gamma \otimes \one \mathscr{U}^*=\gamma
\otimes \one$ by {\bf (G. 1)}).
\begin{Prop}\label{AnniCre}
We have the following:
\begin{itemize}
\item[{\rm (i)}] $\mathscr{U}(-1)^{\mu(x)}(c_{x\uparrow}^{\#}\otimes \one
	     +\gamma\otimes c_{x\uparrow}^{\#})
	     \mathscr{U}^*=\sqrt{2}\mathcal{L}(\phi_x^{\overline{\#}})$.
\item[{\rm (ii)}] $\mathscr{U}(-1)^{\mu(x)}(c_{x\uparrow}^{\#}\otimes \one
	     -\gamma\otimes c_{x\uparrow}^{\#})
	     \mathscr{U}^*=\sqrt{2}\mathcal{L}(\psi_x^{\overline{\#}})$.
\item[{\rm (iii)}] $\mathscr{U}\gamma_{\uparrow}\otimes \gamma_{\uparrow}(c_{x\downarrow}^{\#}\otimes \one
	     +\gamma\otimes c_{x\downarrow}^{\#})
	     \mathscr{U}^*=\sqrt{2}\mathcal{R}(\phi_x^{\overline{\#}})$.
\item[{\rm (iv)}] $\mathscr{U}\gamma_{\uparrow}\otimes \gamma_{\uparrow}(c_{x\downarrow}^{\#}\otimes \one
	     -\gamma\otimes c_{x\downarrow}^{\#})
	     \mathscr{U}^*=\sqrt{2}\mathcal{R}(\psi_x^{\overline{\#}})$.
\end{itemize} 
\end{Prop}

\begin{coro}\label{ElCorr}
Let
\begin{align}
\alpha_{x\sigma; \vepsilon}=c_{x\sigma}\otimes \one+ \vepsilon \gamma \otimes c_{x\sigma}.
\end{align} 
For all $\vepsilon\in \{\pm 1\},\ \#\in \{\pm\}$ and $x\in \Lambda$, we have 
\begin{align}
\mathscr{U}  (-1)^{\mu(x)}
\alpha_{x\uparrow, \vepsilon}^{\#} \gamma_{\uparrow}\otimes
 \gamma_{\uparrow}
\alpha_{x\downarrow, \vepsilon}^{\overline{\#}}
 \mathscr{U}^*\Br 0
\end{align} 
w.r.t. $\mathscr{L}^2(\mathfrak{X})_+$.
\end{coro} 
{\it Proof.} By Proposition \ref{AnniCre}, we have 
\begin{align}
\mathscr{U}  
(-1)^{\mu(x)}
\alpha_{x\uparrow, -1}^{\#} \gamma_{\uparrow}\otimes
 \gamma_{\uparrow}
\alpha_{x\downarrow, -1}^{\overline{\#}}
 \mathscr{U}^*
&=2 \mathcal{L}\Big(\psi_x^{\overline{\#}}\Big)\,  
\mathcal{R}\Big(\psi_x^{\#}\Big) \Br 0 \ \ \mbox{w.r.t. $\mathscr{L}^2(\mathfrak{X})_+$ },\\
\mathscr{U}  
(-1)^{\mu(x)}
\alpha_{x\uparrow, +1}^{\#} \gamma_{\uparrow}\otimes
 \gamma_{\uparrow}
\alpha_{x\downarrow, +1}^{\overline{\#}}\mathscr{U}^{*}
&=2\mathcal{L}\Big(\phi_x^{{\overline \#}}\Big)\,  
\mathcal{R}\Big(\phi_x^{\#}\Big)\Br 0\ \ \mbox{w.r.t. $\mathscr{L}^2(\mathfrak{X})_+$ } .\ \ \ \ \Box
\end{align}

\subsubsection{Completion of proof of  Theorem \ref{HubbardG2}}
\begin{flushleft}
{\it Proof of (\ref{Hubbard1})}
\end{flushleft} 
Let 
$
D_{\vepsilon, \#, x}=(-1)^{\mu(x)}
\alpha_{x\uparrow, \vepsilon}^{\#} \gamma_{\uparrow}\otimes
 \gamma_{\uparrow}
\alpha_{x\downarrow, \vepsilon}^{\overline{\#}}
$.
Then we see that by Corollaries \ref{HubbardPP} and \ref{ElCorr},
\begin{align}
&\mathscr{U}
\Bigg[
\prod_{j=1}^{{n\atop{\longrightarrow}}}(-1)^{\mu(x_j)}
\alpha_{x_j\uparrow; \vepsilon_j}^{\#_j}(s_j)
\gamma_{\uparrow}\otimes \gamma_{\uparrow}
\alpha_{x_j\downarrow; \vepsilon_j}^{\overline{\#}_j}(s_j)
\Bigg]
 \ex^{-\beta \He}\mathscr{U}^*\no
=&
\underbrace{\ex^{-s_1 \hat{H}_{\mathrm{ext}}}}_{\Br 0}
\underbrace{
\mathscr{U}D_{\vepsilon_1, \#_1, x_1}\mathscr{U}^*
}_{\Br 0}
\underbrace{
\ex^{-(s_2-s_1)\hat{H}_{\mathrm{ext}}} 
}_{\Br 0}
\cdots
\underbrace{
\mathscr{U}D_{\vepsilon_n, \#_n, x_n}\mathscr{U}^*
}_{\Br 0}
\underbrace{
 \ex^{-(\beta-s_n)\hat{H}_{\mathrm{ext}}}
}_{\Br 0} 
\Br 0
\end{align} 
w.r.t. $\mathscr{L}^2(\mathfrak{X})_+$.
Thus,  by Theorem \ref{Prototype2}, we obtain (\ref{Hubbard1}).   $\Box$

\begin{flushleft}
{\it Proof of (\ref{Hubbard2})}
\end{flushleft} 
Let $\mathscr{Q}$ be a unitary operator defined by 
$\mathscr{Q}=\one \otimes (-\one )^{N_{\downarrow}}$. Then we see that 
\begin{align}
\mathscr{Q} \He \mathscr{Q}^{-1}=\He,
\ \ \mathscr{Q}\alpha_{x\uparrow, \vepsilon}\mathscr{Q}^{-1}=\alpha_{x\uparrow, \vepsilon},
\ \  \mathscr{Q}\alpha_{x\downarrow, \vepsilon}\mathscr{Q}^{-1}=\alpha_{x\downarrow, -\vepsilon}.\label{UnitaryQ}
\end{align} 
Thus,  (\ref{Hubbard2}) follows from (\ref{Hubbard1}). $\Box$

\subsection{Proof of Corollary \ref{HubbardG3}}
\begin{lemm}\label{HHH1}
Let 
$C_{x, \vepsilon}=
(-1)^{\mu(x)}
\alpha_{x\uparrow, \vepsilon}^{\#} \gamma_{\uparrow}\otimes
 \gamma_{\uparrow}
\alpha_{x\downarrow, -\vepsilon}^{\overline{\#}}
$.
Set $\mathscr{W}=\mathscr{U} \mathscr{Q}$.
Then we obtain
$ \mathscr{W} C_{x, \vepsilon} \mathscr{W}^{-1} \Br 0$ w.r.t. $\mathscr{L}^2(\mathfrak{X})_+$.
\end{lemm} 
{\it Proof.}
By Corollary  \ref{ElCorr} and (\ref{UnitaryQ}), we see that 
\begin{align}
\mathscr{W} C_{x, \vepsilon} \mathscr{W}^{-1} =\mathscr{U} D_{\vepsilon,
 +1, x} \mathscr{U}^{-1} \Br 0\ \ \
 \mbox{w.r.t. $\mathscr{L}^2(\mathfrak{X})_+$.\ \ \ $\Box$}
\end{align} 

\begin{lemm}
For all $\beta \ge 0$, we have 
$\mathscr{W}
\ex^{-\beta \He}
\mathscr{W}^{-1}\Br 0
$
w.r.t. $\mathscr{L}^2(\mathfrak{X})_+$.
\end{lemm} 
{\it Proof.} Since $\mathscr{Q} \He \mathscr{Q}^{-1}=\He$, we see that 
$
\mathscr{W} \ex^{-\beta \He} \mathscr{W}^{-1}=\ex^{-\beta
\hat{H}_{\mathrm{ext}}}\Br 0
$ w.r.t. $\mathscr{L}^2(\mathfrak{X})_+$. $\Box$

\subsubsection{Completion of proof of Corollary \ref{HubbardG3}}

By Lemma \ref{HHH1}, we obtain 
\begin{align}
\mathscr{W} C_{x, \vepsilon}\mathscr{W}^{-1}
=\mathscr{W}A_{x}\mathscr{W}^{-1}+\vepsilon \mathscr{W}B_{x}\mathscr{W}^{-1} \Br 0\ \ \ \mbox{w.r.t. $\mathscr{L}^2(\mathfrak{X})_+$},
\end{align} 
where
\begin{align}
A_{x}&=\mathscr{Q}^{-1}\big(
b_x\otimes \gamma_{\uparrow}-\gamma_{\uparrow}\otimes b_x
\big)\mathscr{Q},\\
B_{x}&= - (-1)^{\mu(x)} \mathscr{Q}^{-1}\big(c_{x\uparrow}^* \gamma_{\downarrow}\otimes
 \gamma_{\uparrow} c_{x\downarrow}
- \gamma_{\downarrow} c_{x\downarrow} \otimes
 c_{x\uparrow}^* \gamma_{\uparrow}
\big)\mathscr{Q}.
\end{align} 
Thus,  we have 
\begin{align}
\mathscr{W}A_{x}\mathscr{W}^{-1}=\frac{1}{2}\underbrace{
\mathscr{W}C_{x, +}\mathscr{W}^{-1}}_{\Br
 0}
+\frac{1}{2}\underbrace{\mathscr{W}C_{x, -}\mathscr{W}^{-1}}_{\Br 0}\Br
 0 \ \ \ \mbox{w.r.t. $\mathscr{L}^2(\mathfrak{X})_+$}. \label{HHH2}
\end{align} 
Finally,   observe that 
\begin{align}
&\mathscr{W} \Bigg[
\prod_{j=1}^{{n}\atop{\longrightarrow}}
A_{x_j}^{\#_j}(s_j)
\Bigg]\ex^{-\beta \He}\mathscr{W}^{-1}\no
=&
\underbrace{
\mathscr{W} \ex^{-s_1 \He}\mathscr{W}^{-1}
}_{\Br 0}
\underbrace{
 \mathscr{W}A_{x_1}^{\#_1}
 \mathscr{W}^{-1}
}_{\Br 0}
\underbrace{
\mathscr{W}\ex^{-(s_2-s_1)\He}\mathscr{W}^{-1}
}_{\Br 0}
\cdots
\underbrace{
 \mathscr{W}\ex^{-(\beta-s_n) \He}\mathscr{W}^{-1}
}_{\Br 0}
\Br 0
\end{align}
w.r.t. $\mathscr{L}^2(\mathfrak{X})_+$. 
By Theorem \ref{Prototype2}, we conclude Corollary \ref{HubbardG3}. $\Box$

\subsection{Proof of Corollary \ref{HubbardG4}}
\subsubsection{First part of the proof}
Note that $
\mathscr{W}B_x\mathscr{W}^{-1}=\frac{1}{2}\mathscr{W}C_{x,
+}\mathscr{W}^{-1}-\frac{1}{2}\mathscr{W}
C_{x,-}
\mathscr{W}^{-1}
$.
Combining this with (\ref{HHH2}), we have 
\begin{align}
&\mathscr{W}\Bigg[\prod_{j=1}^{{2n}\atop{\longrightarrow}}A_{x_j}^{\#_j}(s_j)
\Bigg]\ex^{-\beta \He}
\mathscr{W}^{-1}
-
\mathscr{W}
\Bigg[
\prod_{j=1}^{{2n}\atop{\longrightarrow}}B_{x_j}^{\#_j}(s_j)
\Bigg]
\ex^{-\beta \He}
\mathscr{W}^{-1}\no
=&\sum_{\delta_1, \dots, \delta_n\in \{\pm\}}
X_{\delta_1, \dots, \delta_n}
\underbrace{
\mathscr{W}C_{x_1, \delta_1}(s_1) \cdots C_{x_n,
 \delta_n}(s_n)\, \ex^{\beta \He}\,  \mathscr{W}^{-1}
}_{\Br 0}, \label{HHH3}
\end{align} 
where  each $X_{\delta_1, \dots,\delta_n}$ is a positive constant.
Thus,  the RHS  of (\ref{HHH3})$\Br 0$ w.r.t. $\mathscr{L}^2(\mathfrak{X})_+$.
By Theorem \ref{Prototype2}, we obtain the first inequality in (\ref{GGG}). 

\subsubsection{Second part of the proof }\label{SecondEQN}
We will show the second inequality in (\ref{GGG}).
Let $\Theta$ be an antilinear involution on $\h$ such that 
\begin{align}
\Theta c_{x\sigma} \Theta=c_{x\sigma},\ \ \Theta\Omega_{\h}=\Omega_{\h},
\end{align} 
where $\Omega_{\h}=\Omega\otimes \Omega$. Then by (\ref{Ident}),  
we have  $\mathfrak{H}_{\mathrm{ext}}=\mathscr{L}^2(\h)$ and
\begin{align}
\hat{H}_{\mathrm{ext}}=
\mathcal{L}(\hat{H})+\mathcal{R}(\hat{H}).
\end{align}  
By Corollary \ref{GeneralPP2}, we have the following:
\begin{Prop}\label{HHH6}
For all $\beta \ge 0$, we have $\exp(-\beta \hat{H}_{\mathrm{ext}})\Br
 0$
w.r.t. $\mathscr{L}^2(\h)_+$.
\end{Prop}

Let $S$ be the unitary operator on $\h$ given by
\begin{align}
Sc_{x\uparrow}S^{-1}=c_{x\downarrow},\ \ \ Sc_{x\downarrow}S^{-1}=c_{x\uparrow}.
\end{align} 
Set $\mathscr{R}=\one \otimes S \mathscr{U}$.
Remark that  since  $S\hat{H}S^{-1}=\hat{H}$, we know that
\begin{align}
\mathscr{R} \ex^{-\beta \He} \mathscr{R}^{-1}=\ex^{-\beta
 \hat{H}_{\mathrm{ext}}}\Br 0\ \ \mbox{w.r.t. $\mathscr{L}^2(\h)_+$}\label{HHH7}
\end{align} 
by Proposition \ref{HHH6}.

\begin{Prop}\label{HHH4}
We have the following:
\begin{itemize}
\item[{\rm (i)}] 
$
\mathscr{R}(-1)^{\mu(x)} c_{x\uparrow}\gamma_{\downarrow}\otimes \one
	     \mathscr{R}^{-1}=\mathcal{L}(c_{x\uparrow}^* \gamma_{\downarrow})
$.
\item[{\rm (ii)}]
$
\mathscr{R} c_{x\downarrow}\gamma_{\downarrow}\otimes \one
	     \mathscr{R}^{-1}=\mathcal{L}(c_{x\downarrow} \gamma_{\downarrow})
$.
\item[{\rm (iii)}]
$
\mathscr{R}(-1)^{\mu(x)}\one \otimes  c_{x\uparrow}\gamma_{\uparrow}
	     \mathscr{R}^{-1}=\mathcal{R}( \gamma_{\downarrow}c_{x\downarrow})
$.
\item[{\rm (iv)}]
$
\mathscr{R}\one \otimes  \gamma_{\uparrow} c_{x\downarrow}
	     \mathscr{R}^{-1}=\mathcal{R}( \gamma_{\downarrow}c_{x\uparrow}^*)
$.
\end{itemize} 
\end{Prop} 

\begin{coro}\label{HHH5}
We have the following:
\begin{itemize}
\item[{\rm (i)}]
$
\mathscr{R} (-1)^{\mu(x)}c_{x\uparrow}^* \gamma_{\downarrow} \otimes
	     \gamma_{\uparrow} c_{x\downarrow} \mathscr{R}^{-1} \Br 0
$
w.r.t. $\mathscr{L}^2(\h)_+$.
\item[{\rm (ii)}]
$
-\mathscr{R} (-1)^{\mu(x)}
 \gamma_{\downarrow}c_{x\downarrow} \otimes
	     c_{x\uparrow}^* \gamma_{\uparrow} \mathscr{R}^{-1} \Br 0
$
w.r.t. $\mathscr{L}^2(\h)_+$.
\end{itemize} 
\end{coro} 
{\it Proof.} By Proposition \ref{HHH4}, we see that
\begin{align}
\mathscr{R} (-1)^{\mu(x)}c_{x\uparrow}^* \gamma_{\downarrow} \otimes
	     \gamma_{\uparrow} c_{x\downarrow} \mathscr{R}^{-1}
=&\mathcal{L}(c_{x\uparrow} \gamma_{\downarrow})
 \mathcal{R}(\gamma_{\downarrow} c_{x\uparrow}^*) \Br 0\ \
 \mbox{w.r.t. $\mathscr{L}^2(\h)_+$},\\
-\mathscr{R} 
(-1)^{\mu(x)}
 \gamma_{\downarrow}c_{x\downarrow} \otimes
	     c_{x\uparrow}^* \gamma_{\uparrow} \mathscr{R}^{-1}
=&\mathcal{L}(c_{x\downarrow} \gamma_{\downarrow})
 \mathcal{R}(( c_{x\downarrow}\gamma_{\downarrow})^*) \Br 0\ \
 \mbox{w.r.t. $\mathscr{L}^2(\h)_+$}.
\end{align} 
This completes the proof. $\Box$
\medskip\\

Set 
\begin{align}
K_x=(-1)^{\mu(x)}
\Big(
c_{x\uparrow}^{*}\gamma_{\downarrow}
\otimes \gamma_{\uparrow}c_{x\downarrow }
-\gamma_{\downarrow}c_{x\downarrow}
\otimes c_{x\uparrow}^{*} \gamma_{\uparrow}
\Big).
\end{align} 
By Corollary \ref{HHH5}, we know that $\mathscr{R} K_x\mathscr{R}^{-1}
\Br 0$ w.r.t. $\mathscr{L}^2(\h)_+$.
Thus,  by (\ref{HHH7}), we obtain 
\begin{align}
& \mathscr{R}\prod_{j=1}^{{2n\atop{\longrightarrow}}}
(-1)^{\mu(x_j)}
\bigg[
c_{x_j\uparrow}^{\overline{\#}_j}(s_j)\gamma_{\downarrow}
\otimes \gamma_{\uparrow}c_{x_j\downarrow }^{\#_j}(s_j)
-\gamma_{\downarrow}c_{x_j\downarrow}^{\#_j}(s_j)
\otimes c_{x_j\uparrow}^{\overline{\#}_j}(s_j) \gamma_{\uparrow}
\bigg]\, \ex^{-\beta H_{\mathrm{ext}}} \mathscr{R}^{-1}\no
=& 
\underbrace{
\mathscr{R} \ex^{-s_1 H_{\mathrm{ext}}} \mathscr{R}^{-1}
}_{\Br 0}
\underbrace{
\mathscr{R} K_{x_1}^{\#_1}\mathscr{R}^{-1} 
}_{\Br 0}
\underbrace{
\mathscr{R} \ex^{-(s_2-s_1)H_{\mathrm{ext}}} \mathscr{R}^{-1}
}_{\Br 0}
\cdots
\underbrace{
 \mathscr{R} \ex^{-(\beta-s_n) H_{\mathrm{ext}}} \mathscr{R}^{-1}
}_{\Br 0}\Br 0
\end{align}
w.r.t. $\mathscr{L}^2(\h)_+$. Hence,  by Theorem \ref{Prototype2}, we
obtain the second inequality in (\ref{GGG}). $\Box$

\subsection{Proof of Corollary \ref{HubbardG5}}
Let 
\begin{align}
A_{x, +}=\overline{n}_{x\uparrow}n_{x\downarrow}\otimes \one +\one
 \otimes \overline{n}_{x\uparrow}n_{x\downarrow},\ \ \ 
A_{x, -}=\overline{n}_{x\uparrow}\otimes n_{x\downarrow}
 +n_{x\downarrow}
\otimes \overline{n}_{x\uparrow}.
\end{align} 
Observe that 
\begin{align}
\mathscr{U} A_{x, +}\mathscr{U}^*&=\frac{1}{2}
\Big\{
\mathcal{L}(\mathcal{N}_x) \mathcal{R}(\mathcal{N}_x)
+
\mathcal{L}(\mathcal{M}_x) \mathcal{R}(\mathcal{M}_x)
\Big\},\no
\mathscr{U} A_{x, -}\mathscr{U}^*&=\frac{1}{2}
\Big\{
\mathcal{L}(\mathcal{N}_x) \mathcal{R}(\mathcal{N}_x)
-
\mathcal{L}(\mathcal{M}_x) \mathcal{R}(\mathcal{M}_x)
\Big\}.
\end{align} 
Clearly,
$
\mathcal{L}(\mathcal{N}_x) \mathcal{R}(\mathcal{N}_x)\Br 0,\ \ 
\mathcal{L}(\mathcal{M}_x) \mathcal{R}(\mathcal{M}_x)\Br 0
$
w.r.t. $\mathscr{L}^2(\mathfrak{X})_+$.
Thus,  we have 
\begin{align}
\mathscr{U}
\Bigg(
\prod_{j=1}^{{n}\atop{\longrightarrow}} A_{x_j, +}
-
\prod_{j=1}^{{n}\atop{\longrightarrow}} A_{x_j, -}
\Bigg)
\mathscr{U}^* \Br 0\ \ \mbox{w.r.t. $\mathscr{L}^2(\mathfrak{X})_+$}.
\end{align} 
By applying Theorem \ref{Prototype2}, we obtain the first inequality in
(\ref{EQN}). Proof of the second inequality in (\ref{EQN}) is similar
to that of Section \ref{SecondEQN}. $\Box$

\section{Concluding remarks}\label{CRemark}

Let $\Cone$ be a self-dual cone in the Hilbert space $\h$. Let $H_0$ and
$V$
be self-adjoint operators in $\h$. 
For simplicity, we assume that $H_0$ and $V$ are bounded.\footnote{This
assumption can be relaxed \cite{Miyao2}.}
$H_0$ is the free Hamiltonian and $V$
is the interaction.
The  system's  Hamiltonian is given by $H=H_0-V$.   
Through our studies of the quantum Griffiths  inequality, we
recognize that the following   are  model-independent
properties:\footnote{
 Even when we show the second
Griffiths inequality, the properties ($\Cone$ i) and ($\Cone$ ii) are essential for our proof.
Namely, ($\Cone$ i) and ($\Cone$ ii) still hold true for the extended
Hamiltonian acting in the doubled Hilbert space $\h\otimes \h$, see
Sections \ref{Sec1}--\ref{Sec4}.  }
\begin{itemize}
\item[($\mathfrak{P}$ i)] $\ex^{-\beta H_0} \unrhd 0$ w.r.t. $\Cone$ for all $\beta
	   \ge 0$.
\item[($\mathfrak{P}$ ii)] $V\unrhd 0$ w.r.t. $\Cone$.

\end{itemize} 
($\mathfrak{P}$ ii) is equivalent to $-V\unlhd 0$ w.r.t. $\Cone$. 
Thus, if ($\mathfrak{P}$ ii) is satisfied,    we say that $-V$ is {\it
attractive} w.r.t. $\Cone$.
As we  have discussed in the previous sections,
when we construct the Griffiths   inequality,
it is   most important   to find a self-dual cone $\Cone$ such that
$-V$ becomes attractive w.r.t. $\Cone$.
In this step, we  are faced with the following difficulty:
in general,  there are infinitely many  self-dual cones in a single
Hilbert space.
Let us assume that  ($\Cone$ i) and ($\Cone$ ii) are satisfied by choosing some self-dual
cone $\Cone$.
Now let   us choose another self-dual cone $\Cone'$.
Even if ($\Cone$ i) and ($\Cone$ ii) are satisfied,  we can never conclude that ($\Cone'$ i)
and ($\Cone'$ ii) are fulfilled.
Therefore,   to apply our theory,  we have to choose a proper
self-dual cone  $\Cone$ such that ($\Cone$ i) and ($\Cone$ ii) are satisfied.  In other
words,
a suitable choice of  a self-dual cone  makes the interaction $-V$ attractive.
In this sense, our theory is a kind of  representation theory of attraction.

We remark upon some additional conclusions  from  ($\Cone$ i) and ($\Cone$ ii).
First, we obtain the positivity of a ground state.

\begin{Thm}{\rm \cite{Miyao2}}
Assume {\rm ($\Cone$ i)} and {\rm ($\Cone$ ii)}. Assume that $E=\inf \mathrm{spec}(H)$
 is an eigenvalue of $H$. Then there exists a nonzero  vector $\psi\in \ker(H-E)$
such that $\psi\ge 0$ w.r.t. $\Cone$. Namely,  among all the ground states
 of $H$, there exists at least one ground state that  is positive
 w.r.t. $\Cone$.
\end{Thm} 

Theorem \ref{Energy} claims that the attractive interaction makes the system more stable.
\begin{Thm}\label{Energy}{\rm \cite{Miyao2}}
Assume {\rm ($\Cone$ i)} and {\rm ($\Cone$ ii)}. Let  $E_0=\inf
 \mathrm{spec}(H_0)$.
Then $E\le E_0$.
\end{Thm}

To describe further effects  of ($\Cone$ i) and ($\Cone$ ii), we define
the following:

\begin{define}{\rm 
\begin{itemize}
\item[{\rm (i)}]A vector $y\in \mathfrak{H}$ is called strictly positive
w.r.t. $\Cone$, whenever $\la x| y\ra>0$ for all $x\in
\Cone\backslash \{0\}$. We write this as $y>0 $
w.r.t. $\Cone$.
\item[{\rm (ii)}]
We write  $A\rhd 0$ w.r.t. $\Cone$, if  $Ax >0$ w.r.t. $\Cone$ for all $x\in
\Cone \backslash \{0\}$. 
 In this case, we say that $A$ improves the
positivity w.r.t. $\Cone$. $\diamondsuit$
\end{itemize} 
}
\end{define}

\begin{Thm}\label{UniqG}{\rm \cite{Faris, Miyao1}}
Assume {\rm ($\Cone$ i)} and {\rm ($\Cone$ ii)}. Assume that $\ex^{-\beta H} \rhd 0$
w.r.t. $\Cone$ for all $\beta >0$. If $E=\inf \mathrm{spec}(H)$ is an
 eigenvalue, then  $\dim \ker(H-E)=1$
(equivalently,  if $H$ has a ground state, then it is unique).
Moreover,  the unique ground state is strictly positive w.r.t. $\Cone$.
\end{Thm} 
\begin{rem}
{\rm 
If we impose additional conditions on $V$, we can prove $E<E_0$
 \cite{Miyao2}.
$\diamondsuit$
}
\end{rem} 

As a corollary of Theorem \ref{UniqG}, we obtain information about
structure  of the ground state.
\begin{coro}\label{InvCoro}
Let $G$ be a group and let $\pi$ be an irreducible representation of $G$
 on $\mathfrak{H}$. Assume that $\pi_g\unrhd 0$ w.r.t. $\Cone$ for all
 $g\in G$. Under the same assumptions as in Theorem \ref{UniqG}, 
let $\vphi$ be the ground state of $H$, i.e., $\vphi\in \ker(H-E)$.
Then we have 
$\pi_g\vphi=\vphi$ for all $g\in G$.
\end{coro} 

In the theory of strongly correlated electron systems,
we can investigate the magnetic properties of  the ground state by Theorem
\ref{UniqG} and Corollary \ref{InvCoro} \cite{FrankLieb,FreLieb,LiebMattis,Lieb,LiebNach,LiebSchupp,Miyao1,Miyao3,
Miyao4,Miyao6,Miyao7,Nagaoka,Shen,STU, Tian}. 
Furthermore,  we can  find the same structures in several areas, e.g., 
 in the quantum field theory \cite{JFroehlich1,GJ, Gross, Miyao1,Miyao4,
Miyao5, OS},    open quantum
systems \cite{Lindblad}, topological orders \cite{JaffeP1,JaffeP2},  and the  theory of phase
transitions \cite{ALSSY,AFFS, Biskup,DLS,FILS, FSS,GJS,HL,KLS}.
These facts indicate that 
 {\it {\rm ($\Cone$ i)} and {\rm ($\Cone$ ii)} are universal
expressions  of
the notion of  correlations.}
If this  hypothesis is correct, then several areas could be described by 
the same language and a new discovery in some areas  would  automatically
influence  other areas.
To reinforce this vision of unification, we must continue to
collect  evidence.

\appendix
\section{Fundamental properties of operator inequalities associated with
  self-dual cones} \label{Appendix}
\setcounter{equation}{0}
\subsection{Positivity preserving operators}
In this appendix, we review   useful operator inequalities
studied in \cite{Miyao1}.

Let $\mathfrak{H}$ be a complex Hilbert space and $\Cone $ be a
self-dual cone in $\mathfrak{H}$.

\begin{Prop}\label{Prop2}
Let $\{x_n\}_{n\in \BbbN}$ be a  CONS of
 $\h$.
Assume that $x_n\in \Cone$ for all $n\in \BbbN$. Assume that $A\unrhd 0$
 w.r.t. $\Cone$. Then we have $\Tr[A]\ge 0$.
\end{Prop} 
{\it Proof.} Since $x_n\in \Cone$, we see that $\la x_n|Ax_n\ra \ge 0$
for all $n\in \BbbN$. Thus,  we arrive at $\Tr[A]=\sum_{n=1}^{\infty}\la
x_n|Ax_n\ra \ge 0$.\ $\Box$

\begin{Prop}\label{Prop3}
Let $N=\dim \h\in \BbbN \cup \{\infty\}$. Let $\{x_n\}_{n=1}^N$ be a CONS of $\h$.
Assume that $x_n\in \Cone$ for all $n\in \{1,\dots, N\}$.\footnote{In the 
 case  where $N=\infty$, the symbol $\{1, \dots, N\} $ denotes  $\BbbN$.}
Then the following {\rm (i)} and {\rm (ii)} are equivalent.
\begin{itemize}
\item[{\rm (i)}] $A\unrhd 0$ w.r.t. $\Cone$.
\item[{\rm (ii)}] $A_{mn}=\la x_m|Ax_n\ra\ge 0$ for all $m,n \in \{1,
	     \dots, N\}$.
\end{itemize} 
\end{Prop} 
{\it Proof.} (i) $\Longrightarrow$ (ii): Trivial.

(ii) $\Longrightarrow$ (i): Let $w, z\in \Cone$. Then we can write 
\begin{align}
w&=\sum_{n=1}^N c_n x_n,\ \ \ c_n=\la w|x_n\ra,\\
z&=\sum_{n=1}^N d_n x_n,\ \ \  d_n=\la z|x_n\ra.
\end{align}  
Since $w,z\ge 0$ w.r.t. $\Cone$, we see that $c_n\ge 0, d_n \ge 0$ for
all $n\in \BbbN$.  Thus,  we have 
\begin{align}
\la w|Az\ra=\sum_{m,n=1}^N c_m d_n A_{mn} \ge 0.
\end{align} 
Since $\Cone$ is self-dual, we have $Az\ge 0$ w.r.t. $\Cone$. Thus,  we
conclude that $A\unrhd 0$ w.r.t. $\Cone$.
 $\Box$

\begin{Prop}\label{SumPP}
Assume that $A \unrhd 0$ w.r.t. $\Cone$. Then $\ex^{\beta A} \unrhd 0$
w.r.t. $\Cone$ for all $\beta \ge 0$.
\end{Prop} 
{\it Proof.} Since $A \unrhd 0$ w.r.t. $\Cone$, it holds that
$A^n\unrhd 0$ w.r.t. $\Cone$ for all $n\in \BbbN$. Thus ,
\begin{align}
\ex^{\beta A} =\sum_{n\ge 0}\underbrace{\frac{\beta^n}{n!}}_{\ge 0}
 \underbrace{A^n}_{\unrhd 0} \unrhd 0\ \ \ \mbox{w.r.t. $\Cone$ for
 all $\beta \ge 0$.\ \ \ \ \ $\Box$}
\end{align}

\begin{Prop}\label{TK}
Assume that $\ex^{\beta A} \unrhd 0$ and $\ex^{\beta B} \unrhd 0$
 w.r.t. $\Cone$ for all $\beta \ge 0$. Then $\ex^{\beta (A+B)} \unrhd 0$
 w.r.t. $\Cone$ for all $\beta \ge 0$.
\end{Prop} 
{\it Proof.} Note that $\ex^{\beta A} \ex^{\beta B} \unrhd 0$
w.r.t. $\Cone$ for all $\beta \ge 0$. Thus,  $(\ex^{\beta A/n}\ex^{\beta
B/n})^n \unrhd 0$ w.r.t. $\Cone $ for all $\beta \ge 0$ and $n\in
\BbbN$. By the Trotter--Kato product formula, we obtain the desired
assertion. $\Box$
\medskip\\
 
The following proposition is  repeatedly used in this study. 
\begin{Prop}\label{BasicPP}
Assume the following:
\begin{itemize}
\item[{\rm (i)}] $\ex^{\beta  A} \unrhd 0$ w.r.t. $\Cone$ for all $\beta
	     \ge 0$.
\item[{\rm (ii)}] $B\unrhd 0$ w.r.t. $\Cone$.
\end{itemize} 
Then we have $\ex^{\beta (A+B)} \unrhd 0$ w.r.t. $\Cone$ for all $\beta
 \ge 0$.
\end{Prop} 
{\it Proof.} By (ii) and Proposition \ref{SumPP}, it holds that
$\ex^{\beta B} \unrhd 0$ w.r.t. $\Cone$ for all $\beta \ge 0$. Thus, 
applying Proposition \ref{TK}, we conclude the assertion. $\Box$

\begin{Prop}\label{GSP}
Let $A$ be a positive self-adjoint operator. Assume that $\ex^{-\beta A}
 \unrhd 0$ w.r.t. $\Cone$ for all $\beta \ge 0$. Assume that $E=\inf
 \mathrm{spec}(A)$ is an eigenvalue of $A$. Then there exists a nonzero vector
 $x\in \ker(A-E)$ such that $x\ge 0$ w.r.t. $\Cone$. 
\end{Prop} 
{\it Proof.} {\bf STEP 1.} Let $J$ be an antilinear involution given by
Proposition \ref{Haagerup} below.  Set $\mathfrak{H}_J=\{x\in \h\, |\, Jx=x\}$.
We will show that $\ker(A-E)\cap \h_J\neq \{0\}$.

To see this, let $x\in \ker(A-E)$. Then we have the decomposition 
$x=\Re x+i \Im x$ with $\Re x=\frac{1}{2}(\one +J)x$ and $\Im x
=\frac{1}{2i}(\one-J)x $. Clearly , $\Re x, \Im x\in \h_J$. Since $x\neq 0$, it holds that $\Re x\neq 0$
or $\Im x\neq 0$. Since $\ex^{-\beta A}\unrhd 0$ w.r.t. $\Cone$ for all
$\beta \ge 0$, $A$ commutes with $J$.
Thus,  $\Re x, \Im x\in \ker(A-E)\cap \h_J$.

{\bf STEP 2.} Take $x\in \ker(A-E)\cap \h_J$. By Proposition
\ref{Haagerup} (iii), we have a unique decomposition $x=x_+-x_-$, where 
$x_{\pm}\in \Cone$ and $\la x_+| x_-\ra=0$. Let $|x|=x_++x_-$. Then we
have 
\begin{align}
\ex^{-\beta E}\|x\|=\la x|\ex^{-\beta A}x\ra \le \la |x||\ex^{-\beta
 A}|x|\ra
\le \ex^{-\beta E}\underbrace{\||x|\|}_{=\|x\|}.
\end{align} 
Thus,  $|x|\in \ker(A-E)$. Clearly,  $|x|\ge 0$ w.r.t. $\Cone$. $\Box$

\begin{Prop}\label{Haagerup}
A self-dual cone $\Cone$ has the following properties:
\begin{itemize}
\item[{\rm (i)}] $\Cone\cap (-\Cone)=\{0\}$.
\item[{\rm (ii)}] There exists a unique antilinear involution $J$ in $\h$ such that
                 $Jx=x$ for all $x\in \Cone$.
\item[{\rm (iii)}] Each element $x\in \h$ with $Jx=x$ has a unique
                 decomposition $x=x_+-x_-$ where $x_+,x_-\in\Cone$ and
                 $\la x_+| x_-\ra=0$.
\item[{\rm (iv)}] $\h$ is linearly spanned by $\Cone$.
\end{itemize}
\end{Prop}
{\it Proof.} See, e.g., \cite{Bos}. $\Box$

\subsection{Reflection positive operators}
To apply Theorem \ref{BasicPPP}, it is crucial to show that $
\ex^{-\beta H} \Br 0$ w.r.t. $\mathscr{L}^2(\h)_+$ for all $\beta
>0$. The following proposition  is often useful in proving this condition:

\begin{Prop}\label{GeneralPP}
Let $H_0$ be a self-adjoint operator on $\mathscr{L}^2(\h)$ bounded from below.
Let $V \in \mathscr{B}(\mathscr{L}^2(\h))$ be self-adjoint. Assume the following:
\begin{itemize}
\item[{\rm (i)}] $\ex^{-\beta H_0} \Br 0$ w.r.t. $\mathscr{L}^2(\h)_+$ for all
	     $\beta \ge 0$.
\item[{\rm (ii)}] $V\Br 0$ w.r.t. $\mathscr{L}^2(\h)_+$. 
\end{itemize} 
Let $H=H_0-V$. We have   $\ex^{-\beta H} \Br 0$
 w.r.t. $\mathscr{L}^2(\h)_+$
 for all $\beta \ge 0$.
\end{Prop} 
{\it Proof.}
Note that 
\begin{align}
\ex^{\beta V}=\sum_{n\ge 0} \underbrace{\frac{\beta^n}{n!}}_{\ge 0}
 \underbrace{V^n}_{\Br 0} \Br 0\ \ \ \mbox{w.r.t. $\mathscr{L}^2(\h)_+$}.
\end{align} 
Thus,  by the Trotter--Kato product formula, we obtain
\begin{align}
\ex^{-\beta H}=\slim\Big(
\underbrace{\ex^{-\beta H_0/n}}_{\Br 0} \underbrace{ \ex^{\beta
 V/n}}_{\Br 0}
\Big)^n\Br 0\ \ \ \mbox{w.r.t. $\mathscr{L}^2(\h)_+$ for all $\beta \ge 0$},
\end{align} 
where $\slim$ means the strong limit.
 $\Box$

\begin{coro}\label{GeneralPP2}
Let $H_0=\mathcal{L}(A)+\mathcal{R}(A)$, where $A$ is self-adjoint and 
 bounded from below. 
Let
\begin{align}
V=\sum_{j=1}^{\infty}\mathcal{L}(B_j) \mathcal{R}(B_j),\label{VInt}
\end{align} 
where $B_j\in \mathscr{B}(\h)$  is self-adjoint and the right hand side of (\ref{VInt}) is
 a weak convergent sum. Define $H=H_0-V$. Then we obtain  
$
\ex^{ -\beta H} \Br 0
$ w.r.t. $\mathscr{L}^2(\h)_+$ for all $\beta \ge 0$.
\end{coro} 
{\it Proof.} Observe that  $\ex^{-\beta
H_0}=\mathcal{L}(\ex^{-\beta A}) \mathcal{R}(\ex^{-\beta A})\Br 0$
w.r.t. $\mathscr{L}^2(\h)_+$ for all $\beta \ge 0$.
 Since $V\Br 0$ w.r.t. $\mathscr{L}^2(\h)_+$, we obtain the desired assertion
by Proposition
 \ref{GeneralPP}. 
$\Box$
\medskip\\

The following lemma will be often useful:
\begin{lemm}\label{PSM}
Let $A_j,\ j=1, \dots, N$ be a  bounded operator acting in $\h$.
Let $M=(M_{ij})$ be a positive semidefinite $N\times N$ matrix.
Then we have 
\begin{align}
\sum_{i, j=1}^N M_{ij} \mathcal{L}(A_i^*) \mathcal{R}(A_j) \Br 0\ \
 \mbox{w.r.t. $\mathscr{L}^2(\h)_+$}. \label{PMat}
\end{align} 
\end{lemm} 
{\it Proof.}
There exists a unitary matrix $U$ such that 
$M=U^* DU$, where $D=\mathrm{diag}(\lambda_j)$ is a diagonal matrix with
$\lambda_j\ge 0$. Set $\tilde{A}_i=\sum_{j=1}^N U_{ij} A_j$. Then we see  
\begin{align}
\mbox{LHS  of (\ref{PMat})}=\sum_{j=1}^N \lambda_{j}
 \mathcal{L}(\tilde{A}_j^*) \mathcal{R}(\tilde{A}_j) \Br 0\ \
 \mbox{w.r.t. $\mathscr{L}^2(\h)_+$}.
\end{align} 
This completes the proof. $\Box$

\end{document}